\documentclass[]{aa}

\usepackage{txfonts}
\usepackage{float}
\usepackage{graphicx}
\usepackage{subfigure}
\usepackage{longtable}
\usepackage{tabularx}
\usepackage{natbib}
\usepackage{epsfig}
\bibpunct{(}{)}{;}{a}{}{,} 

\newcommand{\eg}{e{.}g{.}~}
\newcommand{\fg}{Fig{.}~}
\newcommand{\fgs}{Figs{.}~}
\newcommand{\sct}{Sect{.}~}

\newcommand{\resp}{resp{.}}
\newcommand{\degree}{\ensuremath{^\circ}}

\begin{document}


  \title{The EFIGI catalogue of 4458 nearby galaxies with detailed morphology}
  \author{A. Baillard\inst{1,2} \and E. Bertin\inst{1,2} \and 
    V. de Lapparent\inst{1,2} \and P. Fouqu\'e\inst{3,4} \and
    S. Arnouts\inst{5,6,7} \and Y. Mellier\inst{1,2} \and 
    R. Pell\'o\inst{3,4} \and J.-F. Leborgne\inst{3,4} \and 
    P. Prugniel\inst{8,9,10} \and D. Makarov\inst{8,9,10,11} \and L. Makarova\inst{8,9,10,11} \and
    H.J. McCracken\inst{1,2} \and A. Bijaoui\inst{12,13} \and
    L. Tasca\inst{5,6}}

  \institute{Universit\'e Pierre et Marie Curie-Paris 6, Institut d'Astrophysique de Paris, 98 bis Boulevard Arago, F-75014 Paris, France \and
    CNRS, UMR 7095, Institut d'Astrophysique de Paris, 98 bis Boulevard Arago, F-75014 Paris, France \and
    Universit\'e Paul Sabatier-Toulouse III, Laboratoire d'Astrophysique de Toulouse Tarbes, 14 Avenue Edouard Belin, F-31400 Toulouse, France \and
    CNRS UMR 5572, Laboratoire d'Astrophysique de Toulouse Tarbes, 14 Avenue Edouard Belin, F-31400 Toulouse, France \and
    Universit\'e de Provence Aix-Marseille 1, Laboratoire d'Astrophysique de Marseille, Traverse du Siphon-Les trois Lucs, BP 8, F-13376 Marseille Cedex 12, France \and
    CNRS UMR 6110, Laboratoire d'Astrophysique de Marseille, Traverse du Siphon-Les trois Lucs, BP 8, F-13376 Marseille Cedex 12, France \and
    Canada-France-Hawaii Telescope Corporation, Kamuela, HI-96743, USA \and
    Universit\'e Claude Bernard Lyon 1, Centre de Recherche Astronomique de Lyon, Observatoire de Lyon, St. Genis Laval, F-69561, France \and
    CNRS UMR 5574, Centre de Recherche Astronomique de Lyon, Observatoire de Lyon, St. Genis Laval, F-69561, France \and
    ENS-L, Ecole Normale Sup\'erieure de Lyon, Centre de Recherche Astronomique de Lyon, Observatoire de Lyon, St. Genis Laval, F-69561, France \and
    Special Astrophysical Observatory, Nizhnij Arkhyz, Zelenchukskij region, Karachai-Cirkassian Republic, Russia 369167 \and
    Universit\'e de Nice Sophia-Antipolis, Observatoire de la C\^ote d'Azur, BP 4229, F-06304 Nice Cedex 4, France \and
    CNRS UMR 6202, Observatoire de la C\^ote d'Azur, BP 4229, F-06304 Nice Cedex 4, France}

  \date{Received December 31, 2010; accepted March 22, 2011}
  
  \abstract{Now that modern imaging surveys have produced large
    databases of galaxy images advanced morphological studies have
    become possible. This has driven the need for well-defined
    calibration samples.}  {We present the EFIGI catalogue, a
    multi-wavelength database specifically designed to densely sample
    all Hubble types. The catalogue merges data from standard surveys
    and catalogues (Principal Galaxy Catalogue, Sloan Digital Sky
    Survey, Value-Added Galaxy Catalogue, HyperLeda, and the NASA
    Extragalactic Database) and provides detailed morphological
    information.}{Imaging data were obtained from the SDSS DR4 in the
    \textit{u}, \textit{g}, \textit{r}, \textit{i}, and \textit{z}
    bands for a sample of 4458 PGC galaxies, whereas photometric and
    spectroscopic data were obtained from the SDSS DR5
    catalogue. Point-spread function models were derived in all five
    bands. Composite colour images of all objects were visually
    examined by a group of astronomers, and galaxies were assigned
    positions in the Hubble sequence and classified according to 16
    morphological attributes describing their structure, texture,
    environment and appearance on a five-level scale.}  {The EFIGI
    Hubble sequence is in remarkably good agreement with the RC3
    Revised Hubble Sequence. The main characteristics and reliability
    of the catalogue are examined, including photometric completeness,
    type mix, systematic trends and correlations.}  {The final EFIGI
    database is a large sub-sample of the local Universe which densely
    samples Sd, Sdm, Sm and Im types compared to magnitude-limited
    catalogues. We estimate that the photometric catalogue is more
    than $\approx 80$\% complete for galaxies with $10<g<14$. More
    than 99.5\% of EFIGI galaxies have known redshifts in the
    HyperLeda and NED databases.}


  \maketitle
  \section{Introduction}
 
  The morphological analysis of galaxies is a problem of interest in
  astronomy because it provides us with important clues concerning the
  evolutionary processes of these objects \citep{gray}.  Morphology is
  strongly correlated with physical properties
  \citep{bertinlin,colpi,seigar} and morphological classification is
  often considered as a convenient way to distinguish between galaxies
  that have different physical properties \citep[see][ for a recent
  review]{buta11}.  The first widely used system was Hubble's ``tuning
  fork'' \citep{hubble}, where galaxies are assigned a visual
  classification (elliptical, lenticular, spiral or irregular). Among
  other improvements, the revised Hubble system \citep[RHS,
  ][]{devaucouleurs5} added a numerical ``stage'' $-6\le T\le +10$
  which allows intermediate states to be considered.
  
  Apart from the issue of classification schemes, many properties of
  galaxies have been explored to describe morphology.  The
  bulge-to-disk ratio \citep{dejong} and the degree of azimuthal
  variation of surface brightness are often used as discriminant
  parameters along the Hubble sequence. Other features such as rings,
  dust, or colour are also useful for understanding how galaxies
  evolve \citep{buta96}.  Some analyses based on particular features
  (\eg \citealt{naim}, \citealt{buta06}) have shown that the Hubble
  type is not sufficient for describing all galaxy
  properties. Complementary descriptors of galaxy shapes are
  particularly important at redshifts $\ga 1$, where a higher fraction
  of objects appear highly distorted and cannot be easily classified
  according to the Hubble scheme in place at lower redshifts
  \citep{abraham}.

  The volume of well-resolved galaxy images provided by modern imaging
  surveys such as the Sloan Digital Sky Survey (SDSS) or the
  CFHTLS\footnote{{\tt http://cfht.hawaii.edu/Science/CFHTLS/}}
  (Canada-France-Hawaii Telescope Legacy Survey) is too large to be
  analysed by eye by individual astronomers, and must be carried out
  by automatic software tools or through {\it crowdsourcing} initiatives
  like the ``Galaxy Zoo'' project \citep{lintott}. Whether crowdsourcing
  or automatic morphometry techniques are used, a well-defined
  calibration set is essential.

  This paper describes the EFIGI catalogue, a catalogue of 4458
  galaxies with digital images and detailed morphological
  information. The aim of the EFIGI\footnote{http://www.efigi.org}
  (``Extraction de Formes Id\'ealis\'ees de Galaxies en Imagerie'')
  project is to build automated morphometry measurement and
  classification systems operating on resolved galaxy images. The
  direct motivation for creating the EFIGI catalogue is the need for
  suitable morphological data to train supervised learning
  machines. The EFIGI morphological description therefore includes the
  Hubble Type and 16 attributes estimated by eye, which are
  specifically designed to describe the various components of a
  galaxy: bulge, arms and other dynamical features (bars, rings);
  textures (dust, flocculence, hotspots, etc.); inclination along the
  line of sight; and environment (contamination, multiplicity). We
  also provide FITS images, Point-Spread Function (hereafter PSF)
  estimates, and spectroscopic data extracted from existing
  catalogues.

  This paper is organised as follows: Section \ref{sec:data}
  introduces the data, the catalogues used as a reference and the
  whole compilation process. In Section \ref{sec:cfigi}, we describe
  in detail the morphological sequence and the various attributes and
  check for the absence of obvious trends in the final sample, and
  study correlations among attributes.  Section \ref{sec:psf} presents
  the PSF estimates obtained with our software.  Section
  \ref{sec:coverage} gives an overview of the properties of the
  resulting catalogue in terms of sky coverage, clustering,
  photometric completeness, morphological fractions and redshift
  distribution. Finally, Section \ref{sec:nair} compares the EFIGI
  catalogue with the recent morphological catalogue by \citet{nair}.
  A summary and an prospects for future uses are finally addressed in
  Section \ref{sec:conclusion}.

  \section{\label{sec:data}Data and selection process}

  Data are extracted from the Third Reference Catalogue of Bright Galaxies
  (RC3), the Principal Galaxy Catalogue (PGC), the
  Sloan Digital Sky Survey (SDSS), the New York University Value-Added
  Galaxy Catalogue (NYU-VAGC), HyperLeda, and the NASA Extragalactic
  Database (NED).
  
  \subsection{\label{chap:selproc}Master list}

  The master list of EFIGI catalogue galaxies is a subset of the Third
  Reference Catalogue of Bright Galaxies
  \citep[RC3,][]{devaucouleurs}. We extract from the RC3 15 columns
  (see Table \ref{PGCfields}), including the sky coordinates, position
  angle, $R_{25}$ aspect ratio, $D_{25}$ diameter\footnote{The
    description of all catalogue fields is given in Appendix
    \ref{apx:tables}.}, Hubble type $T$, and the associated error
  $e_T$. Galaxies are designated by their PGC \citep{paturel}
  number. Since 1995, the RC3 has been enlarged but neither renamed
  nor revised. Consequently, it is reasonably complete for galaxies
  with an apparent diameter larger than 1 arcmin at the $\mu_B =
  25$~mag.arcsec$^{-2}$ isophotal level, a total B-band magnitude
  ($B_T$) brighter than about 15.5, and a recession velocity lower
  than 15,000 km s$^{-1}$ \citep{paturel2}. In addition to objects
  already listed in the RC2 \citep{rc2}, the RC3 includes objects of
  special interest, such as compact galaxies smaller than 1 arcmin or
  fainter than magnitude 15.5.  The number of RC3 objects meeting both
  diameter and magnitude conditions is 11,897. Adding objects meeting
  only the diameter or the magnitude criterion, and objects of
  interest smaller than 1 arcmin, fainter than 15.5, or with a
  velocity $>$ 15,000 km s$^{-1}$, brings the total number of galaxies
  to 23,011.

  A column of particular interest is this work is the error estimate
  on the Hubble type ($e_T$), which, when present, indicates that a
  galaxy has been classified by several astronomers. To ensure that
  the Hubble stages $T$ indicated in our master list are sufficiently
  reliable, the EFIGI catalogue contains only RC3 objects with known
  $e_T$, or with a Hubble type 90 (non-Magellanic irregulars).

  \subsection{Imaging data\label{chap:sdssima}}

  The fourth release of the SDSS \citep[DR4,][]{adelman} provides the
  imaging material for the EFIGI database. The SDSS DR4 offers
  \textit{ugriz} CCD imaging for about $6851$ deg$^{2}$ of the
  Northern Galactic sky complemented with spectroscopy for galaxies
  with $r < 17.77$ magnitude over $4681$ deg$^{2}$ .  The EFIGI
  catalogue contains only galaxies for which there is imaging in all 5
  bands in the SDSS DR4.

  For each  galaxy, up to nine DR4 survey (``{\tt fpC}'') frames
  closest to the coordinates are obtained directly from the SDSS DAS
  server\footnote{\tt http://das.sdss.org} in all five \textit{u}, \textit{g}, 
  \textit{r}, \textit{i} and \textit{z} bands.  
  The original catalogue
  coordinates often appear to be too imprecise for unambiguous
  identification, hence for each PGC galaxy satisfying the criteria
  listed in \sct\ref{chap:selproc}, better J2000 coordinates are
  obtained through the CDS Sesame name resolver service
  \citep{bonnarel,schaaff}. We corrected the coordinates of 47 galaxies
  that appear to be off by more than $\sim 10$ arcseconds, based on the
  SDSS images (the corrected coordinates are listed in Table \ref{EFIGIfields}).
  Finally, we exclude a few objects, for which no non-ambiguous
  match can be made, and we only keep one galaxy in pairs of identical objects with
  a different PGC name. 

  Each set of {\tt fpC} images is automatically background-subtracted,
  rescaled, and combined using the {\sc SWarp} software
  \citep{bertinswarp}. SDSS images are properly sampled, which allows us to
  use a Lanczos3 interpolant to minimise resampling artifacts. Two sets of data
  are produced at this stage: the first data set is meant for visual inspection;
  the second one, with a slightly larger field of view, contains the actual
  EFIGI science images released in FITS format.
  For the ``visual'' dataset, galaxy images are scaled and framed to
  255x255 pixels in such a way that the RC3 blue isophote at $\mu = 25$ mag
  arcsec$^{-2}$ fits exactly
  onto the central area of 169x169 pixels. This is done by setting the
  output pixel scale to $9\times {\rm dexp}\,(\log{D_{25}}) / 256$ arcsec, where
  $\log{D_{25}}$ is taken from the RC3 catalogue. Science dataset images have
  the same number of pixels, but the angular pixel scale is 33\% larger pixel
  on each axis. We do not apply any flux
  correction after rescaling, as our goal is to build a collection of
  galaxy images with  aconstant zero-point in apparent surface brightness in each
  band.  We check all images by eye, and discard from the sample all
  galaxies for which at least one of the visual dataset images is completely
  ruined by artifacts or missing data, ending up with a final EFIGI selection
  of 4458 galaxies. Galaxy images partly
  contaminated by bright stars or satellite trails are intentionally kept
  to provide a realistic sampling of survey conditions. 

  Finally, all \textit{i}, \textit{r}, and \textit{g} frames are use
  to make composite ``true colour'' RGB images in PNG format with the
  {\sc STIFF} software\footnote{\tt
    http://astromatic.net/software/stiff}, using the same intensity
  mapping parameters for all RGB images. In order to provide an
  optimal rendering of objects under typical screen viewing
  conditions, a gamma correction of 1.3 is applied to the luminance
  component (supplementing the regular RGB gamma of 2.2), and colour
  saturation is exaggerated by a factor 2.0.

  \subsection{\label{chap:SDSS}SDSS photometric and spectroscopic cross-identifications}

  Identifying EFIGI galaxies in the SDSS catalogue allows us to find a
  unique identifier necessary to unambiguously extract both
  photometric and spectroscopic data from any existing or future
  release of the SDSS catalogues. Although the EFIGI imaging data come
  from SDSS DR4, SDSS photometric and spectroscopic catalogue data are
  extracted from the DR5.

  At the start of the project, we matched the positions of the objects
  by querying the SDSS DR5
  database\footnote{http://cas.sdss.org/astrodr5/en/tools/search/IQS.asp}
  for all objects within a radius of 0.1 arcmin from the list of 4458
  corrected coordinates. 3942 objects IDs were found in the DR5
  photometric data. To complete the set, a manual match was made using
  the SDSS DR5 finding tool and 423 additional objects are found. The
  automatic search often fails because of deblending problems which
  results in the galactic nucleus being classified as a star. The SDSS
  detection is sometimes located far from the galaxy
  barycentre and tens of other objects (real of false sources) are
  closer and thus detected first. In the end, 93 objects were left
  without DR5 photometric data. We used the corrected coordinates of
  these objects to locate them within the DR7 using the ``Navigate''
  SDSS visual tool
  \footnote{http://cas.sdss.org/astrodr7/en/tools/chart/navi.asp}
  which allowed us to find additional photometric identifications for
  37 galaxies.  The remaining 56 galaxies without SDSS photometric
  data are all sources located outside the sky region scanned by the
  SDSS pipeline.

  The SDSS photometric identifications allow one to retrieve the
  airmasses, zero points ($aa$) and extinction coefficients ($kk$),
  which are necessary to correct flux measurements.  We also use the
  unique SDSS object ID to perform SQL queries on the SDSS
  DR7 \textit{SpecObj} table; this yields 3136 EFIGI galaxies with
  spectroscopic data from the SDSS DR7. The SDSS retrieved fields are
  listed in Table \ref{SDSSfields}.

  \subsection{VAGC}

  The NYU-VAGC \citep{blanton} is a cross-matched collection of galaxy
  catalogues extracted from the SDSS, which includes carefully
  constructed large-scale galaxy structure samples.  We use the
  version based on the SDSS DR4.  The most useful component of the
  catalogue for our purposes is the low-redshift sample ($10 < d <
  150$ Mpc h$^{-1}$), which provides redshifts corrected for the Local
  Group motion, and k-corrections in all SDSS \textit{ugriz} bands
  \citep{blanton07}.  Because the VAGC is extracted from the SDSS,
  objects can easily be cross-identified in both catalogues using
  \textit{plate, fiberID} and \textit{MJD} fields.

  \subsection{HyperLeda}

  HyperLeda \citep{paturel2} is a merging of several catalogues, and
  is originally based on the PGC. It also provides  morphological data
  (bar, ring and compactness).
  Cross-identification with HyperLeda is simply done using the PGC
  number. All EFIGI objects are found. The columns retrieved from HyperLeda are
  listed in Table \ref{Hyperledafields}. Particularly useful are velocities 
  corrected for the Virgocentric infall of the Local Group. All other names for the PGC 
  objects are also extracted from HyperLeda, in which 50 different catalogues are
  referenced. 

  \subsection{NED}

  NED is a database obtained by merging data from many sources. Its
  most appealing features for the EFIGI catalogue are redshift measurements and cluster
  searches (see \sct\ref{chap:clusters}). Searching NED using PGC names yields a total of
  4404 objects with known redshifts (see Table \ref{NEDfields}).

\begin{table*}
    \caption{\label{hubbletype}The EFIGI Hubble Sequence (EHS). The
      left column lists the class (elliptical, lenticular, spiral,
      irregular, Dwarf) and the second column the intermediate
      stage within each class. The third and fourth columns give
      respectively the literal type and the EHS code for each
      corresponding EFIGI type. The last column briefly describes the class.}
  \begin{tabularx}{\textwidth}{p{15mm}p{15mm}p{8mm}p{8mm}X}
    \hline \hline
    Class & Stage & Literal type & EHS type & Description\\
    \hline
    Ellipticals & & & & Ellipse or sphere. Structureless, smooth intensity distribution with relatively steep gradient.\\
    \hline
    Elliptical & Compact & cE & -6 & Compact elliptical.\\
    Elliptical & 0-6 & E & -5 & More or less elongated.\\
    Elliptical & cD & cD & -4 & Giant elliptical. Sharp central profile and very extended 
    low surface brightness halo.\\
    \hline
    Lenticulars & & & & Spheroidal bulge and disk but no visible spiral arms in the disk.\\
    \hline
    Lenticular & Early & S0$^-$ & -3 & Dominant bulge, no sign of structure in disk nor dust.\\
    Lenticular & Intermediate & S0$^0$ & -2 & Some structure in disk but no arms, low
    amounts of dust.\\
    Lenticular & Late & S0$^+$ & -1 & Clear structure in disk but no arms, thin 
    dust lanes.\\
    \hline
    Spirals & & & & Central bulge and disk with spiral arms. May harbour a bar.\\
    \hline
    Spiral & 0/a & S0/a & 0 & Very tightly wound arms, very prominent bulge, low amounts of dust\\
    Spiral & a & Sa & 1 & Tightly wound arms, very prominent bulge, low amounts of dust\\
    Spiral & ab & Sab & 2 & Quite tightly wound arms, prominent bulge, low amounts of dust\\
    Spiral & b & Sb & 3 & Quite tightly wound arms, prominent bulge, strong dust lanes\\
    Spiral & bc & Sbc & 4 & Quite loosely wound arms, medium bulge, dust lanes\\
    Spiral & c & Sc & 5 & Grand design spiral, fairly weak bulge, dust lanes\\
    Spiral & cd & Scd & 6 & Loosely wound and weak arms, weak bulge, scattered dust\\
    Spiral & d & Sd & 7 & Loosely wound and very weak arms, weak bulge, scattered dust\\
    Spiral & dm & Sdm & 8 & Very loosely wound arms, very weak bulge, low amounts of dust\\
    Spiral & m & Sm & 9 & Some indication of spiral arms, very weak bulge, low amounts of dust\\
    \hline Irregular & Magellanic & Im & 10 & No arms, no
    bulge. Irregular
    profile. Low surface brightness. May host a bar.\\
    \hline
    Dwarf & Dwarf spheroid elliptical & dE & 11 & Regular low surface brightness profile,
    no arms. May contain a tight nucleus.\\
    \hline
  \end{tabularx}
\end{table*}

  \subsection{EFIGI catalogue}

  To summarise, the EFIGI catalogue is composed of 4458 galaxies with:

  \begin{itemize}
  \item PGC data
  \item SDDS images in 5 bands and object identifiers
  \item HyperLeda data
  \item NED data
  \item VAGC data (incomplete)
  \end{itemize}

  Among these objects:
  \begin{itemize}
  \item 4402 have SDSS DR7 photometric catalogue data
  \item 3136 have SDSS DR7 spectroscopic catalogue data
  \item 4415 have redshifts from HyperLeda
  \item 4404 have redshifts from NED
  \item 1729 have NYU-VAGC low-z data
  \end{itemize}
  Comparison of the redshifts of each object among the different
  catalogues allows us to find some rare erroneous redshifts in the
  HyperLeda, NED et SDSS catalogues (see \citealt{lapparent1}).

  \section{\label{sec:cfigi}The EFIGI morphological description}

  \subsection{Morphological sequence}

  The EFIGI morphological classification is defined by slightly
  modifying the RC3 Hubble classification: the main galaxy sequence is
  identical to that of the RC3, but peculiar galaxies and special
  features are no longer considered as separate stages. The resulting
  classification, which we call the EFIGI Hubble Sequence (EHS) is
  summarised in Table \ref{hubbletype}.

  There are many similarities between the RC3 and the EHS.  Compact
  elliptical galaxies are rare and difficult to distinguish, and cD
  galaxies exhibit a more peaky light profile with more diffuse wings
  than common elliptical galaxies. Lenticular galaxies usually look
  more elongated than elliptical galaxies and their Bulge/Total flux
  ratio (\textit{B/T} hereafter) decreases going from S0$^-$ to
  S0$^+$.  Among spiral galaxies, Sa types have the highest
  \textit{B/T} whereas Sm have no bulge. Also, Sa arms are the most
  tightly wound and Sm arms are the most open. The major distinction
  between Sm and Im galaxies is that the former have indications of a
  spiral arm pattern that is not visible in the latter.

  Table \ref{hubbletype} shows that the one difference between the
  RC3 and EHS is that the RC3 elliptical galaxy stages 0 to 6,
  indicating the elongation of the elliptical galaxy, are not specified
  in the EHS.  Other differences are for irregular and dwarf
  elliptical galaxies. Whereas the RC3 Magellanic irregular type 
  matches the Im type in the EFIGI classification sequence, the RC3 
  includes the additional class of non-Magellanic irregular galaxies 
  (type 90, literal type I0) that is not used in the EHS. These objects are 
  considered as a ``perturbed'' version of some ``regular'' stage, whose amount of
  perturbation is measured by the {\tt perturbation} attribute (see next
  sub-section).

  Moreover, in contrast to the RC3, the EFIGI dwarf elliptical
  galaxies are kept in a separate class from ellipticals.  Dwarf
  ellipticals have smooth elliptical isophotes reminiscent of early
  type galaxies but their bluer colours, lower surface brightness and
  closer to exponential profiles are more typical of later type
  galaxies \citep{ferguson94,kormendy2}; many of these galaxies are
  also nucleated \citep{binggeli91}. This class also includes dwarf
  lenticular galaxies (dS0), with indications of a lens or a bar
  feature \citep{sandage84,binggeli91}.  We also include in the dE
  class dwarf spheroidal galaxies (dSph), the lowest luminosity
  galaxies known, and which have a flatter surface brightness profile
  than dE galaxies, and less regular shapes.  \citet{kormendy09} also
  merge all these objects into the single class of ``Spheroidal''
  galaxies.

  Finally, galaxies that have some abnormality in shape, size, or
  content which sets them apart from the normal ellipticals, spirals,
  irregulars and dwarf ellipticals are said to be
  ``peculiar'' in the RHS. Peculiar galaxies often result from galaxy
  interactions or galaxy mergers, and thus may have distorted isophotes.
  They may also show some other distinctive
  feature such as jets emerging from the nucleus, or unusual amounts of
  dust. In the RHS, they are denoted by the
  addition of ``p'' or ``pec'' to their main classification type and are
  classified separately (type 99). In the EFIGI classification,
  peculiarity is described by a set of attributes (see next sub-section) 
  rather than by stage.

  \subsection{Attribute definition\label{chap:att}}

  Attributes are defined primarily
  according to their ability to characterise the morphology of galaxies and
  their relevance to the physical properties of the objects.
  Attribute measurements (morphometry) must
  nevertheless be able to describe the morphological properties of the known
  types along the Hubble sequence and any deviation from those types
  \citep{roberts,baillard08}.\\

  The EFIGI attributes can be divided into six groups:
  \begin{itemize}
  \item Appearance: {\tt inclination/elongation}
  \item Environment: {\tt multiplicity}, {\tt contamination}
  \item Bulge: {\tt B/T ratio}
  \item Spiral arm properties: {\tt arm strength}, {\tt arm curvature},
	 {\tt rotation}
  \item Textural aspect: {\tt visible dust}, {\tt dust dispersion},
	{\tt flocculence}, {\tt hot spots}
  \item Dynamical features: {\tt bar length}, {\tt inner ring}, {\tt outer ring},
   {\tt pseudo-ring}, {\tt perturbation}
  \end{itemize}

  Most attributes are not binary (for instance {\tt inclination/elongation},
  {\tt B/T} or {\tt dust dispersion}), hence it is important to set a scale
  describing their strength. We choose a scale with five steps 
  (0 to 1, by steps of 0.25), which provides two
  intermediate values between the median and each of the extreme values;
  a higher number of steps would have been irrelevant and confusing
  in a context of visual identification.

  In order to facilitate comparison with automatic techniques, we
  decided to make attribute strength a monotonically increasing
  function of the fraction of flux held in each feature relative to
  that of the whole galaxy, whenever applicable ({\tt contamination},
  {\tt arm strength}, {\tt hot spots}, {\tt inner ring}, {\tt outer
    ring}, {\tt pseudo-ring}); it is also evidently the case for {\tt
    B/T}.  Flux fractions are visually estimated from the composite
  \textit{irg} 3-colour image, and the function may not be linear. For
  attributes {\tt B/T}, {\tt inclination/elongation}, {\tt
    multiplicity}, and {\tt rotation}, the strength scale ranges
  between the extreme possible values.  For all other attributes, the
  strength scale ranges between the most extreme cases found in the
  EFIGI catalogue.

  For each attribute of each galaxy, a 70\% confidence interval is also
  estimated by setting a lower and upper limit among the five possible
  attribute values of 0, 0.25, 0.5, 0.75 and 1.0. The narrowest possible
  confidence interval is 0.25, the widest is 1.0. When an attribute is
  not defined for a given galaxy, it is set to 0.5 with lower and upper
  confidence limits at 0 and 1 respectively.

  \subsubsection{\tt Inclination/elongation}

  For galaxies with a visible disk (lenticulars and spirals), this
  attribute measures the disk inclination $f=1-cos\theta$, where
  $\theta$ is the angle between the rotation axis of the disk and the
  line-of-sight of the observer, or equivalently between the galaxy
  disk and the plane of the sky; $\theta$ then varies from 0$\degree$
  (face-on) to 90$\degree$ (edge-on). For galaxies with no evident
  disk (ellipticals, cD, cE, Im and dE), this attribute provides an
  estimate of the apparent elongation of the object, that is $f = 1 -
  b/a$ with $a$ and $b$ the apparent major and minor axes lengths
  respectively. The scale of elongation is matched to that of
  inclination so that a circular disk of a given inclination attribute
  value would have an identical elongation attribute value. Values of
  the attribute are:

  \begin{tabularx}{0.95\columnwidth}{lX}
  0 & 0$\degree$-35$\degree$ disk inclination angle (face-on) or very low elongation $f<0.2$\\
  0.25 & 35$\degree$-50$\degree$ disk inclination angle or low elongation $0.2<f<0.4$\\
  0.5 & 50$\degree$-70$\degree$ disk inclination angle or moderate elongation $0.4<f<0.7$\\
  0.75 & 70$\degree$-80$\degree$ disk inclination angle or strong elongation $0.7<f<0.8$\\
  1 & 80$\degree$-90$\degree$ disk inclination angle (edge-on) or very strong elongation $f>0.8$\\
  \end{tabularx}

  \subsubsection{\tt Multiplicity}
  This attribute quantifies the abundance of galaxies in the vicinity of the
  main galaxy. Only galaxies that have magnitudes less than that of the main
  galaxy plus $\sim 5$ mag, and whose centre lies within 0.75 $D_{25}$ from the
  image centre are counted. Values are:

  \begin{tabularx}{0.95\columnwidth}{lX}
  0 & no other galaxy\\
  0.25 & one neighbouring galaxy\\
  0.5 & two neighbouring galaxies\\
  0.75 & three neighbouring galaxies\\
  1 & four or more neighbouring galaxies\\
  \end{tabularx}

  \subsubsection{\tt Contamination}
  This attributes indicates the severity of the contamination by
  bright stars, overlapping galaxies or image artifacts (diffraction
  spikes, star halos, satellite trails, electronic defects). Values are:

  \begin{tabularx}{0.95\columnwidth}{lX}
    0 & no overlapping source visible on the galaxy isophotal footprint\\
    0.25 & only faint sources overlapping the galaxy isophotal footprint
    (negligible effect on photometry and morphometry)\\
    0.5 & overlapping sources on the galaxy isophotal footprint or faint
    gradient/background light pollution (some impact on photometry and morphometry)\\
    0.75 & bright sources overlapping the galaxy isophotal footprint or strong
    gradient/background light pollution (large impact on photometry and morphometry)\\
    1 & most of the galaxy image footprint dominated by light from a very
    bright contaminant (unreliable photometry or morphometry).\\
  \end{tabularx}

  \subsubsection{\tt Perturbation}
  This attribute measures the amplitude of distortions in the galaxy profile,
  due to tidal effects for instance. 0 indicates a 
  light profile with rotational symmetry (hence regular spiral arms
  when applicable), while 1 corresponds to the most distorted galaxies observed in
  the catalogue. Values are:

  \begin{tabularx}{0.95\columnwidth}{lX}
    0 & No distortion\\
  0.25 & Slight distortion of the object profile and/or the spiral arms \\
  0.5 & Moderate distortion of the object profile and/or the spiral arms\\
  0.75 & Strong distortion of the object profile, with profile components 
  (bulge, disk, spiral arms) still identifiable\\
  1 & Completely distorted profile, components can be barely distinguished\\
  \end{tabularx}

  \subsubsection{\tt Bulge/Total ratio}
  This attribute measures the relative contribution of the bulge
  (spheroidal) component to the total flux of the galaxy. For
  instance, Im galaxies have $B/T = 0$ and elliptical galaxies have $B/T
  = 1$. Values are:

  \begin{tabularx}{0.95\columnwidth}{lX}
  0 & no bulge\\
  0.25 & very weak bulge comprising $\sim 25\%$ of the total flux\\
  0.5 & medium bulge comprising $\sim 50\%$ of the total flux\\
  0.75 & strong bulge comprising $\sim 75\%$ of the total flux\\
  1 & all flux within bulge, no disk nor spiral arms\\
  \end{tabularx}

  \subsubsection{\tt Arm strength}
  This attribute measures the relative strength of spiral arms, in terms
  of the flux fraction relative to the whole galaxy. 0 indicates no
  visible arms, while 1 corresponds to the highest fraction observed in the
  EFIGI sample. When no spiral arms are visible (either due to the
  galaxy structure or to a high inclination that prevents any existing arm to
  be seen), this attribute is undefined. Defined values
  are:

  \begin{tabularx}{0.95\columnwidth}{lX}
  0 & Very weak or no spiral arms\\
  0.25 & weak contribution of the spiral arms to the galaxy flux\\
  0.5 & moderate contribution of the spiral arms to the galaxy flux\\
  0.75 & significant contribution of the spiral arms to the galaxy flux\\
  1 & highest contribution of the spiral arms to the galaxy flux\\
  \end{tabularx}

  \subsubsection{\tt Arm curvature}
  This attribute measures the average intrinsic curvature of the spiral
  arms, that is for the same galaxy seen face-on, as inclination
  increases the variations of the arm curvature along the arms. 
  It is subordinate to the {\tt arm strength} attribute: for all EFIGI galaxies
  for which {\tt arm strength} is undefined, and for the majority 
  of EFIGI galaxies with {\tt arm strength} equal to 0, specifically
  those with no visible spiral arms, {\tt arm curvature} has no meaning 
  and the attribute is undefined. Defined values are:

  \begin{tabularx}{0.95\columnwidth}{lX}
  0 & wide open spiral arms, with pitch angles of 40 degrees or more\\
  0.25 & open spiral arms, with pitch angles of 30 to 40 degrees\\
  0.5 & moderately open spiral arms, with pitch angles of 20 to 30 degrees\\
  0.75 & closed-in spiral arms, with pitch angles of 10 to 20 degrees\\
  1 & tightly wound spiral arms, with pitch angles of 10 degrees or
       less\\
  \end{tabularx}

  \subsubsection{\tt Rotation}
  This attribute indicates whether the spiral pattern appears to rotate
  clockwise (East of North negative) or counterclockwise (East of North
  positive). It is subordinate to the
  {\tt arm strength} attribute: for the vast majority of EFIGI galaxies 
  with {\tt arm strength} equal to 0, specifically those with no visible 
  spiral arms, {\tt rotation} has no meaning and the attribute is undefined.
  Defined values are:

  \begin{tabularx}{0.95\columnwidth}{lX}
  0 & definitely clockwise\\
  0.25 & probably clockwise\\ 
  0.5 & no preferred direction\\
  0.75 & probably counterclockwise\\
  1 & definitely counterclockwise\\
  \end{tabularx}

  \subsubsection{\tt Visible dust}
  This attribute measures the strength of features revealing the
  presence of dust: obscuration and/or diffusion of star light by a
  dust lane or molecular clouds. Values are:

  \begin{tabularx}{0.95\columnwidth}{lX}
  0 & no dust\\
  0.25 & indications of dust, but dust cannot be located\\
  0.5 & low to moderate amounts of dust, can be located\\
  0.75 & significant amounts of dust covering $< 50\%$ of the surface of the galaxy\\
  1 & high amounts of dust covering $> 50\%$ of the surface of the galaxy\\
  \end{tabularx}

  \subsubsection{\tt Dust dispersion}
  This attribute measures the ``patchiness'' of the dust distribution
  in terms of whether the dust is smoothly distributed in sharp
  lanes (or rings) or distributed in strongly irregular
  patches. This attribute is subordinate to the {\tt visible dust}
  attribute : {\tt dust dispersion} is undefined when {\tt visible dust} is equal
  to 0 or 1. Defined values are:

  \begin{tabularx}{0.95\columnwidth}{lX}
  0 & Thin lane(s) of dust with smooth outline\\
  0.25 & Thin lane(s) of dust with patchy outline\\
  0.5 & Patchy lane(s) of dust and some other small patches\\
  0.75 & Very patchy lane(s) of dust and many other patches\\
  1 & Extremely patchy distribution of the dust\\
  \end{tabularx}

  \subsubsection{\tt Flocculence}
  This attribute measures the importance of ``flocculent'' features
  due to scattered HII regions relative to the strength of spiral arms
  and the underlying smooth profile components. Values are:

  \begin{tabularx}{0.95\columnwidth}{lX}
    0 & no visible flocculence\\
    0.25 & weak/barely visible flocculence/patchiness limited to small parts of the galaxy disk\\
    0.5 & some flocculence visible in parts of the galaxy disk\\
    0.75 & significant flocculence over most of the galaxy disk\\
    1 & strong flocculence over most of the galaxy disk\\
  \end{tabularx}
	
  \subsubsection{\tt Hot spots}
  This attribute indicates the presence of ``hot spots'', that is
  regions with a very high surface brightness such as giant regions of
  star formation, active nuclei, or stellar nuclei of dwarf galaxies.
  It is not to be confused with
  scattered HII regions of normal intensity contributing to the
  ``flocculent'' aspect of some disks. Values are:

  \begin{tabularx}{0.95\columnwidth}{lX}
  0 & no hot spot\\
  0.25 & small part of the galaxy flux included in one or several hot spots\\
  0.5 & moderate part of the galaxy flux included in one or several hot spots\\
  0.75 & significant part of the galaxy flux included in one or several hot spots\\
  1 & one or several hot spots dominate the galaxy flux\\
  \end{tabularx}

  \subsubsection{\tt Bar length}
  This attribute quantifies the presence of a central bar component in
  the galaxy. Contrary to other attributes, the flux fraction may not be the
  most physically-relevant quantifier for bar strength. Another possible
  quantifier is bar elongation which is found to correlate strongly with
  bar torque in some samples \citep{block,laurikainen2}. Unfortunately,
  elongation cannot always be estimated reliably on ground-based images,
  especially for short bars which can be very thin. Bar length (relative to
  the galaxy $D_{25}$) is a more convenient quantifier, and has been chosen here
  as the criterion to quantify bar strength. This attribute is often undefined
  for highly inclined disks in which one cannot assess the presence of a bar.
  Defined values are:

  \begin{tabularx}{0.95\columnwidth}{lX}
  0 & no visible bar\\
  0.25 & short, barely visible bar feature\\
  0.5 & short bar, with a length about one third of $D_{25}$ \\
  0.75 & long bar, that extends over about half of $D_{25}$\\
  1 & very long, prominent bar that extends over more than half of $D_{25}$ \\
  \end{tabularx}

  \subsubsection{\tt Inner ring}

  This attribute measures the presence of a circular or elliptical
  ring-like overdensity that is within the disk and/or spiral arm
  pattern, and at the end of the bar when present (contrary to nuclear
  rings that occur well within the bar). This attributes also includes
  inner lenses, that is regions of nearly constant surface brightness
  with radius, as well as the inner pseudo-rings, which are not
  completely closed and more spiral-like.  Strength is proportional to
  the fraction of galaxy light contained within the ring. A value of 1
  corresponds to the highest fraction observed in the catalogue. This
  attribute is undefined for highly inclined disks in which one cannot
  assess the presence of a ring.  Defined values are:

  \begin{tabularx}{0.95\columnwidth}{lX}
  0 & no inner ring\\
  0.25 & low ring contribution to the galaxy flux\\
  0.5 & intermediate ring contribution to the galaxy flux\\
  0.75 & significant ring contribution to the galaxy flux\\
  1 & highest ring contribution to the galaxy flux\\
  \end{tabularx}

  \subsubsection{\tt Outer ring}
  This attribute measures the presence of a circular or elliptical
  ring-like over-density that lies mostly outside the disk and/or spiral arm pattern.
  This attributes also includes outer lenses, that
  is regions of nearly constant surface brightness with radius.
  Strength is proportional to the fraction of galaxy light contained within
  the ring. A value of 1 corresponds to the highest fraction
  observed in the catalogue. This attribute is undefined for highly
  inclined disks in which one cannot assess the presence of a ring.
  Defined values are:

  \begin{tabularx}{0.95\columnwidth}{lX}
  0 & no outer ring\\
  0.25 & low ring contribution to the galaxy flux\\
  0.5 & intermediate ring contribution to the galaxy flux\\
  0.75 & significant ring contribution to the galaxy flux\\
  1 & highest ring contribution to the galaxy flux\\
  \end{tabularx}

  \subsubsection{\tt Pseudo-Ring}
  This attribute measures the presence of outer pseudo-rings as defined 
  by \citet{buta96}: the $R_1^\prime$ ring, having a
  dimpled ``eight'' shape due to a 180$\degree$ winding of the spiral arms
  with respect to the end of a bar; the $R_2^\prime$ feature, with a higher winding
  of 270$\degree$ of the spiral arms with respect to the bar; and the intermediate 
  $R_1R_2^\prime$ pattern.

  Values of 0.75 and 1 are assigned to galaxies having two closed-loop 
  arms of type $R_1^\prime$, and grades the fraction of the total galaxy light that is included 
  in the pseudo-ring feature. Values of 0.25 and 0.5 are attributed to galaxies for which one or 
  both of the arms do not form a closed-loop of type $R_1^\prime$, thus including types 
  $R_2^\prime$ and $R_1R_2^\prime$ from \citet{buta96}, and also grade the 
  fraction of the total galaxy light that is included in the pseudo-ring feature. 
  This attribute is undefined for highly
  inclined disks in which one cannot assess the presence of a pseudo-ring.
  Defined values are:

  \begin{tabularx}{0.95\columnwidth}{lX} 
  0 & no visible pseudo-ring feature \\
  0.25 & $R_2^\prime$ and $R_1R_2^\prime$ pseudo-rings containing a low fraction of the galaxy flux \\
  0.5 & $R_2^\prime$ and $R_1R_2^\prime$ pseudo-rings containing a higher fraction of the galaxy flux \\
  0.75 & $R_1^\prime$ pseudo-ring feature containing a low fraction of the galaxy flux\\
  1 & $R_1^\prime$ pseudo-ring feature containing a higher fraction of the galaxy flux\\
  \end{tabularx}

 \subsection{Classification process}

  \begin{figure}
    \includegraphics[width=\columnwidth]{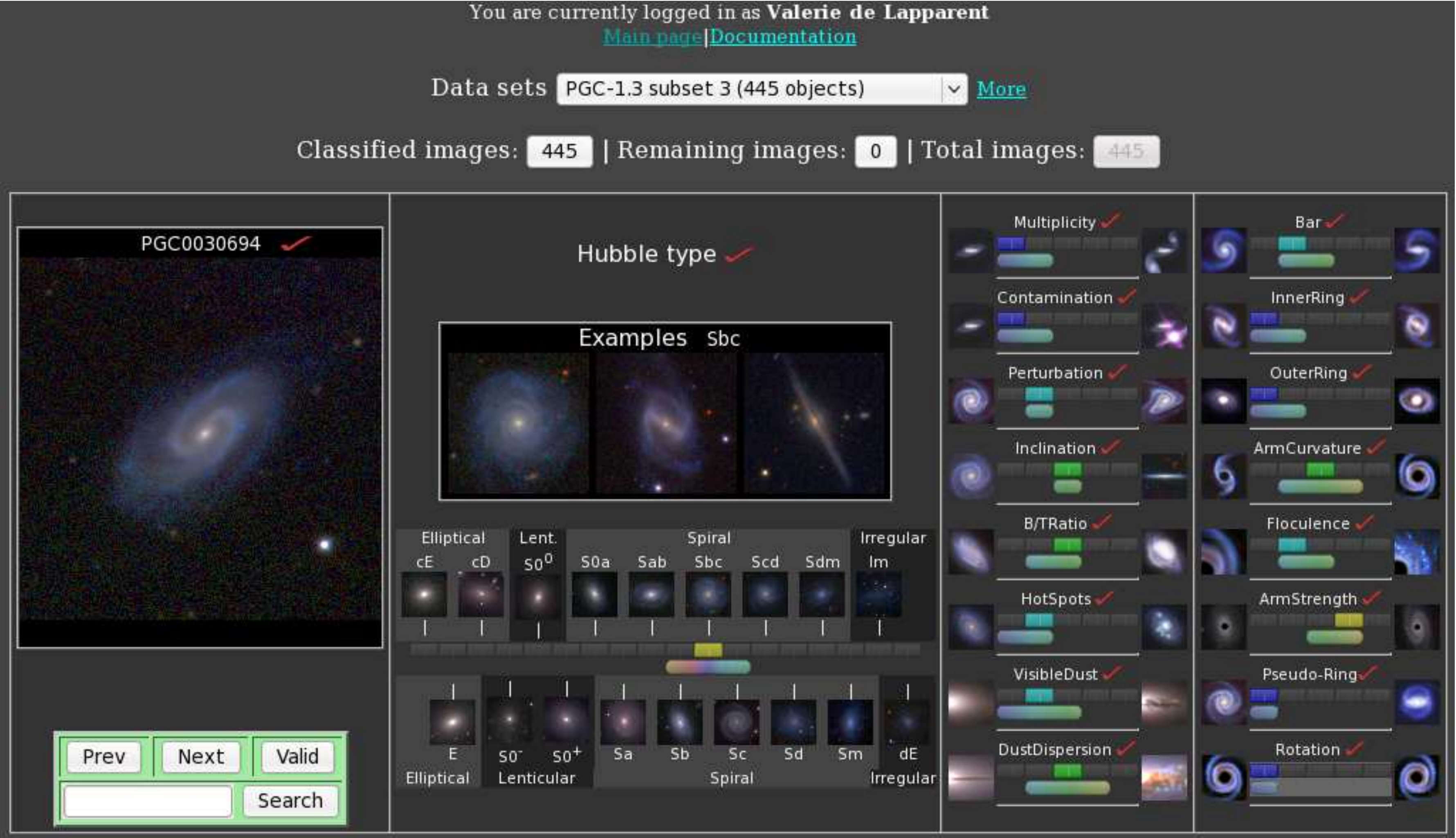}
    \caption{\label{manclass}Screenshot of \textit{Manclass}, the
      classification interface. The galaxy to be classified is shown on
      the left. The EHS sliders and three prototypes (face-on, edge-on and barred) are
      displayed in the middle. The attribute sliders are located in the rightmost
      portion of the frame. }
  \end{figure}

  In order to test the EFIGI morphological classification and
  attribute definition, a common set of 100 galaxies is extracted from
  the whole sample and classified by 11 astronomers among the authors
  of this article (S. Arnouts, A. Baillard, E. Bertin, P. Fouqu\'e,
  V. de~Lapparent, J.-F. Le~Borgne, D. Makarov, L. Makarova,
  H.J. McCracken, Y. Mellier, R. Pell\'o).  Statistics are
  automatically computed for each astronomer measuring his/her mean
  confidence, and comparing his/her classification with those of the
  other astronomers and with the RC3 morphological type. This step
  allows each astronomer to estimate his/her relative biases and
  adjust his/her classification with that of others.  Classifying this
  small sample also allows the astronomers to adjust the final EHS.
  Then the full sample is divided into 10 sub-samples of 445 galaxies
  which are all, with one exception, classified by one astronomer (one
  sample is classified by several astronomers).
  
  An interactive interface called ``Manclass'' has been designed to
  ease the visual classification process. A snapshot is shown
  \fg\ref{manclass}.  This interface displays for each galaxy the
  \textit{irg} colour image, and 17 double sliders corresponding to
  the EHS and the 16 EFIGI attributes. The interface allows the
  astronomer to define for all examined galaxies a triplet of values
  for each attribute: the mode, defined as the value considered to be
  the closest to the correct stage or attribute strength, and the two
  boundaries of the 70\% confidence interval (see \sct\ref{chap:att}),
  which is not necessarily centred on the mode. This is particularly
  useful for defining a plausible EHS interval.  For both the EHS and
  the attributes, the upper slider adjusts the mode and the lower bar
  sets the confidence interval.

  Using the \textit{irg} ``true colour'' images created from the SDSS
  imaging data, a subset of 3 images has been selected as reference
  for each EHS stage: a face-on non barred galaxy, a face-on barred
  galaxy, and an edge-on galaxy. These are shown when the mouse slides 
  across the EHS slider, and the face-on non barred galaxy image is shown
  as a fixed ``thumbnail'' next to each position of the EHS slider.
  Moreover, 2 ``true colour'' comparison images have been
  selected as templates for each extreme strength values of the
  attributes and are displayed as ``thumbnails'' at the limit of each
  corresponding slider.

\begin{figure}
    \includegraphics[angle=-90,width=\columnwidth]{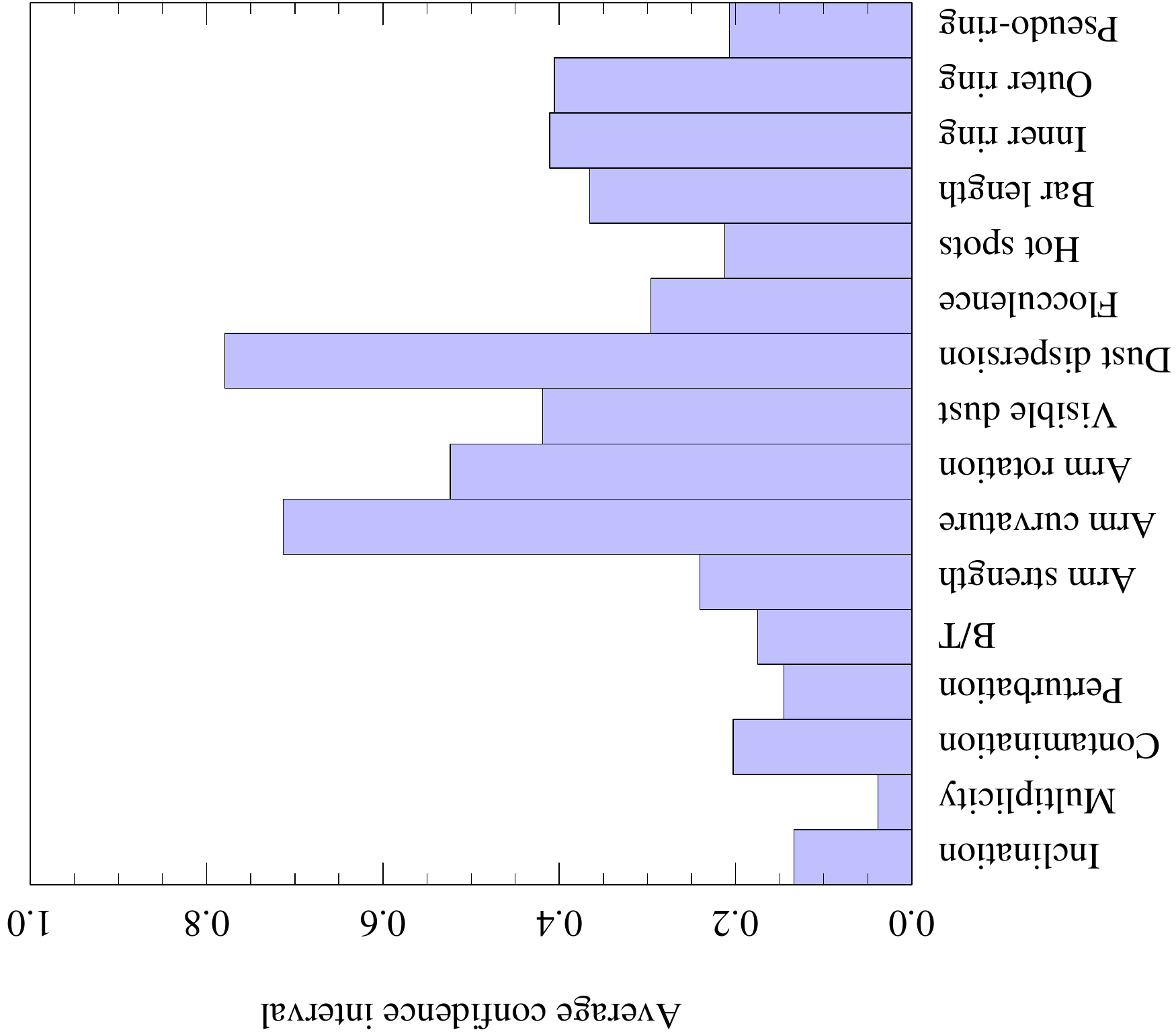}
    \caption{\label{meanu}Average confidence intervals of EFIGI attributes 
                in the homogenised data set (only
		confidence intervals $<1$ are considered).}
  \end{figure}

  Once the 4458 galaxies have been classified, a long process of visual homogenisation
  of attribute strengths takes place. Contrary to the classification of the sub-samples 
  of 445 galaxies by each astronomer, homogenisation is performed separately for each
  attribute, and by two
  astronomers: E. Bertin for {\tt multiplicity}, {\tt contamination}, {\tt B/T ratio}, 
  {\tt bar length},
  {\tt hot spots,} {\tt flocculence}, {\tt rotation} and V. de Lapparent for
  {\tt perturbation}, {\tt inclination/elongation}, {\tt visible dust}, {\tt dust dispersion}, 
  {\tt internal ring}, 
  {\tt external ring}, {\tt eight shape}, {\tt arm strength}, {\tt arm curvature} and
  the {\tt EHS}. 
  A variant of the ``Manclass'' interface is used for the homogenisation, which
  displays mosaics of galaxy images, together with the slider of the attribute
  under consideration below each image. 
  Homogenisation is particularly useful for attributes that are difficult
  to estimate. {\tt Arm Curvature} and {\tt Dust dispersion} have the
  widest confidence intervals in \fg\ref{meanu} and were the most difficult to
  grade, contrary to {\tt Multiplicity} and {\tt Inclination}, which have the
  lowest error estimates.

  The homogenisation process requires several passes through the data
  in order to first define for each attribute a well-defined scale,
  then to train the observer's eye in order to have a stable
  classification, and finally to check for biases and spurious
  correlations with other attributes.  Inter-user statistics are used
  to derive useful hints about attributes and astronomers that require
  more attention, because of a high discrepancies rate.  To this end,
  the homogenisation interface adds the possibility to display mosaics
  of image and attribute values for arbitrary intervals of as many
  attributes as desired, an option which was extensively used for
  checking the stability of each attribute versus time and versus
  other attributes.  The homogenisation process establishes the final
  and definitive classification of the EFIGI catalogue.

  \subsection{Representative images of attributes strength}

\begin{figure*}
    \includegraphics[width=1.98\columnwidth]{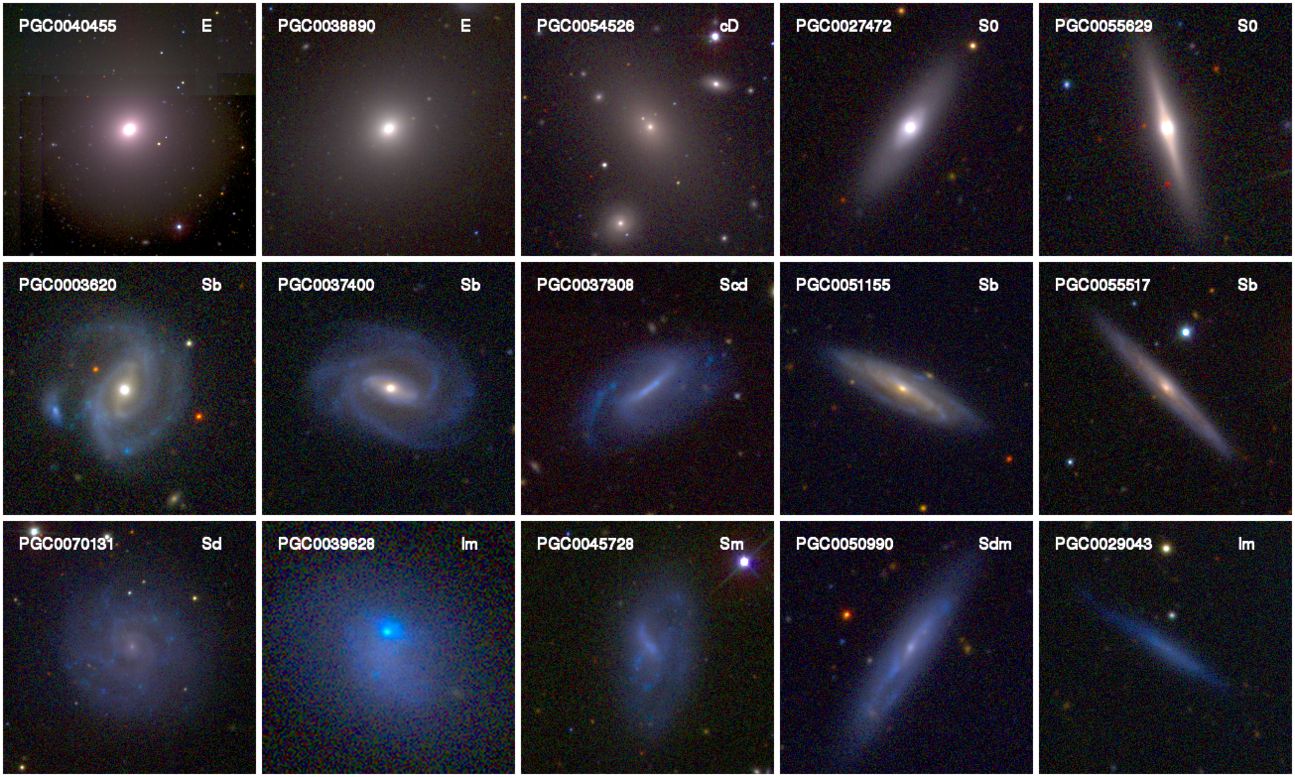}
    \caption{\label{att_Inc}Examples of EFIGI galaxies with the five
      possible values of the {\tt Inclination/Elongation} attribute: 0,
      0.25, 0.5, 0.75, and 1 from left to right. The PGC number and the EFIGI Hubble
      type are indicated inside each image. For each attribute value,
      3 types are shown: pure bulge (cE, cD, E), lenticular and S0a
      galaxies in the top row, early spirals in the middle row (from
      Sa to Scd), later spirals (Sd to Sm), Im and dE in the bottom row. Note the 
      very low surface brightness of the Scd, Sd, Im and Sm galaxies in the 
      central panel and the 2 left panels in the bottom row, typical of such 
      objects which are numerous in the EFIGI catalogue. In all images,
      north is up and east to the left.}
  \end{figure*}

\begin{figure*}
    \includegraphics[width=1.98\columnwidth]{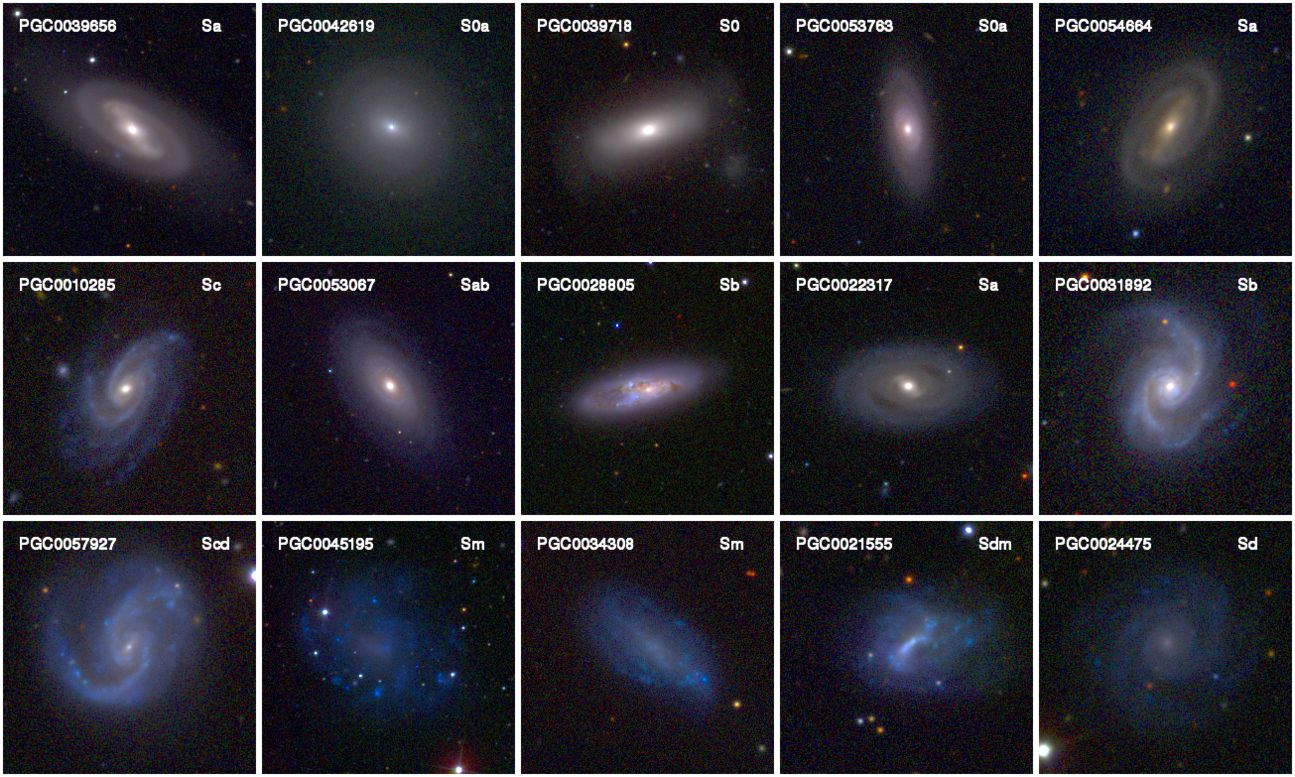}
    \caption{\label{att_Rot}Same as \fg\ref{att_Inc} for {\tt
        Rotation} attribute. The central column containing galaxies
      having arm-like features but for which no rotation direction can
      be defined contains unusual galaxies.}
  \end{figure*}

\begin{figure*}
    \includegraphics[width=2\columnwidth]{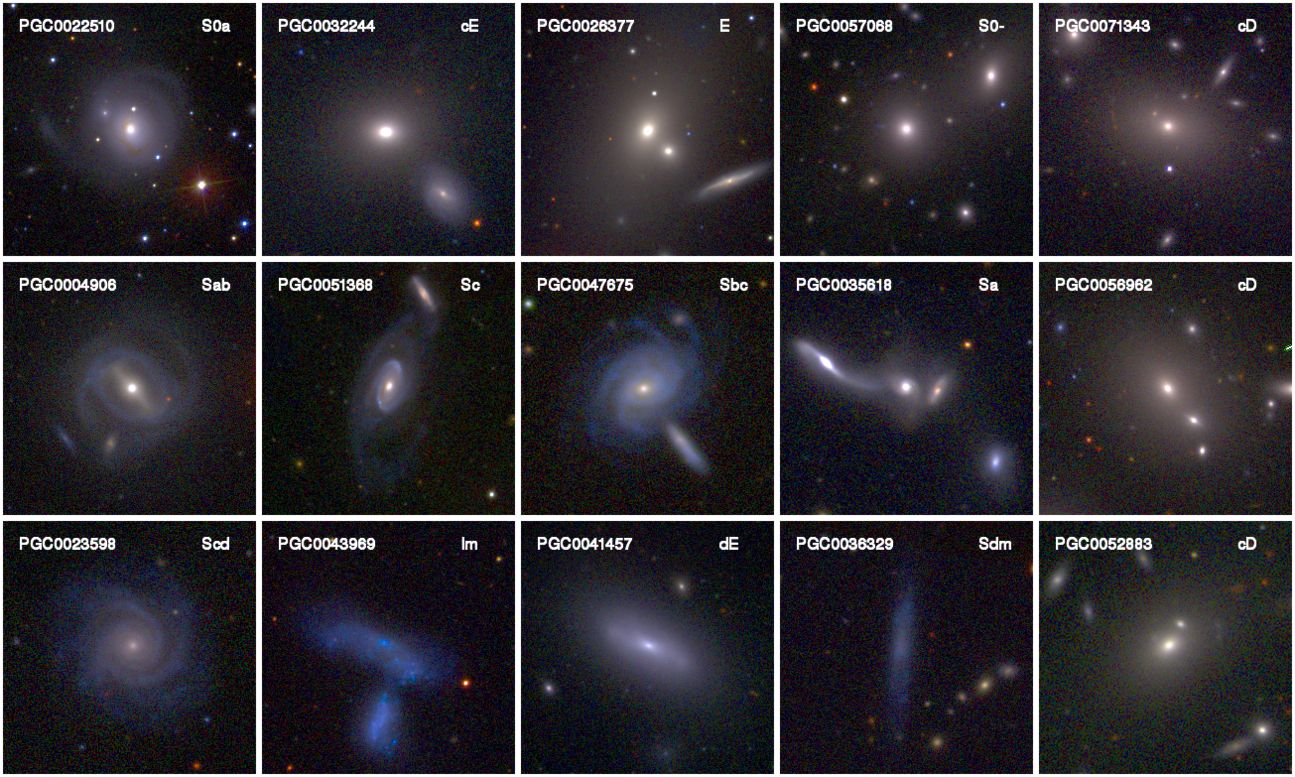}
    \caption{\label{att_Mult}Same as \fg\ref{att_Inc} for {\tt
        Multiplicity} attribute. Because EFIGI galaxies with an
      attribute value of 0.25 are all cD galaxies, we only display
      these types in the right panel of each line.  As cD galaxies
      have an extended halo and lie in dense regions of clusters, they
      have both high {\tt Multiplicity} attribute and low to moderate
      {\tt Contamination}.  Note that although PGC0004906 (left panel
      of central row) has 2 neighbouring galaxies, these are too faint
      in comparison to the central galaxy to be accounted for in the
      {\tt Multiplicity}.}
  \end{figure*}

\begin{figure*}
    \includegraphics[width=2\columnwidth]{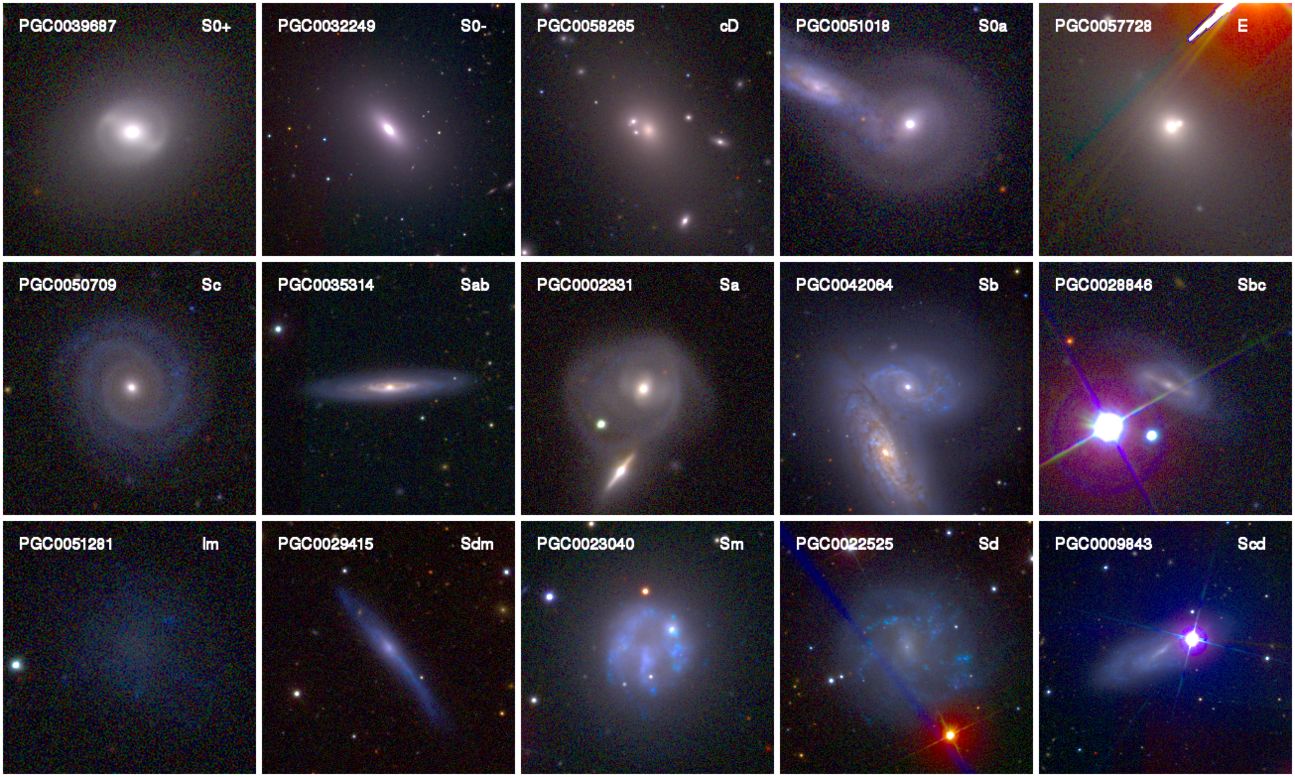}
    \caption{\label{att_Cont}Same as \fg\ref{att_Inc} for {\tt
        Contamination} attribute. Contamination can be due to
      neighbouring stars, satellite tracks, as well as interacting or
      overlapping foreground or background galaxies.}
 \end{figure*}

\begin{figure*}
    \includegraphics[width=2\columnwidth]{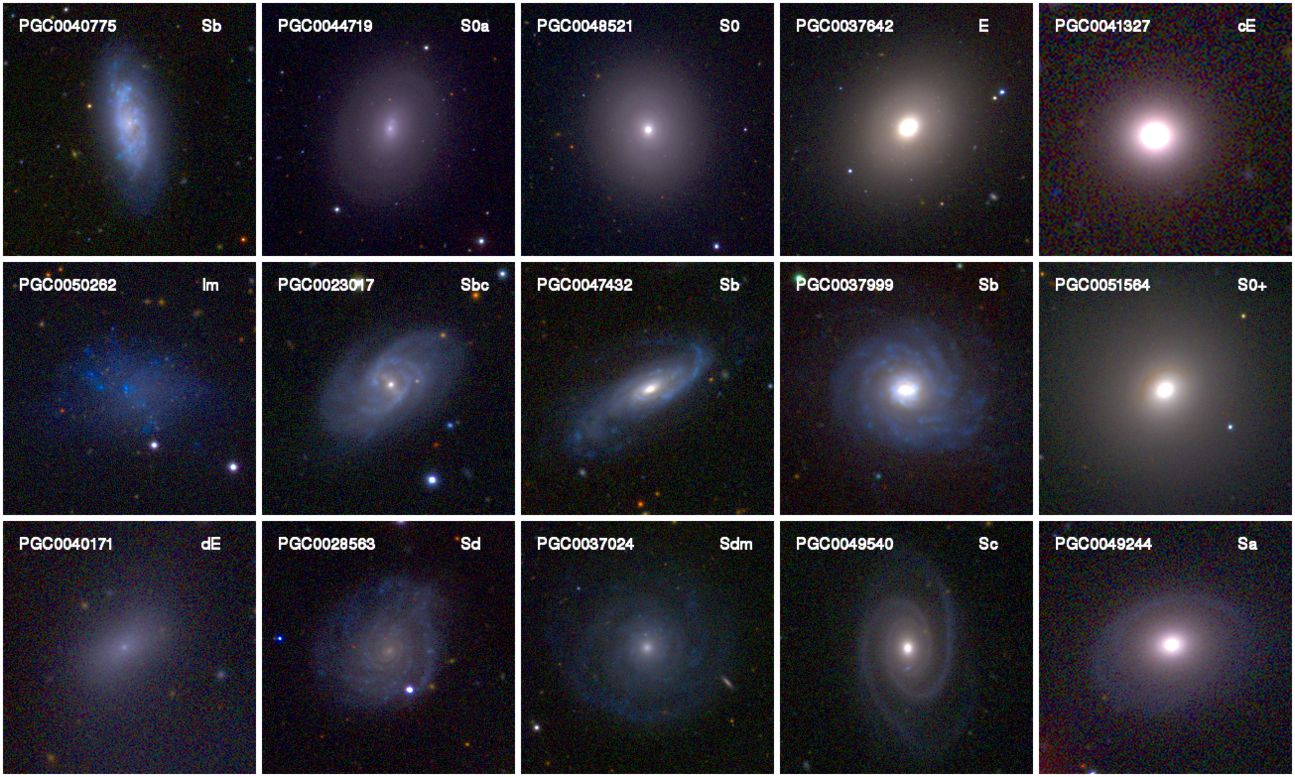}
    \caption{\label{att_BT}Same as \fg\ref{att_Inc} for {\tt
        Bulge/Total ratio} attribute. Earlier types than Scd
      are shown in the bottom row for attribute values 0.75 and 1,
      because later types have lower values of this attribute.}
  \end{figure*}

\begin{figure*}
    \includegraphics[width=2\columnwidth]{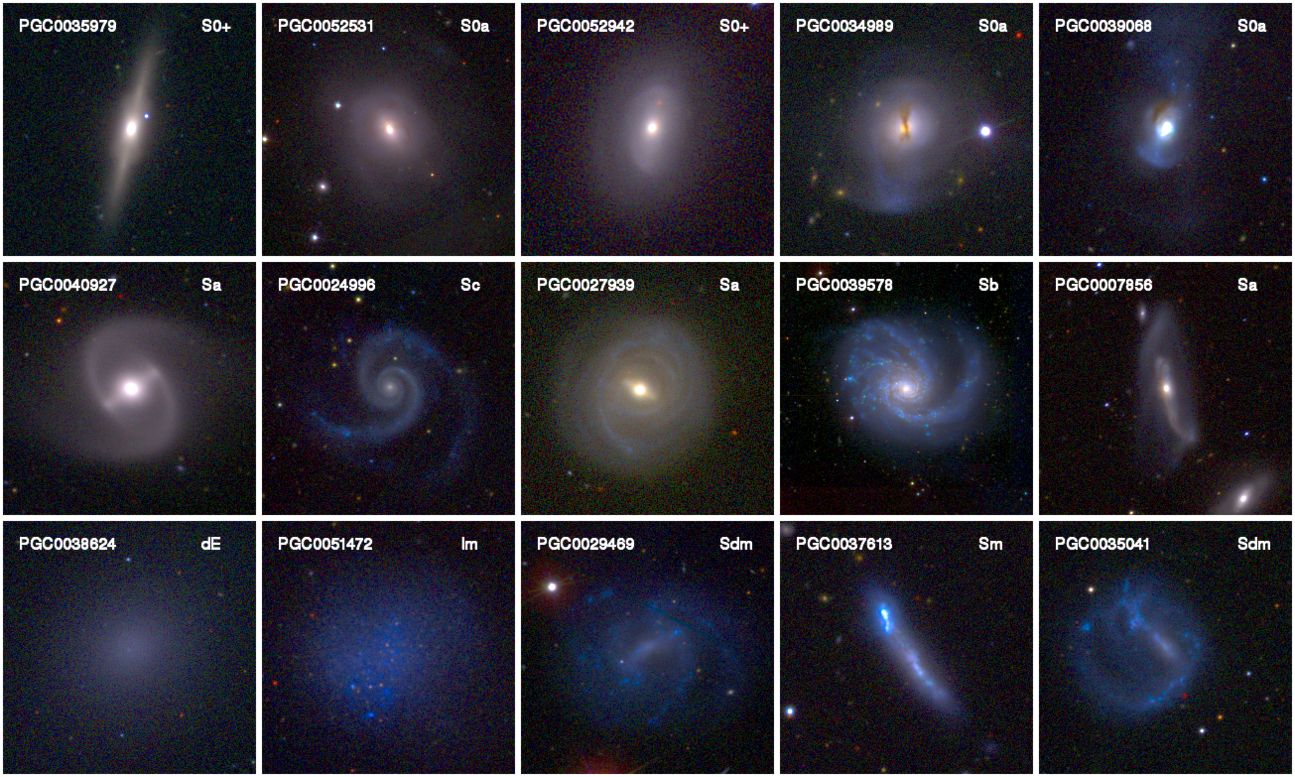}
    \caption{\label{att_Pert}Same as \fg\ref{att_Inc} for {\tt
        Perturbation} attribute. Most dE are unperturbed, whereas many
      spirals and irregulars are weakly or moderately distorted
      (attributes values 0.25 and 0.5).  Note however the extremely
      regular nature of the Sa galaxy PGC0040927 (left panel of
      central row).}
  \end{figure*}

\begin{figure*}
    \includegraphics[width=2\columnwidth]{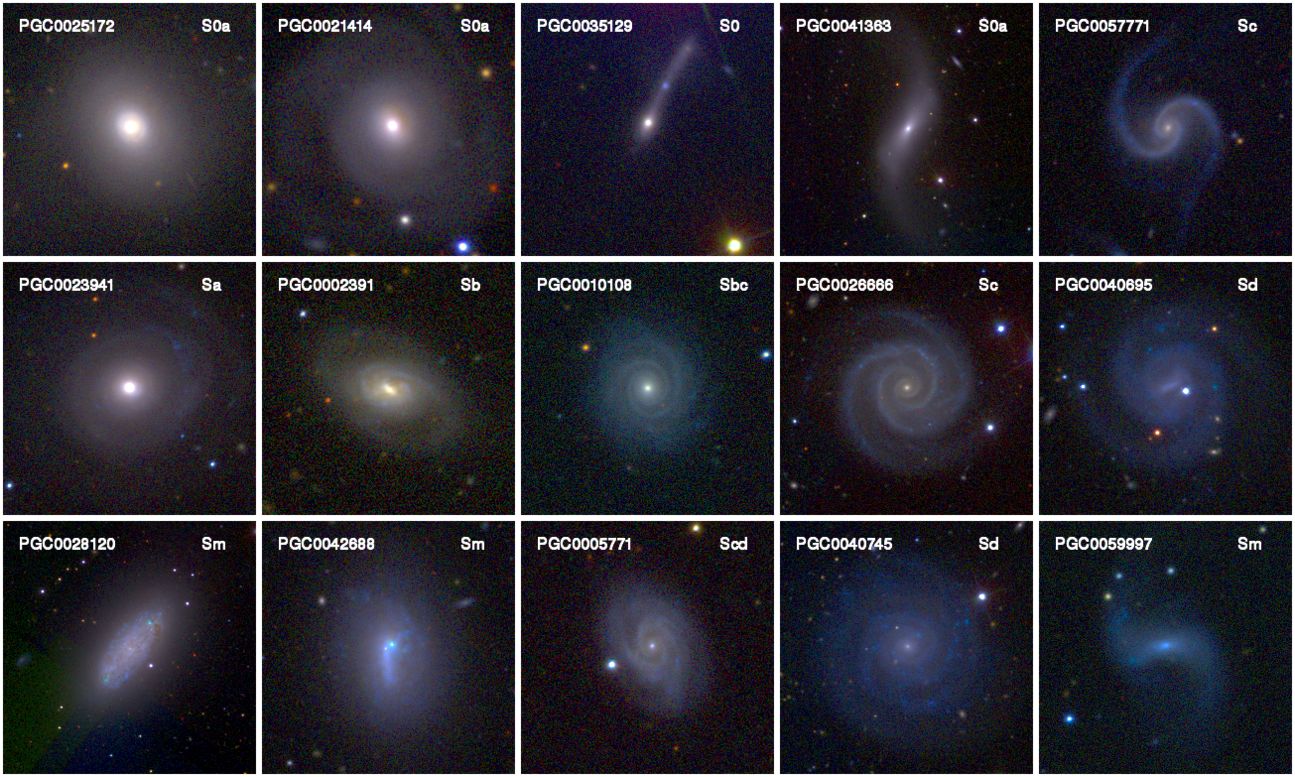}
    \caption{\label{att_Arm}Same as \fg\ref{att_Inc} for {\tt Arm
        strength} attribute. Later types than 
      \fg\ref{att_Inc} are shown in the right panels of the top and
      central row because all earlier type galaxies have values less or
      equal to 0.75 for this attribute.}
  \end{figure*}

\begin{figure*}
    \includegraphics[width=2\columnwidth]{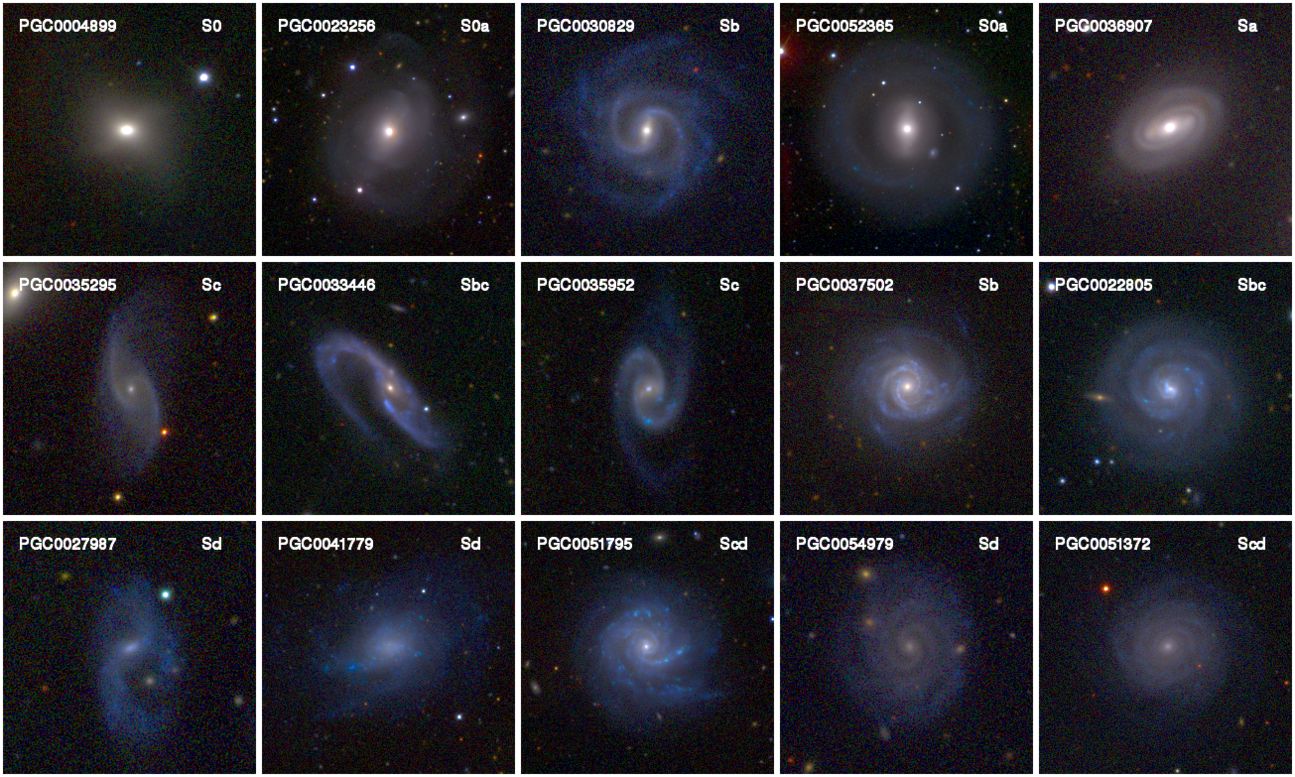}
    \caption{\label{att_Curv}Same as \fg\ref{att_Inc} for {\tt Arm
        curvature} attribute. For most spiral arms, the curvature
      varies along the arms, and a mean value is estimated for the
      attribute (see for example PGC0033446, in the central row). 
      Note also the peculiar cross-like shape
      of lenticular galaxy PGC004899 in the left panel of upper row. }
  \end{figure*}

\begin{figure*}
    \includegraphics[width=2\columnwidth]{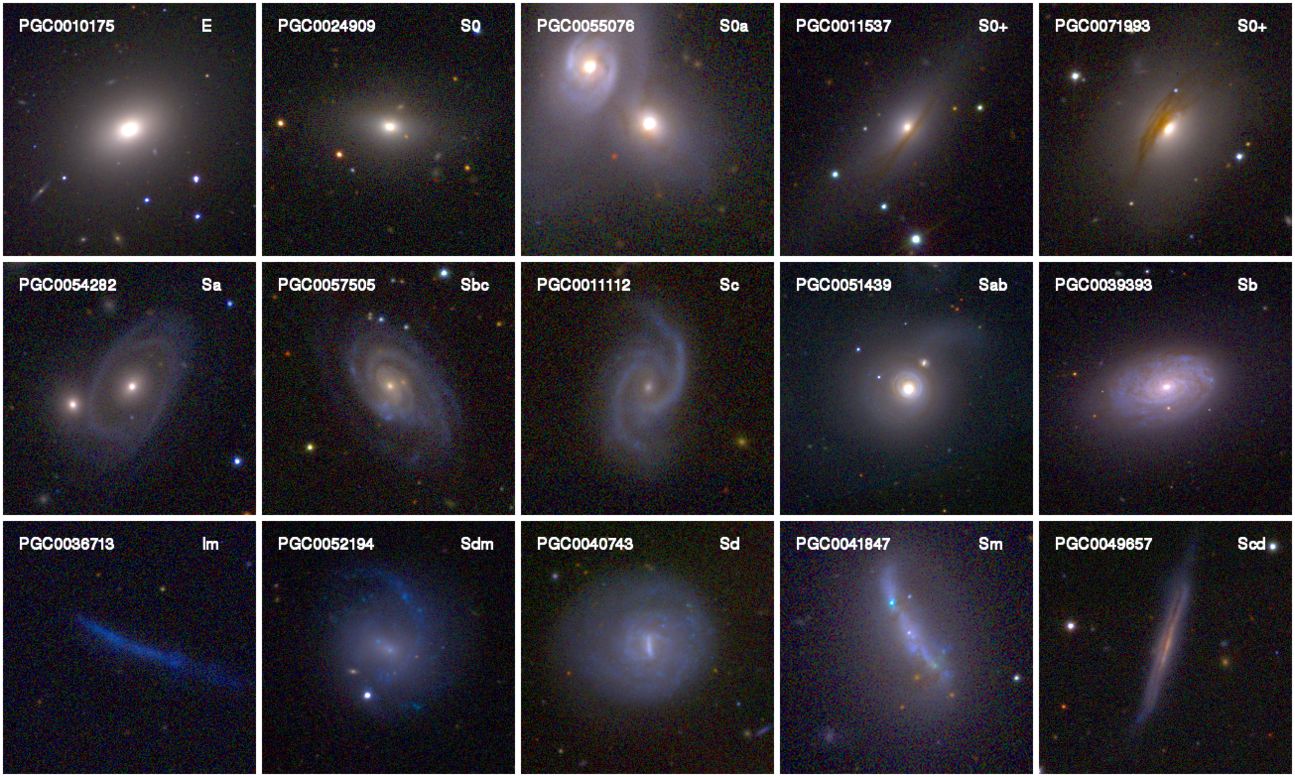}
    \caption{\label{att_Dust}Same as \fg\ref{att_Inc} for {\tt Visible
        dust} attribute. Most spiral galaxies contain dust, even in
      low amounts, and galaxies with high amounts of dust are often
      perturbed, as illustrated by PGC0071993 (top row), PGC0039393 
      (central row) and PGC0041847 (bottom row).}
  \end{figure*}

\begin{figure*}
    \includegraphics[width=2\columnwidth]{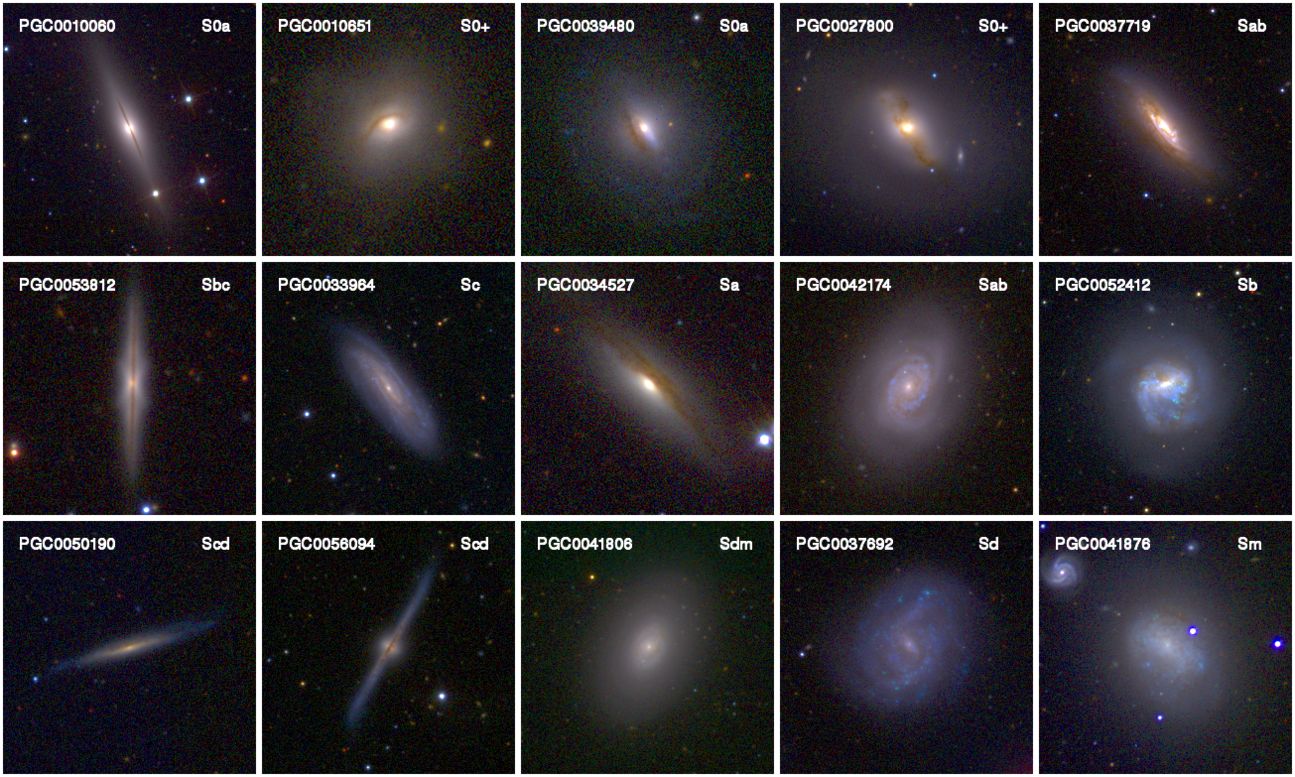}
    \caption{\label{att_Disp}Same as \fg\ref{att_Inc} for {\tt Dust
        dispersion} attribute. PGC0037719 in the top right panel is
      the earliest type galaxy (Sab) with an attribute value of
      1. The low values of the {\tt Dust dispersion}
      attribute are preferentially seen in highly inclined galaxies,
      as indicated by the anti-correlation between both attributes in
      \fg\ref{fig:corrmat}. Note the unusually faint bulge of Sab galaxy 
      PGC0042174 (central row, second from the right).}
  \end{figure*}

\begin{figure*}
    \includegraphics[width=2.00\columnwidth]{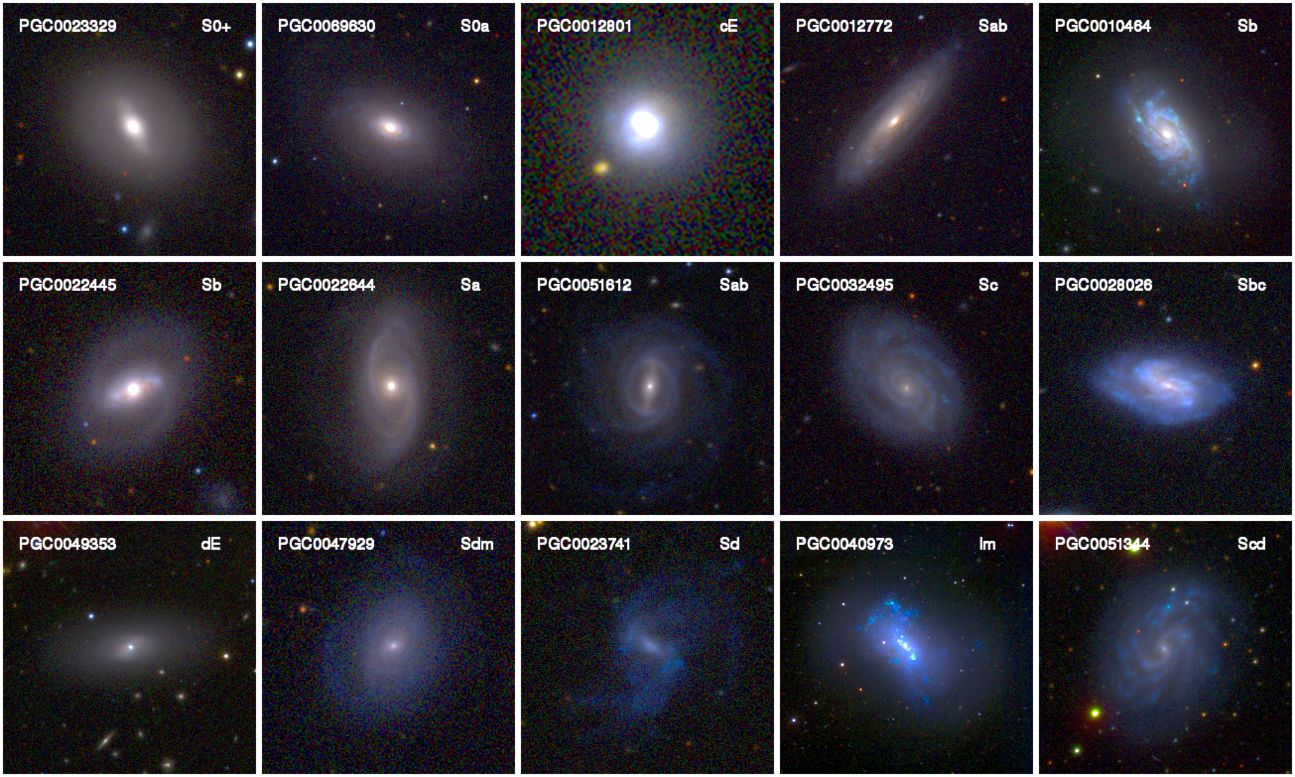}
    \caption{\label{att_Floc}Same as \fg\ref{att_Inc} for {\tt
        Flocculence} attribute. Sab and Sb galaxies (PGC0012772 and
      PGC0010464) are shown in the top row because there are no
      galaxies of earlier type with an attribute value of 0.75 or 1.
      We show in the left panel of the bottom row a nucleated dwarf
      lenticular galaxy classified as dE in the EFIGI catalogue, which
      has no flocculence; it however has an {\tt Hot Spot} attribute
      value of 0.5 due to an unusually strong nucleus.}
  \end{figure*}

\begin{figure*}
    \includegraphics[width=2.00\columnwidth]{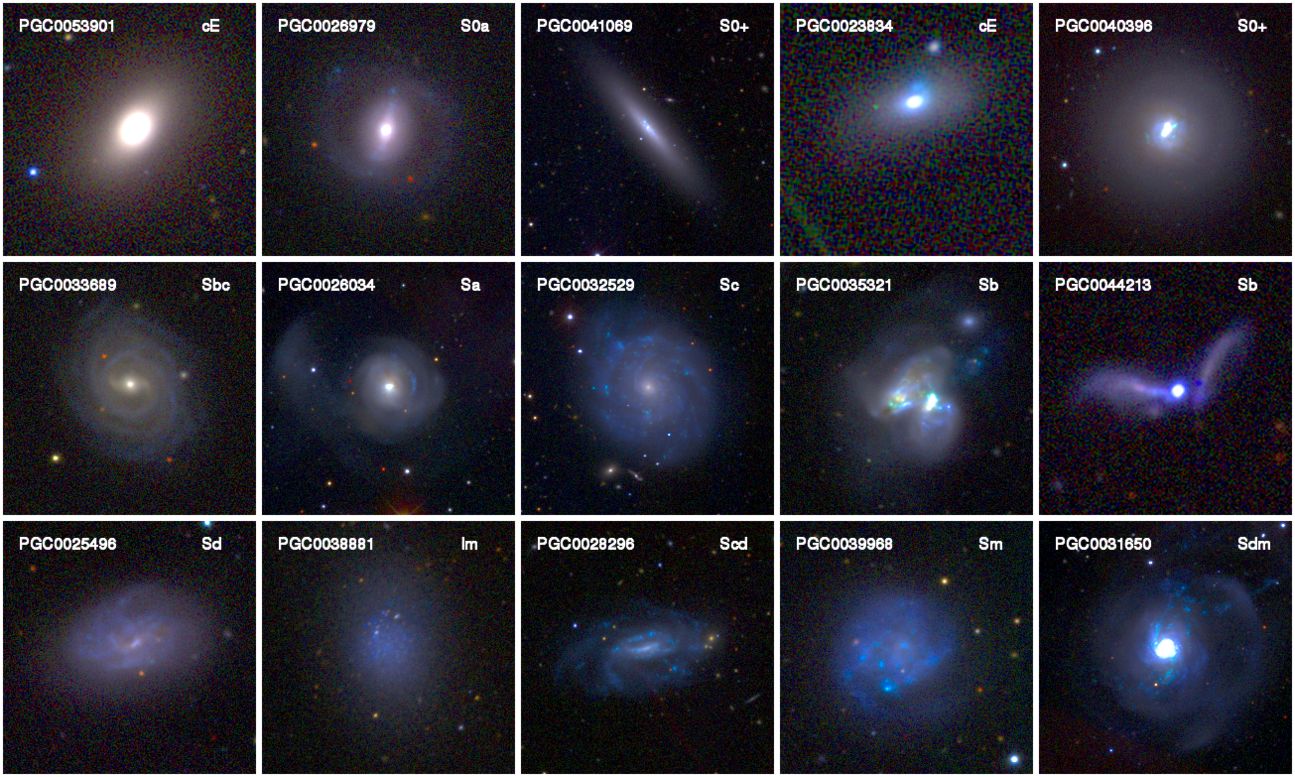}
    \caption{\label{att_Spot}Same as \fg\ref{att_Inc} for {\tt Hot
        spots} attribute. As seen here, strong values of the attribute (2 right
      columns) often implies strong values of the {\tt Perturbation}
      attribute; this is also measured by the strong correlation between both
      attributes in \fg\ref{fig:corrmat}.}
  \end{figure*}

\begin{figure*}
    \includegraphics[width=2\columnwidth]{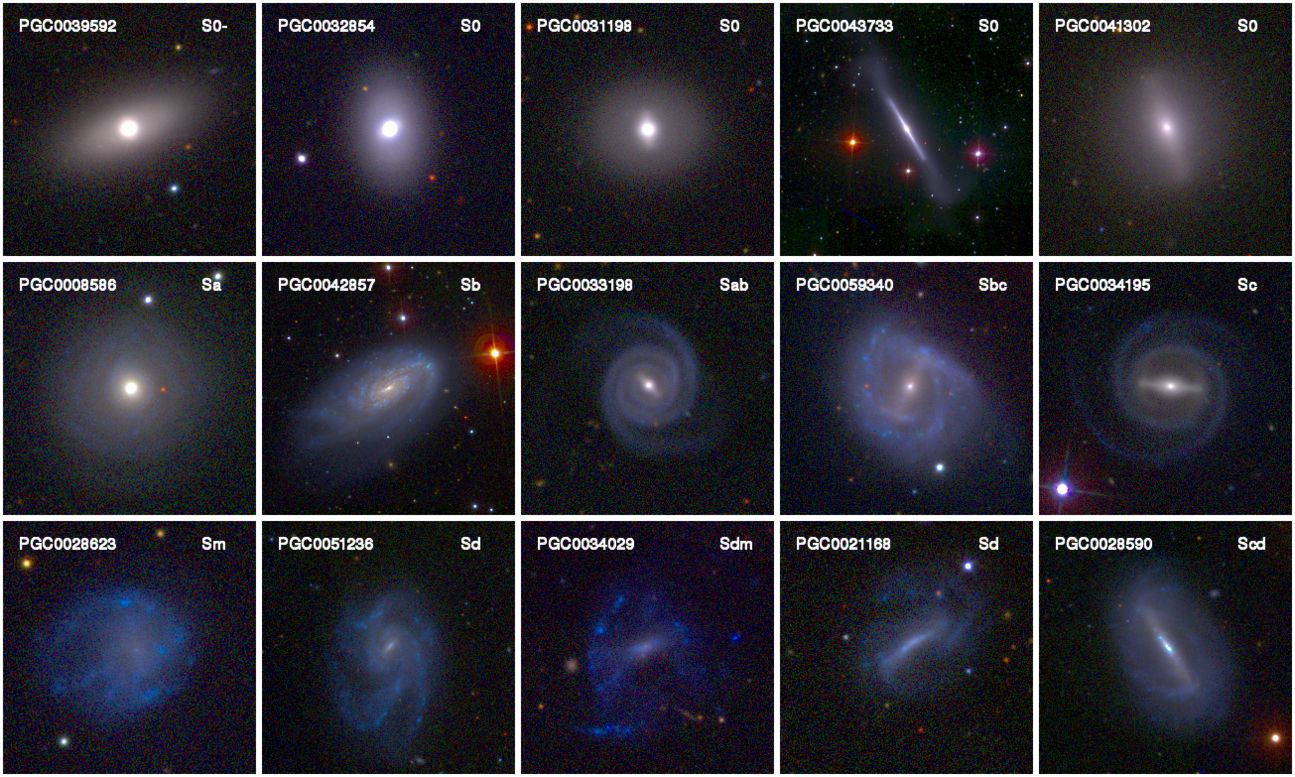}
    \caption{\label{att_Bar}Same as \fg\ref{att_Inc} for the {\tt Bar
        length} attribute. Bars are frequent in all galaxy types except
      cE, cD, ellipticals and dE. PGC0059340 (second panel from the right in the
      central row) illustrates that a bar can be long, hence a high
      attribute value of 0.75, but with a low surface brightness.}
  \end{figure*}

\begin{figure*}
    \includegraphics[width=2\columnwidth]{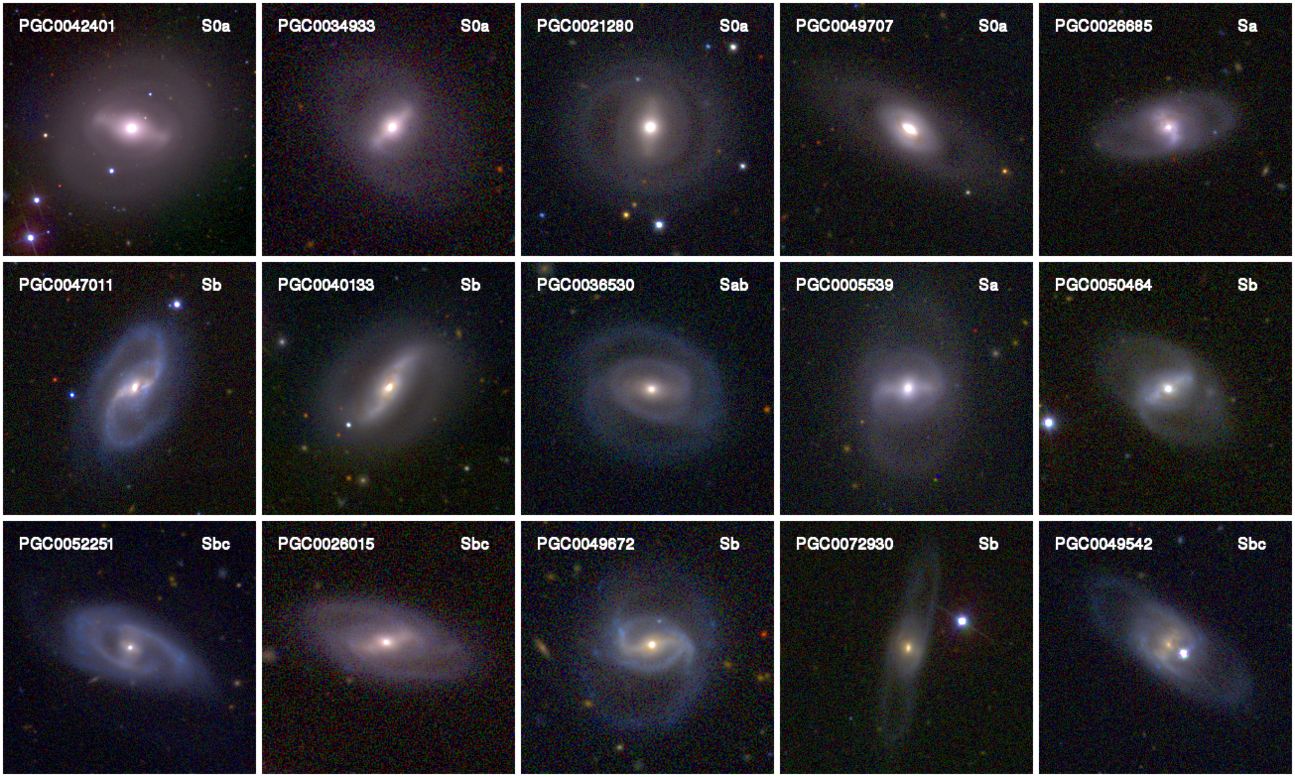}
    \caption{\label{att_Pseu}Same as \fg\ref{att_Inc} for {\tt
        Pseudo-Ring} attribute. As this is a low surface brightness
      feature, high-contrast pseudo-rings are preferentially shown
      here. The galaxies shown in the bottom row are the latest type galaxies 
      (Sbc, Sb, Sb, Sbc \resp) with attributes values of 0.25, 0.5, 0.75, 1.0
      respectively (for a fair comparison, an Sbc galaxy is also shown
      in the bottom row for attribute value 0).}
  \end{figure*}

\begin{figure*}
    \includegraphics[width=2\columnwidth]{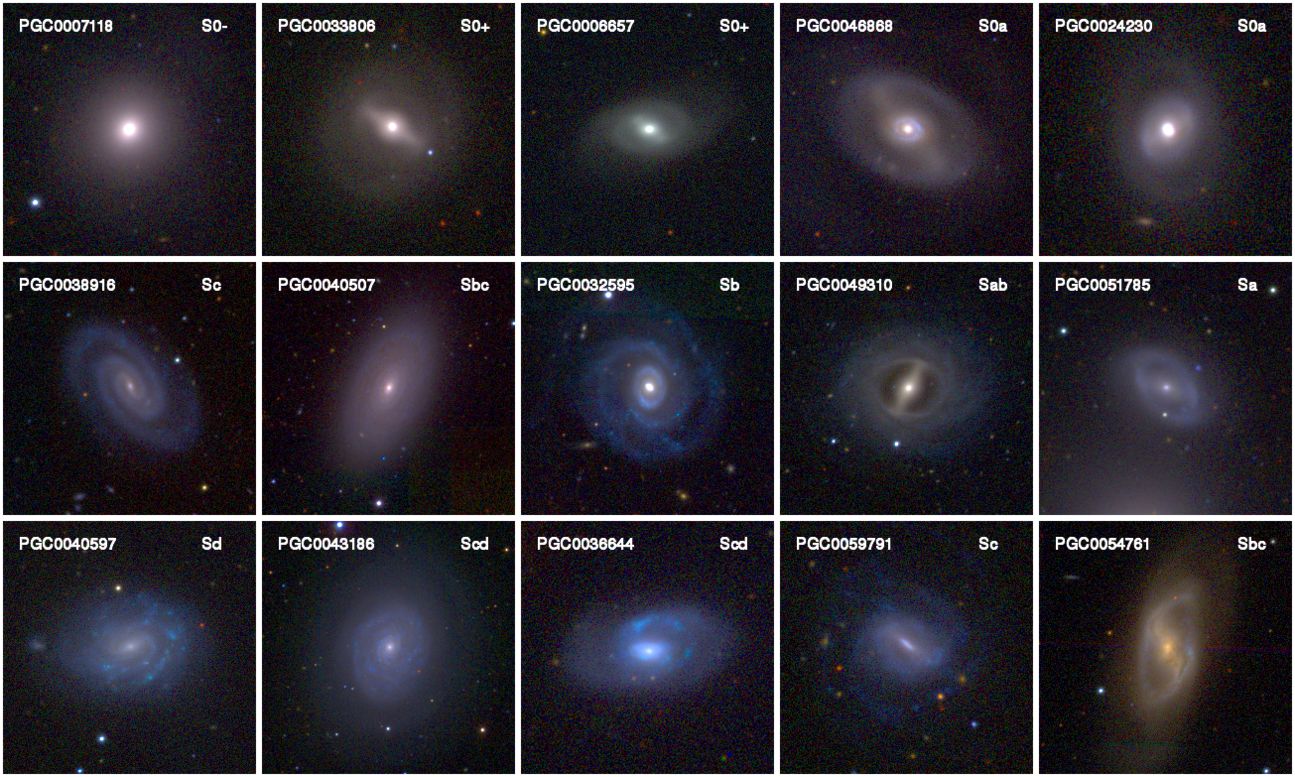}
    \caption{\label{att_Iring}Same as \fg\ref{att_Inc} for {\tt Inner
        Ring} attribute. PGC0059791 and PGC0054761 (2 right panels in
      the bottom row) are the latest type galaxies (Sc and Sbc \resp) with a
      strong and very strong {\tt Inner Ring} respectively. In PGC0046868
      (second panel from the right in the top row) the {\tt Inner
      Ring} attribute points to the ring encircling the bar; the additional
      blue ring inscribed within the bar is a ``nuclear ring'', not
      characterised here.}
  \end{figure*}

\begin{figure*}
    \includegraphics[width=2\columnwidth]{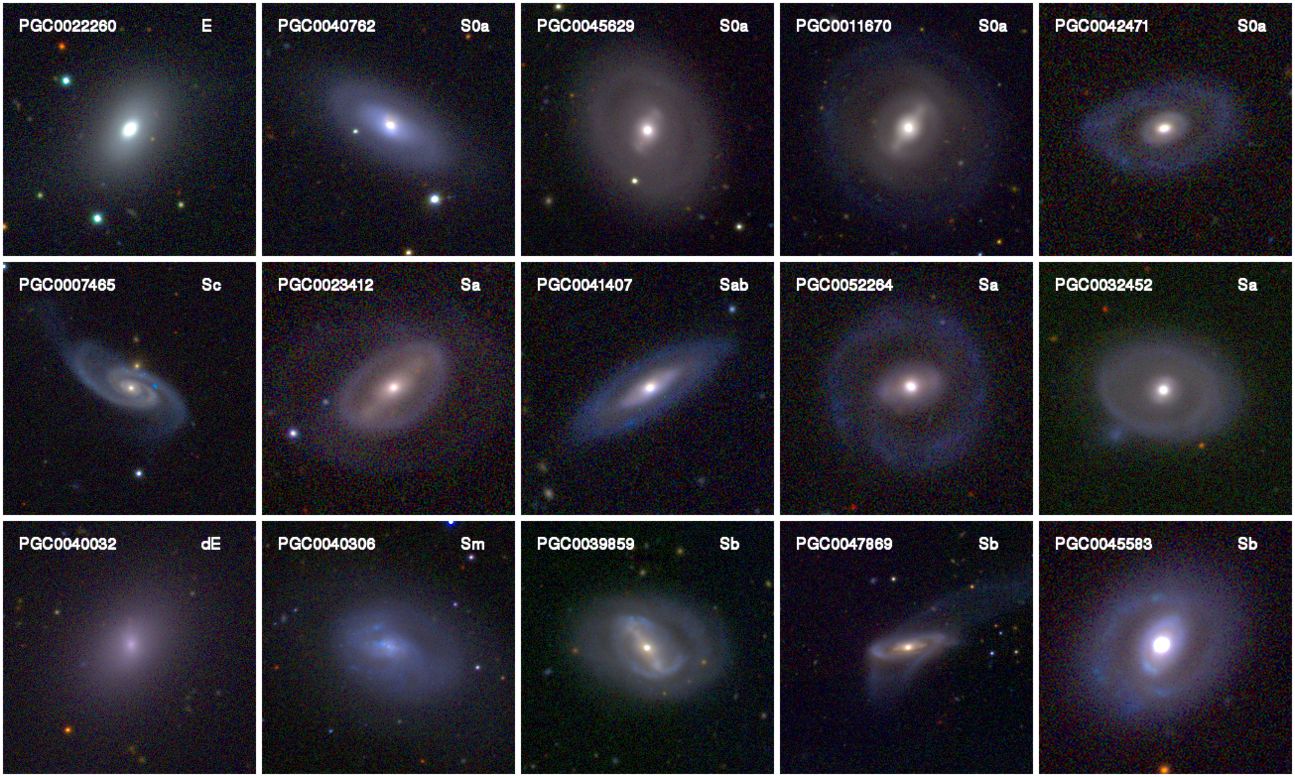}
    \caption{\label{att_Ering}Same as \fg\ref{att_Inc} for {\tt Outer
        Ring} attribute.  As some outer rings can be of very low surface 
      brightness, we preferentially show here galaxies with
      high-contrast rings. PGC0039859, PGC0047869 and PGC0045583 (3
      right panels in the bottom row) are among the latest type galaxies (Sb)
      with values of the {\tt Outer Ring} attribute between 0.5 and 1.}
  \end{figure*}

  To illustrate the 5 different levels of each attribute, we display
  in \fgs\ref{att_Inc} to \ref{att_Ering} the $irg$ colour images (see
  \sct\ref{chap:sdssima}) of 3 galaxies for each attribute value (from
  0 to 1). For a given attribute value, we display from top to bottom
  galaxies of increasing EFIGI Hubble type: the top row shows the
  commonly called ``early-type'' galaxies, bulge-dominated (cE, cD,
  E), lenticulars and S0a; the middle row contains the
  ``intermediate-type'' galaxies, that is early spirals, from Sa to
  Sc; and the bottom row, the ``late-type'' galaxies containing later
  spirals, from Scd to Sm, Im and dE. For illustrative purposes, or
  when none of the types in the 3 categories exist for a given row and
  attribute value, earlier or later types are shown (in \fgs
  \ref{att_Rot}, \fgs \ref{att_Mult}, \ref{att_BT}, \ref{att_Arm},
  \ref{att_Disp}, \ref{att_Floc}, \ref{att_Iring}, \ref{att_Ering},
  and \ref{att_Pseu}).



  We emphasise that galaxies shown in \fgs\ref{att_Inc} to
  \ref{att_Pseu} are not randomly chosen but visually selected from
  the catalogue, with three goals: 1) to show representative
  galaxies of each attribute value; 2) to ensure that the wide variety
  of typical galaxies observed within each EFIGI morphological type
  are present, and that the different types are equally represented;
  3) to display a number of atypical galaxies with the various EFIGI
  types and attribute values.

  For most attributes (except evidently {\tt Inclination/Elongation}),
  we show preferentially galaxies with low inclination, as this
  implies a higher confidence level in most attribute strengths: for
  example, the presence and strength of spiral arms is difficult and
  impossible to assess in inclined and edge-on galaxies
  respectively. We nevertheless show galaxies with all values of the
  {\tt Inclination/Elongation} attribute for the {\tt Contamination},
  {\tt Multiplicity}, {\tt Visible Dust}, and {\tt Dust Dispersion}
  attributes, as these are as or more reliably determined in highly
  inclined galaxies than in nearly face-on objects.

  The atypical galaxies shown in \fg\ref{att_Inc}-\ref{att_Ering} include 
  peculiar galaxies which 
  often have a high value of the {\tt Perturbation} attribute, but also
  regular galaxies with unusual combinations of attributes, or galaxies 
  in which the attribute is of strikingly high-contrast compared to the 
  rest of the catalogue. For example PGC0042174 in \fg\ref{att_Disp} has 
  a very low {\tt B/T ratio} for an Sab galaxy;
  PGC0043733 in \fg\ref{att_Bar} has a remarkably sharp
  and contrasted bar; PGC0049310 in \fg\ref{att_Iring} has nearly no
  disk emission between the bar and the inner ring.

  \subsection{Validation and reliability}
  
  We explore the main characteristics and the reliability of the homogenised
  EHS and the various attributes by comparing the EHS to the RHS, and by
  examining the trends and dependencies in attributes.
 
\subsubsection{EFIGI-RC3 Hubble sequence comparison}

  \begin{figure*}
\centerline{\includegraphics[angle=-90,width=0.5\textwidth]{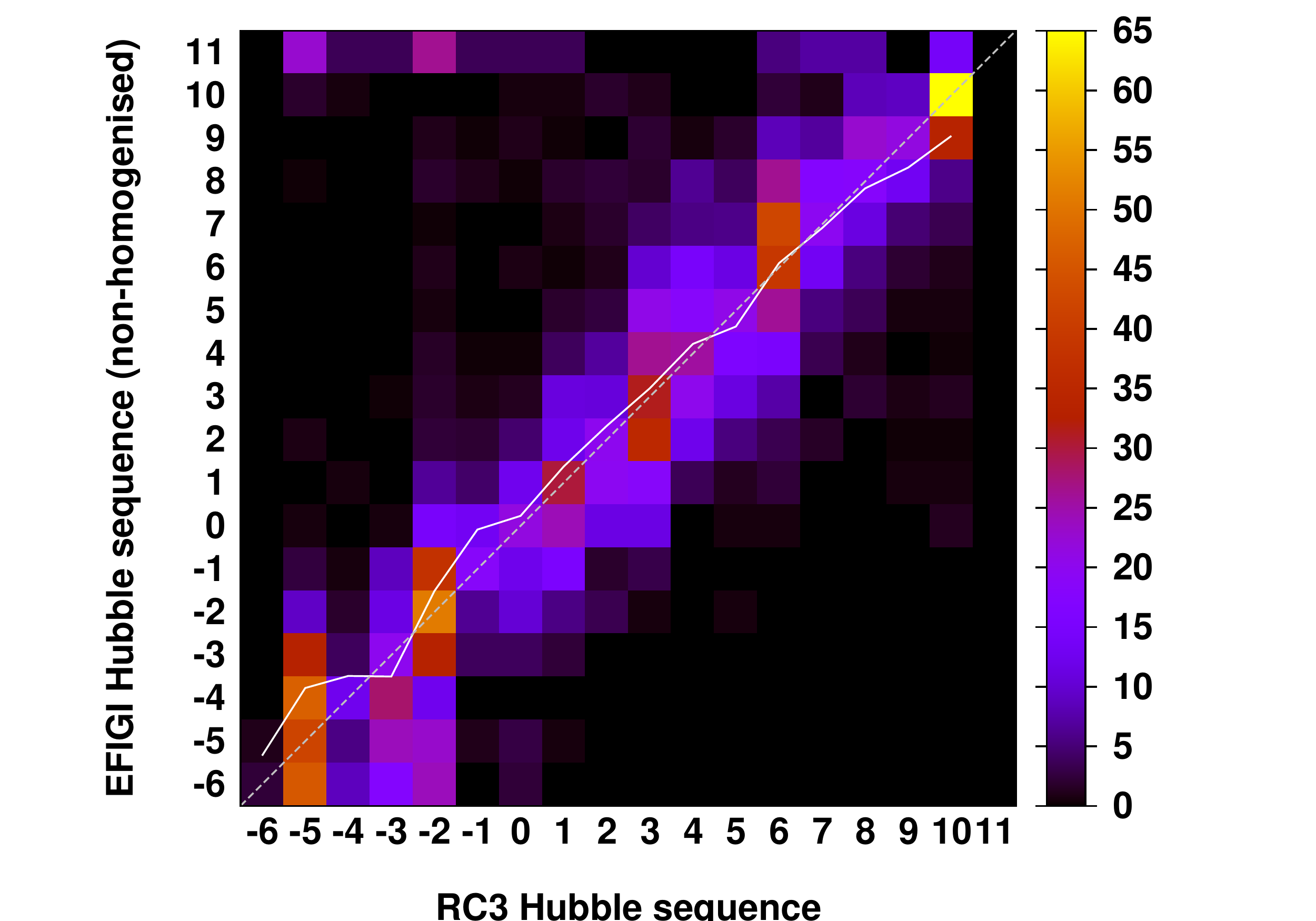}
    \includegraphics[angle=-90,width=0.5\textwidth]{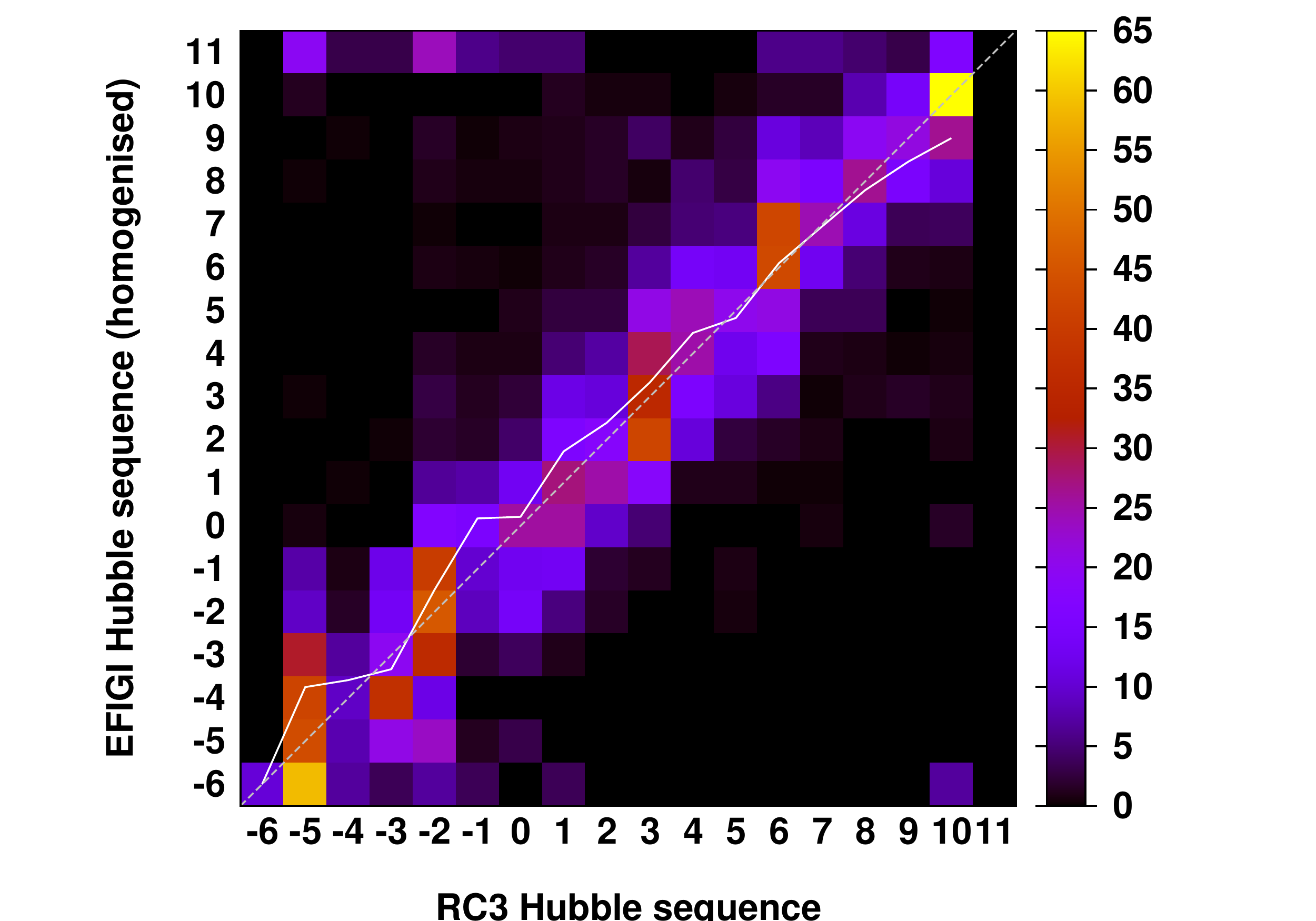}}
  \caption{Confusion matrices between the RC3 Revised Hubble Sequence
    (RHS) and two versions of the EFIGI Hubble sequence: before ({\em left})
    and after homogenisation ({\em right}). The dash-dotted line along the
    diagonal corresponds to a perfect match. The solid white line connects the
    average EFIGI type in each PGC type bin.}
    \label{confHT}
  \end{figure*}

  \fg \ref{confHT} shows the confusion matrices for the RHS, and the
  non-homogenised and homogenised EHS classifications.  The comparison
  between both versions of the EFIGI catalogue shows that the
  non-homogenised catalogue lacks compact galaxies (T=-6), and reveals
  some confusion between ellipticals (T=-5) and lenticulars (T=-3 to
  -1).  The lower dispersion observed for all types (but cE and dE
  galaxies) in the right panel compared to the left is proof of a
  better agreement between the RHS and the EHS after homogenisation of
  the EFIGI types, thus confirming that this process leads to a more
  reliable classification.  We note that the remarkable level of
  overall agreement (with the exception of dE types) between the RHS
  and the EHS classes (despite being produced by different individuals
  and based on different imaging material) is evidence of the
  reliability of the Hubble classification for galaxies in the local
  universe.
 

  \subsubsection{Dependencies among attributes}

  \begin{figure*}
    \includegraphics[angle=-90,width=0.8\textwidth]{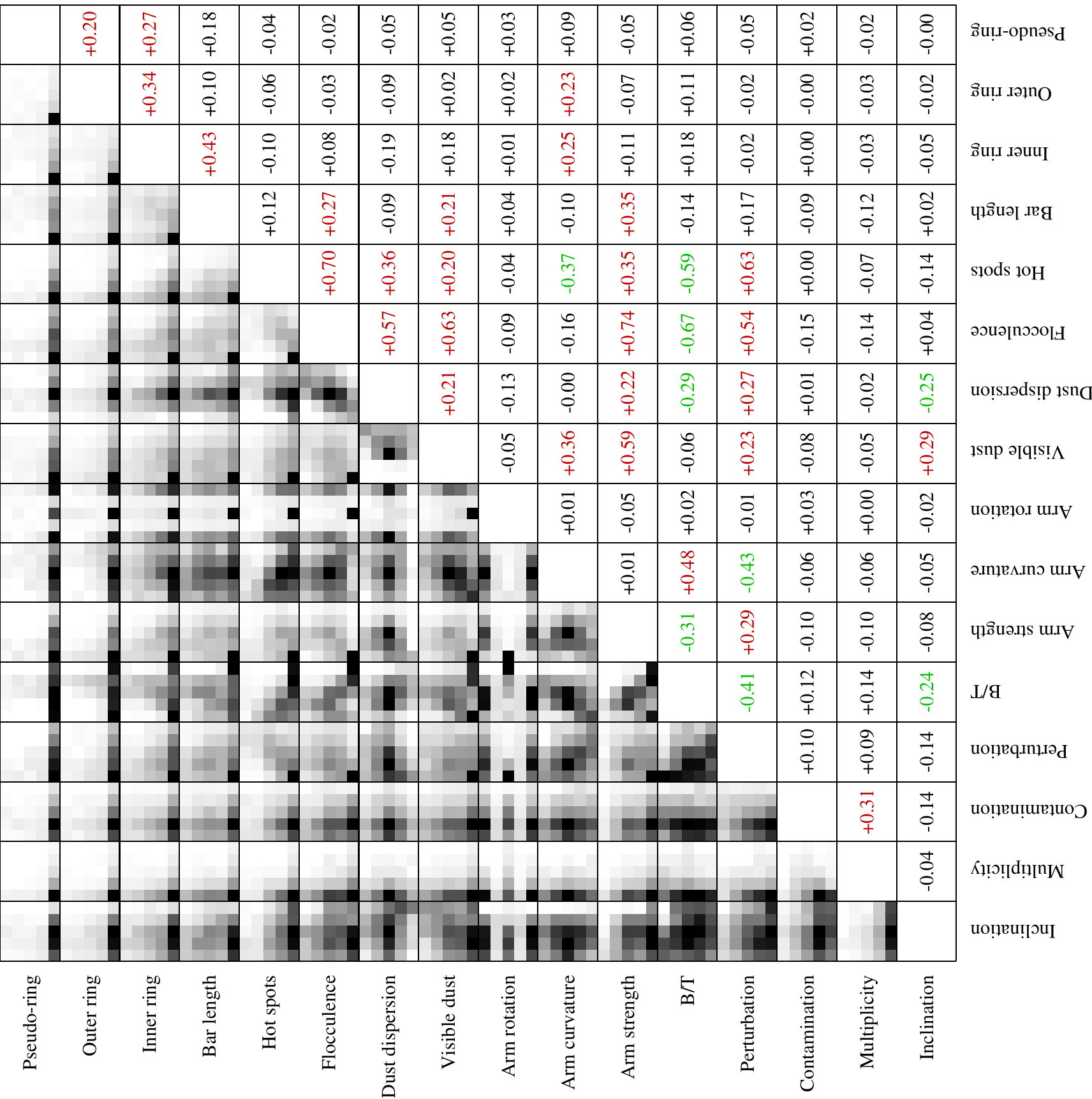}
    \caption{Confusion matrices (in grey levels) and Pearson's correlation
    coefficients for attribute pairs in the homogenised data set. Correlation
    coefficients higher than 0.2 are coloured in red, while those lower than
    -0.2 are in green.}
    \label{fig:corrmat}
  \end{figure*}

  \fg\ref{fig:corrmat} shows the weighted confusion matrices and
  Pearson's correlation coefficients for all pairs of morphological
  attributes.  The strongest correlations are found between {\tt
    flocculence} and the {\tt hot spots} (+0.70) and {\tt arm
    strength} (+0.74) attributes.  {\tt Visible dust} is also strongly
  correlated to {\tt flocculence} (+0.63), and {\tt arm strength}
  (+0.59).  Taken together, these correlations probably reflect the
  fact that {\tt flocculence}, {\tt hot spots} and {\tt visible dust}
  are related indicators of the level of star formation activity in a
  galaxy. These attributes are correlated with {\tt arm strength},
  presumably because star formation is enhanced in the spiral arms due
  to gas compression by density waves.  The {\tt perturbation}
  attribute is also strongly correlated with both {\tt hot spots}
  (+0.63) and {\tt flocculence} (+0.54), hinting that tidal and
  merging processes enhance star formation in galaxies.

  The strongest anti-correlations are found between {\tt B/T ratio} and
  the {\tt flocculence} (-0.67) and {\tt hot spots} (-0.59) attributes.  Strong
  anti-correlations with {\tt B/T ratio} are also measured when
  considering {\tt perturbation} (-0.41), {\tt arm strength} (-0.31),
  and {\tt dust dispersion} (-0.29).  All these anti-correlations
  reflect the increasing strength of the aforementioned attributes
  along the Hubble sequence, along with the {\tt B/T ratio} which
  gradually decreases, as shown in \citet{lapparent1}. The strong
  correlation observed between {\tt B/T ratio} and {\tt arm curvature}
  (+0.48) is consistent with the winding of spiral arms being a major
  criterion for defining the progression of spiral types along the
  Hubble sequence \citep{vandenbergh98}.

  The correlation between the presence of a bar and that of an inner
  ring noticed by \citet{kormendy} and \citet{hunt} is clearly seen in the EFIGI
  catalogue (+0.43). Some level of correlation is also found between
  {\tt inner ring} and {\tt outer ring} (+0.34), and between both attributes and
  {\tt pseudo-ring} (+0.27 and +0.20 \resp). The presence of a
  {\tt pseudo-ring} often implies that of an {\tt inner ring}, and the full
  range of transition stages between an {\tt outer ring} and a {\tt pseudo-ring}
  is present in the EFIGI catalogue.
  These various correlations are further examined in the companion 
  article reporting on the statistical analysis of the EFIGI attributes
  \citep{lapparent1}. These effects were already pointed out by \citet{buta96},
  who discuss rings in terms of barred galaxy models.

  The various correlations and anti-correlations between the EFIGI
  {\tt inclination/elongation} attribute and the {\tt B/T ratio}, {\tt
    visible dust} and {\tt dust dispersion} attributes \resp\ are
  difficult to interpret as they may partly originate from a type
  effect, pure bulge galaxies having higher fractions with {\tt
    inclination/elongation}$ = 0$ or 0.25 than disk galaxies (see
  \citealt{lapparent1}).

\subsubsection{Trends in attribute strength}

  \begin{figure*}
    \includegraphics[width=\textwidth]{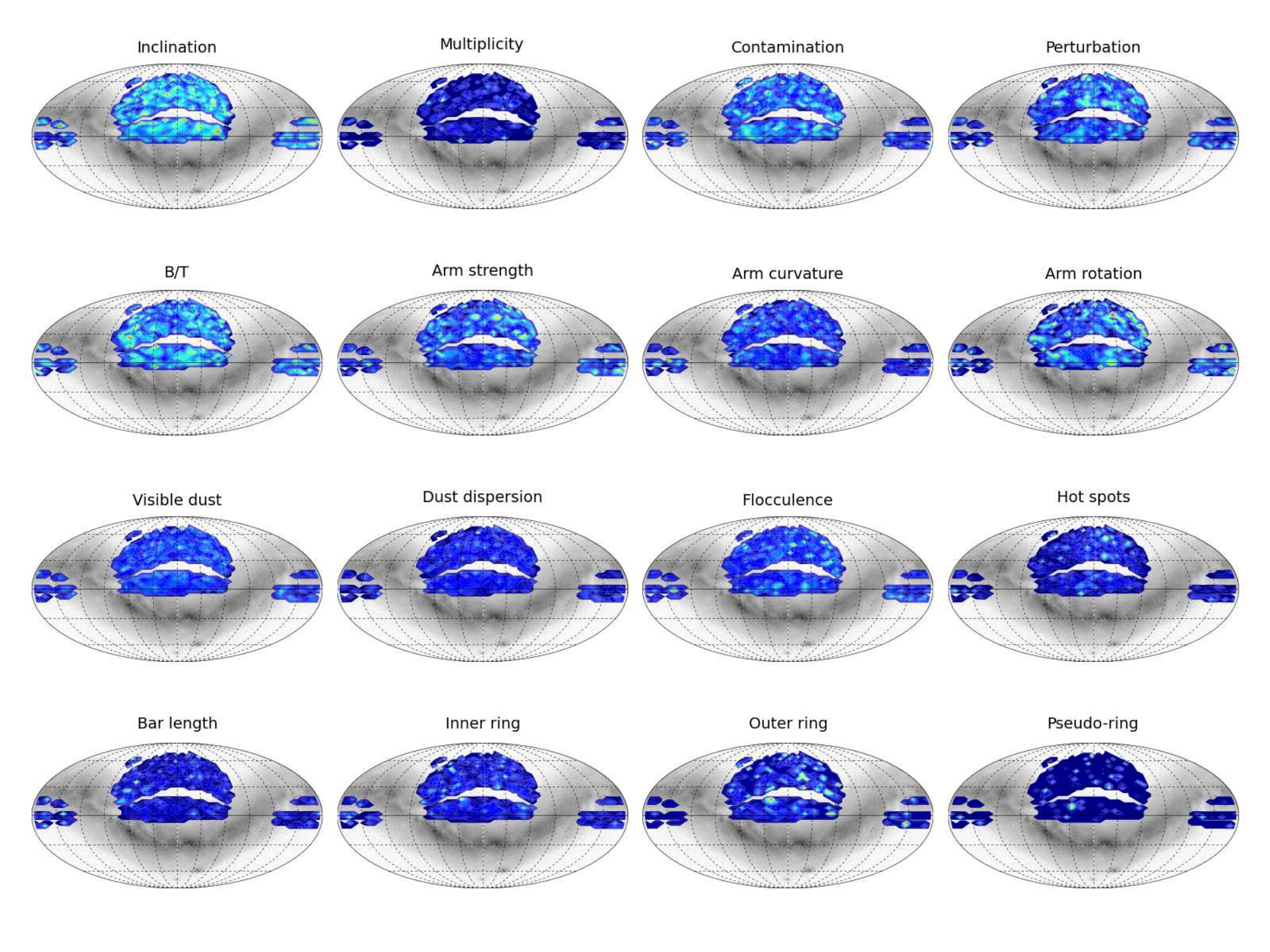}
    \caption{Angular distribution of weighted average attribute strengths in
      the EFIGI catalogue. Right ascension runs right to left from 0$^h$ to 24$^h$
      (see \fg\ref{clusters}).
      The weight is set to 0 for confidence intervals
      $\Delta x = 1$, and $(\Delta x+0.25)^{-2}$ otherwise (see \sct\ref{chap:att}).
      The colour scale is
      relative and runs from dark-blue to green, yellow, and red for the highest
      values.}
    \label{trends}
  \end{figure*}

  The EFIGI catalogue is vulnerable to possible trends arising from a
  modification of the perception of the astronomer during the long
  time periods (weeks) spent at classifying and homogenising
  individual attributes.  Visual examination has often been conducted
  by increasing PGC number, which is itself ordered in right
  ascension. Although this approach makes it much easier to identify
  galaxy pair components and spot duplicates, it also makes the
  detection of trends more ambiguous because of the well-known
  correlation between galaxy morphology and density \citep{dressler80,
    postman84}.

  To examine whether there are systematic drifts with time in the
  homogenised EFIGI attributes, we plot in \fg\ref{trends} the sky
  distribution of the local weighted average of attribute
  strengths. This graph does show some significant peaks, which can be
  attributed to galaxy clusters (\fg\ref{clusters}), but reassuringly
  no significant ``banding'' pattern is visible in right ascension.

  The mean \textit{B/T} increases near to clusters because of the
  local over-density of early type galaxies.  This is particularly
  apparent in the Hercules cluster. The increase is less pronounced in
  Virgo as it is close enough for a number of dwarf elliptical galaxies with
  disk-like profiles to enter the sample and somewhat compensate the
  excess number of early-type galaxies (the majority of the EFIGI dEs
  are located in Virgo). Apart from those peaks, global trends seems
  to be well behaved, at least below the 10\% level.

  \section{\label{sec:psf}PSF estimates}

  Knowledge of the image Point Spread Function (PSF) is essential to
  most morphometric measurements, in particular those involving models
  adjustements.  The PSFs of the EFIGI images were measured on the same
  combined ``fPc'' SDSS DR4 images rescaled around each galaxy, as in
  \sct\ref{chap:sdssima} but without clipping to 255x255 pixels. The
  wider field of view provides typically a hundred point-sources
  suitable to PSF modelling. We used the {\sc PSFex}
  software\footnote{\tt http://astromatic.net/software/psfex}
  Bertin et al{.} (in prep{.}) which generates a tabulated model with adaptive
  sampling and achieves super-resolution when the input data are
  undersampled, which is the case for a large fraction of the EFIGI
  images because of the angular rescaling.  For simplicity, we ignored
  the spatial variations of the PSF around each galaxy. PSF variations
  within the field of view are indeed expected for the most extended
  objects as a result of combining exposures obtained under different
  conditions; however downscaling makes the PSF so sharp for these
  objects that the consequences are negligible.

  \section{\label{sec:coverage}Overall properties of the catalogue}

  \subsection{Sky coverage}

  \fg\ref{clusters} shows the projected distribution of EFIGI galaxies onto the
  sky. The solid angle covered by the EFIGI catalogue corresponds to
  the SDSS photometric DR4, 6670 deg$^2$.

  \subsection{\label{chap:clusters}Cluster detection}

  \begin{figure*}
    \includegraphics[width=\textwidth]{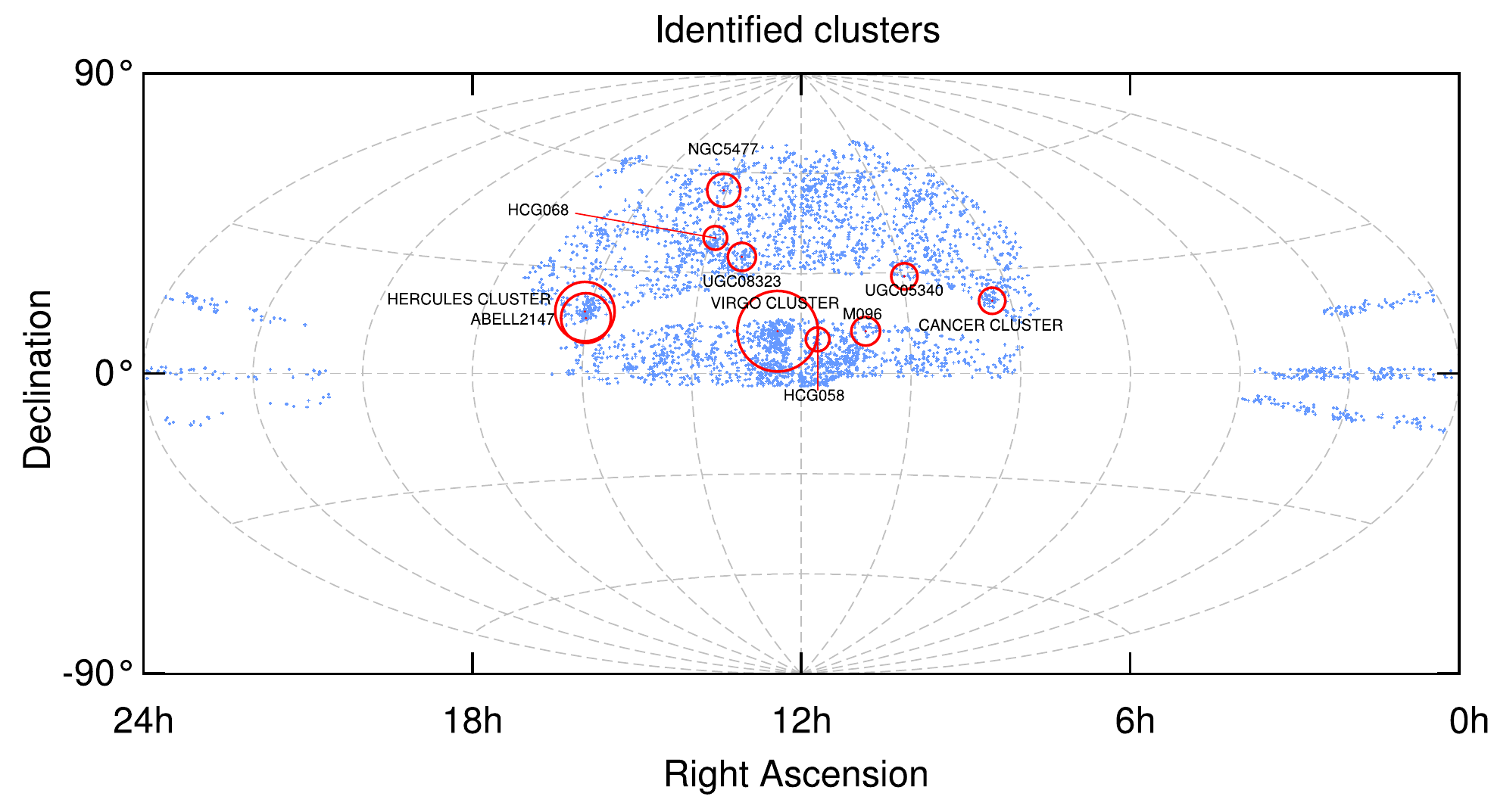}
    \caption{Blue points are EFIGI galaxies. Red circles indicate clusters and are
      centred on NED cluster coordinates. The enclosed area is proportional
      to the number of cluster members as reported by NED.}
    \label{clusters}
  \end{figure*}

  A systematic search for clusters and groups is performed to trace
  environment effects in the present analysis (see \fg \ref{trends}
  and the related section) and in future works. A list of clusters in
  the area covered by EFIGI catalogue is retrieved using the NED
  ``all-sky'' query form for redshifts $0.001 < z < 0.05$ and richness
  above 50. For each cluster listed, we count the number of EFIGI
  galaxies within a limited volume around the cluster. We build a
  ``box'' centred on right ascension, declination and redshift of the
  cluster with dimensions equal to $1$ Mpc $\times 1$ Mpc $\times
  2.000$ km s$^{-1}$. We assume $H_0=73$ km s$^{-1}$ Mpc$^{-1}$
  \citep{freedman01}.  Concentrations with more than 5 EFIGI galaxies
  within the search box are added to the list of EFIGI clusters.
  Clusters identified using this procedure are overplotted as a red
  dot surrounded by red circles onto the sky distribution of EFIGI
  galaxies in \fg\ref{clusters}; the circle size indicates the
  relative cluster richness extracted from NED.  We find 10 clusters
  in the EFIGI catalogue, which indeed correspond to over-densities of
  galaxies in the projected map of \fg\ref{clusters}. Other visible
  over-densities may be due to chance projections.

  \subsection{\label{chap:maghist}Magnitude distributions}

\begin{figure}
    \includegraphics[angle=0,width=\columnwidth]
	{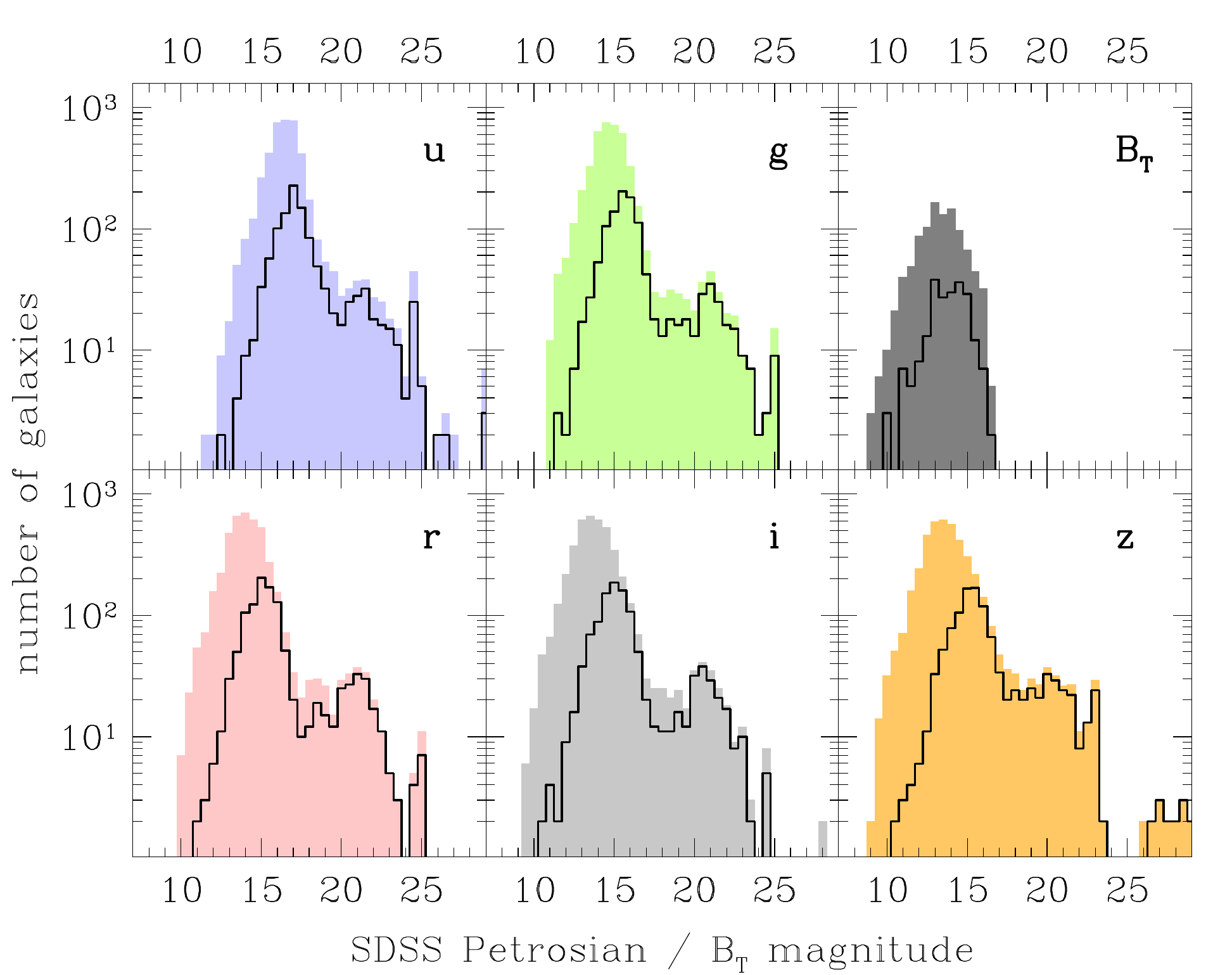}
    \caption{\label{maghist}Shaded histograms show the distribution of SDSS
      Petrosian magnitudes for EFIGI galaxies in the \textit{u}, \textit{g},
      \textit{r}, \textit{i} and \textit{z} bands in intervals of 0.5 magnitude
      (2 upper left panels, and bottom panels), and the distribution of the
      PGC $B_T$ magnitude for the 1006 EFIGI galaxies for which it is available
      (upper right panel). In each panel, the black lines shows the histograms
      restricted to galaxies of types Sd and later. All panels for SDSS
      magnitudes show an extension to faint magnitudes (enhanced by the
      logarithmic scale), caused by the limitations in the processing 
      of large objects by the SDSS photometric pipeline.}
  \end{figure}

  \fg\ref{maghist} shows for each SDSS filter the distribution of SDSS
  Petrosian magnitudes for all EFIGI galaxies by intervals of 0.5
  magnitude.  A bright bell-shaped distribution with an extension
  fainter than magnitude $\sim 17-19$ is seen in all five filters.  We
  interpret this extension as a consequence of erroneous flux
  measurements, likely due to the splitting of large galaxies into
  several fainter and smaller objects by the SDSS pipeline. SDSS
  photometric measurements are unreliable for objects larger
  than $\sim 1-2$ arcmin. Among the 4458 EFIGI galaxies, 257 have no
  measured major isophotal diameter $D_{25}$ in the RC2 system
  \citep{rc2}, 954 have $D_{25}$ below 1 arcmin, 2444 between 1 and 2
  arcmin and 811 above 2 arcmin.  \fg\ref{maghist} also shows that a
  large fraction of objects in the extensions are late spiral types,
  as indicated by histograms restricted to EFIGI galaxies with types
  Sd and later. This magnitude extension is not seen in the
  distribution of PGC $B_T$ magnitudes in \fg\ref{maghist}, which
  shows instead a sharp cut-off.  Additional support for this
  interpretation comes from the two magnitude brighter extension of the
  $B_T$ distribution at bright magnitudes compared to that of SDSS
  \textit{g} measurements (see \fg\ref{maghist}).

\begin{figure}
    \includegraphics[angle=0,width=\columnwidth]{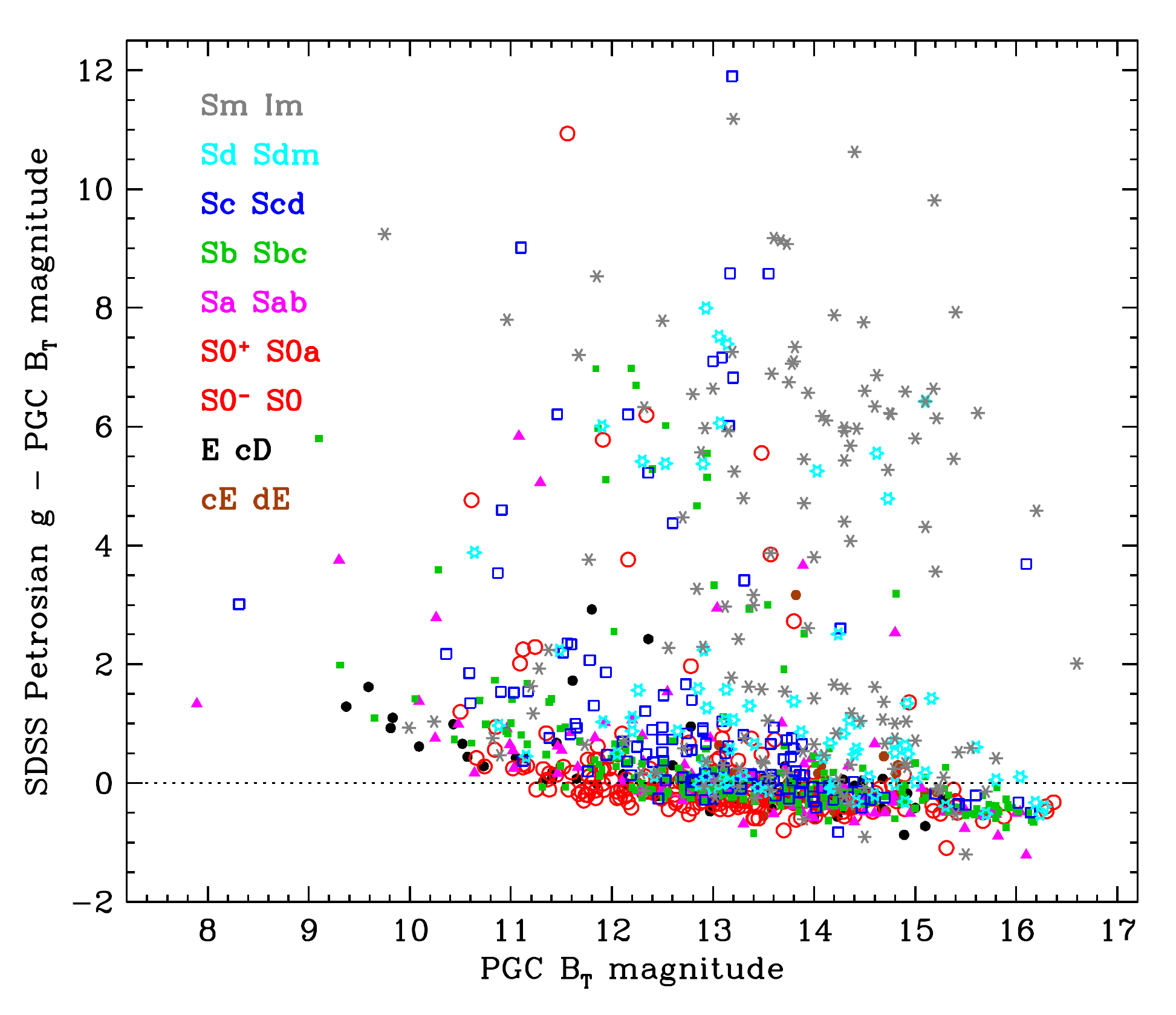}
    \caption{\label{magcomp}Comparison between the PGC $B_T$ and the SDSS
    Petrosian \textit{g} magnitudes, showing a high dispersion from the
    splitting of large galaxies in the SDSS photometric pipeline.}
  \end{figure}

  This photometric discrepancy is further illustrated in \fg\ref{magcomp}, which
  shows the magnitude difference between the PGC $B_T$ band and the SDSS Petrosian
  \textit{g} band for the various EFIGI types of the 1006 galaxies with a $B_T$ 
  magnitude. The graph shows a high
  dispersion in the SDSS \textit{g} magnitudes for all values of PGC $B_T$,
  with a significant tail towards flux underestimation in the SDSS by up
  to 8 magnitudes. Galaxies with magnitude differences of more than 2 magnitudes
  in \fg\ref{magcomp} are mostly of types Sd to Im, in agreement with the
  excess splitting of these objects illustrated in \fg\ref{maghist}.

  \subsection{Photometric completenesses}

  We extracted from the SDSS DR4 all objects in the ``Galaxy''
  catalogue that do not have the ``SATURATED'' flag. This flag
  indicates that an object includes one or more saturated pixels, and
  allows the removal of a large proportion of stars included in the
  ``Galaxy'' catalogue ($\sim30$\% at $g\le18$).  However, we include
  objects located near the edge of the SDSS CCD frames (with the
  ``EDGE'' flag), which is true for $\sim2-3$\% of EFIGI
  galaxies. These SDSS samples are selected in Petrosian magnitude,
  with a limit of 20 in \textit{u} and $18$ in the other bands, in
  order to sample the bright peak of the EFIGI magnitude distributions
  (see \fg\ref{maghist}); beyond these limits, the number of galaxies
  in the SDSS increases steeply and the completeness drops to 0.

\begin{figure}
    \includegraphics[angle=0,width=\columnwidth]
	{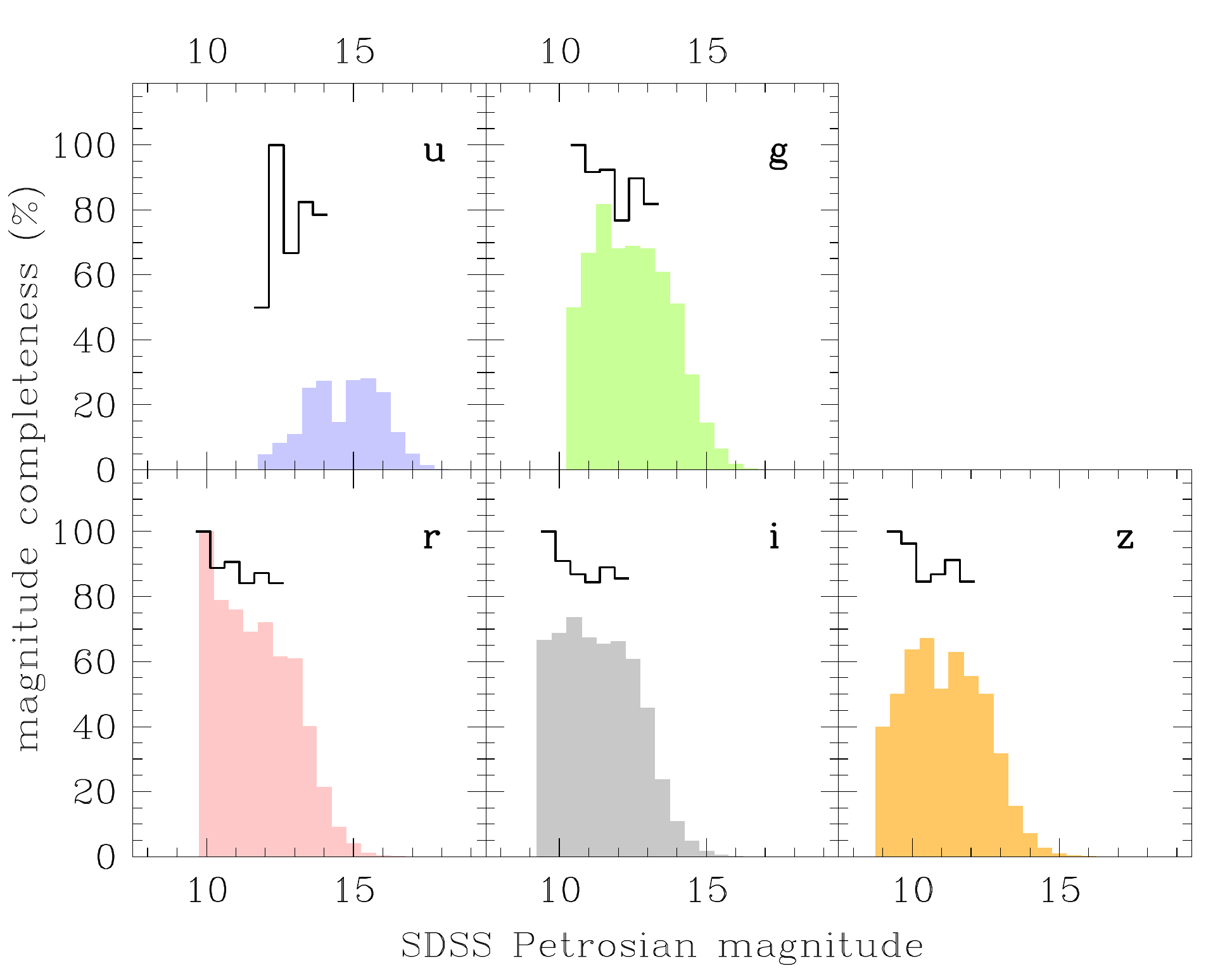}
        \caption{\label{compmag}Shaded histograms show the magnitude
          completeness of the EFIGI catalogue compared to the SDSS DR4
          samples excluding objects flagged as ``SATURATED''. All
          EFIGI galaxies not included in the DR4 comparison samples
          have been removed (see text for details).  The apparent
          30-60\% completeness is largely an underestimate due to
          numerous artefacts (halos of bright stars, satellite trails,
          faint compact objects) included in the DR4 sample. The
          bright part of the corrected completeness curves are shown
          as solid lines; note that the brightest bin of the
          \textit{u} corrected curve with 50\% completeness contains a
          single EFIGI galaxy. }
  \end{figure}

  Before calculating the photometric completeness, we exclude the 56
  EFIGI galaxies that were not identified in the SDSS magnitude limited 
  samples (see \sct\ref{chap:SDSS}). We also exclude EFIGI galaxies flagged as
  ``SATURATED'' in the SDSS database. These galaxies either have a bright 
  core with some likely saturated central pixels, or contain a contaminating or 
  nearby bright star (as confirmed by the EFIGI {\tt contamination} attribute),
  or have the ``COSMIC\_RAY'' flag indicating that the object contains a pixel 
  interpreted to be part of a cosmic ray. This removes another
  570, 252, 754, 634, and 544 EFIGI galaxies that are not included 
  in the $u\le20$, $g\le18$, 
  $r\le18$, $i\le18$, and $z\le18$ SDSS sub-samples respectively.

  The magnitude completeness in each sample is calculated as the ratio
  of the number of EFIGI galaxies to that in the SDSS using 0.5
  magnitude intervals. The resulting curves for the 5 SDSS filters
  are shown as shaded histograms in \fg\ref{compmag}, indicating median
  completenesses of $\sim 30$\% in \textit{u}, and $\sim 60-70$\% in \textit{g},
  \textit{r}, \textit{i}, and \textit{z}.

  These curves however largely underestimate the EFIGI completeness.
  We have examined visually (using the SDSS DR7 ``Explore'' tool) 890
  of the brightest SDSS sources in all 5 filters which are not
  included in the EFIGI catalogue. Only 95 were found to be a bright galaxy with
  the stated Petrosian magnitude in the considered band. One third of
  the other 795 sources are small compact objects (faint stars or
  galaxies), some of which are in the USNO catalogue.  These objects
  have faint SDSS ``psf'' magnitudes of 20 to 22 in the \textit{r} filter, and
  their Petrosian magnitudes are contaminated by the background halo
  of bright stars. The remaining two-thirds are spurious objects
  caused by the halos of bright stars and satellite trails. The
  apparently low completeness ($\le30$\%) in \textit{u} is due to an even higher
  number of spurious sources in this filter than in the other bands. This
  illustrates that the SDSS catalogues are not appropriate for making
  a complete census of nearby galaxies. A clean reanalysis of the
  images is necessary.

  We overplot in \fg\ref{magcomp} as continuous black lines the
  corrected EFIGI completeness curves after removing the aforementioned 795
  spurious sources from the SDSS. Because the number of
  both real and spurious SDSS sources increases steeply at fainter
  magnitudes, these corrected curves are only measured for their first
  5 to 6 bins, and indicate completenesses of $\sim80$\% in the \textit{u}
  and \textit{g} filters, and $\sim85$\% in the \textit{r}, \textit{i}, and
  \textit{z} filters. These findings lead us to conclude that the EFIGI
  catalogue is reasonably complete down to $g\sim 14$.

  We checked that the estimated DR4 completenesses are also valid for
  the SDSS DR7 final release. To this end, we made the same
  extractions from the DR7 as from the DR4 (all objects from the
  ``Galaxy'' catalogue that do not have the ``SATURATED'' flag).
  Depending on the filter, the DR7 contains about 55\% to 75\% more
  galaxies than the DR4 to the same Petrosian magnitude limits. This
  is due to an increased sky coverage: there are about 10 additional
  thin strips and also the part of the northern galactic cap region
  near the declination interval $10^\circ$ to $30^\circ$ missing in
  the DR4 which is now covered (see \fg\ref{clusters}).  In the common
  area of the DR4 and DR7, both releases have similar magnitude
  distributions.  The stated completenesses of the EFIGI catalogue in
  the \textit{ugriz} filters are therefore also valid for the DR7
  within the area covered by the DR4.

  \subsection{Morphological fractions}

  \begin{figure}
    \centering
    \includegraphics[angle=0,width=\columnwidth]{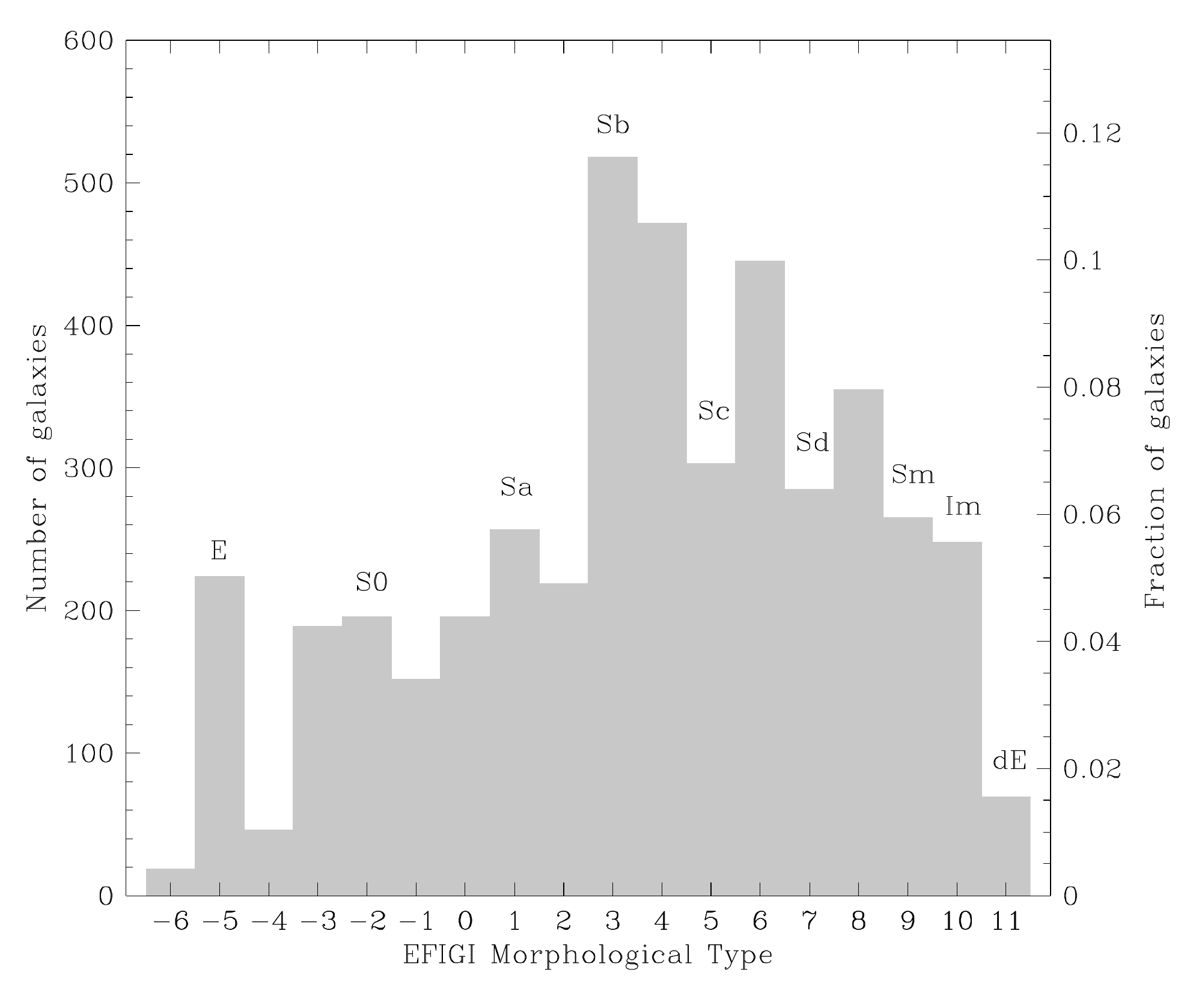}
    \caption{\label{Thist}Distribution of EFIGI morphological types for all
    4458 galaxies.}
  \end{figure}

  \begin{figure}
    \centering
    \includegraphics[angle=0,width=\columnwidth]{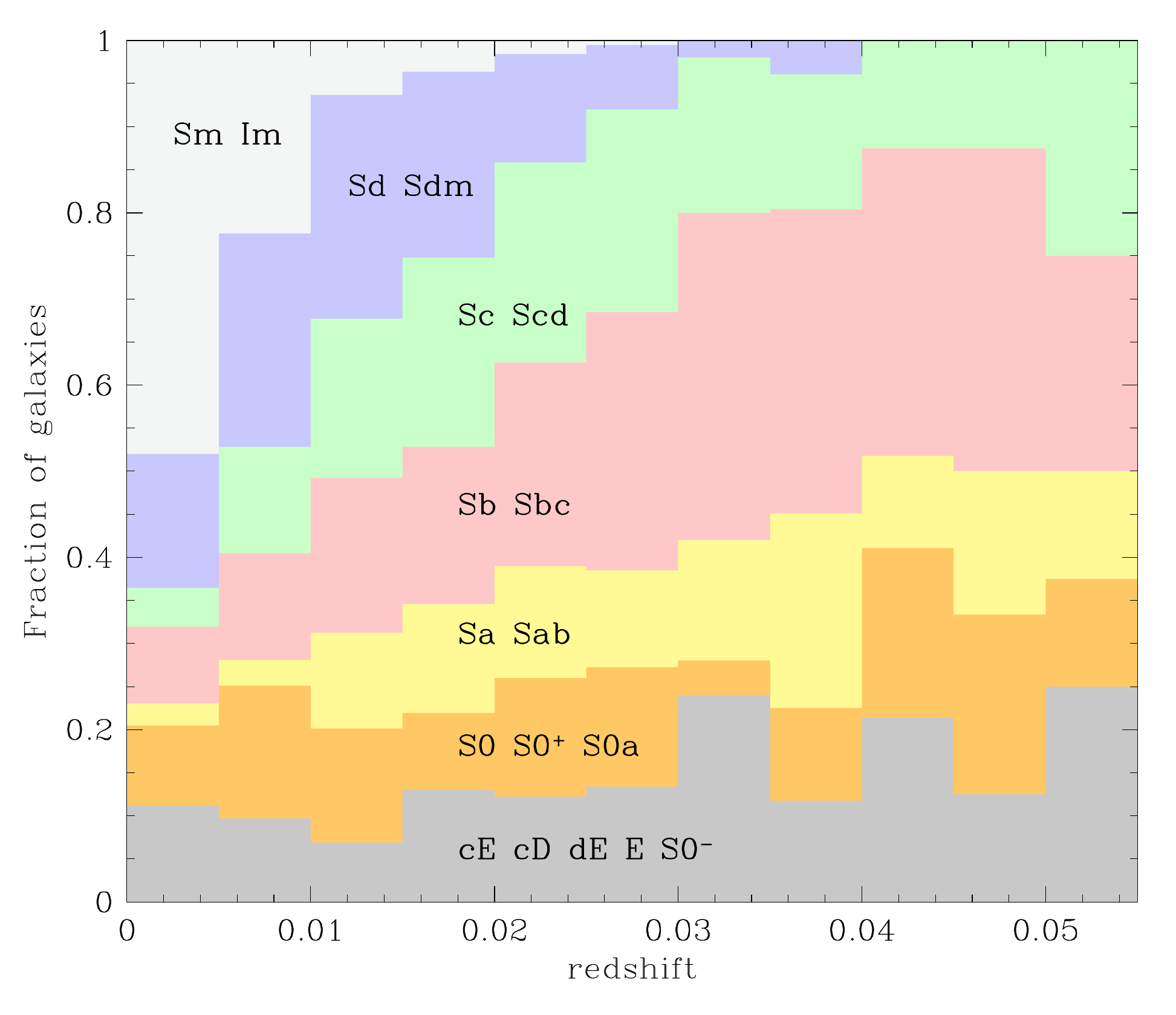}
    \caption{\label{zTfrac}Fraction of EFIGI galaxies with a given morphological
    type as a function of redshift.}
  \end{figure}

  \begin{figure}
    \centering
    \includegraphics[angle=-90,width=\columnwidth]{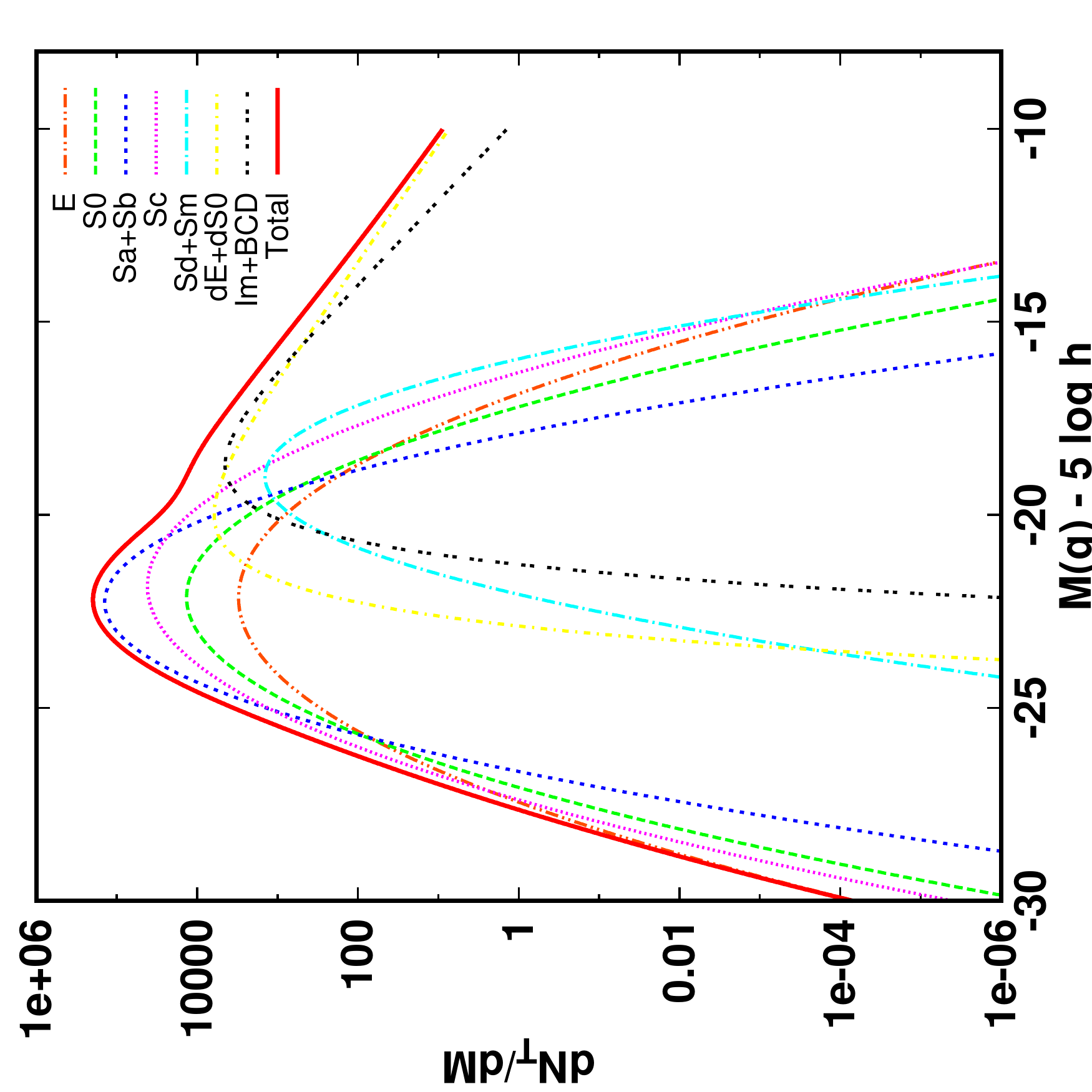}
    \caption{\label{lf}Expected numbers of galaxies with $g<15.75$ per unit of
    absolute V band magnitude within the survey volume of the EFIGI catalogue
    (see text).}
  \end{figure}

  \begin{figure}
    \centering
    \includegraphics[angle=-90,width=\columnwidth]{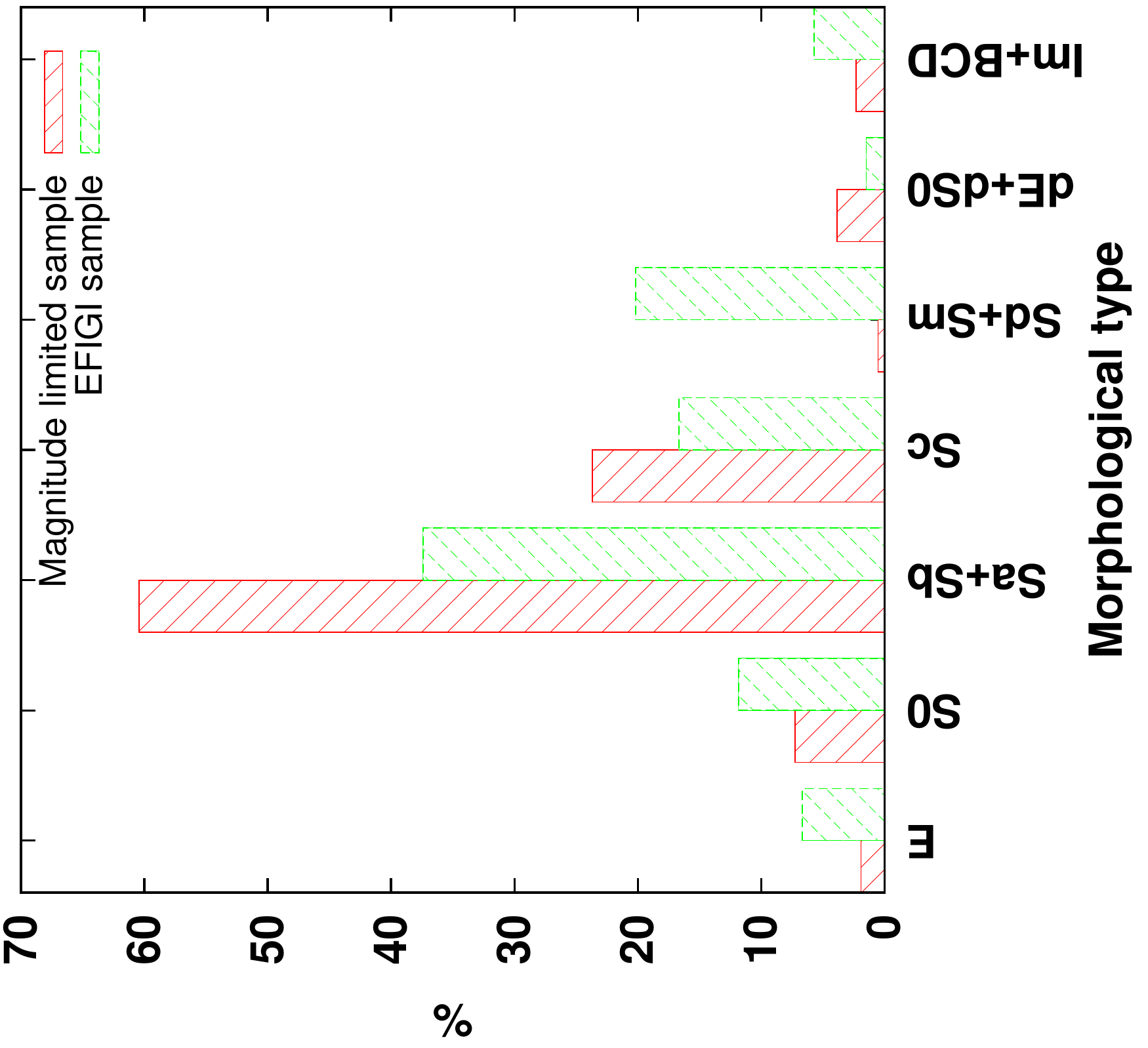}
    \caption{\label{mix}Comparison between the observed EFIGI catalogue type mix
    (in {\em green}) and the expected type mix from a magnitude-limited
    sample with $g<15.75$ (in {\em red}).}
  \end{figure}

  \fg \ref{Thist} shows the numbers and fractions of EFIGI galaxies with
  each EFIGI morphological type. This graph shows that the late-type
  Sd, Sm and Im galaxies are as numerous as Sa and Sc types, with
  $\sim250-300$ galaxies per type; Sb and intermediate types Sbc, Scd,
  and Sdm have even more galaxies per type ($\sim350-450$); only cD,
  cE and dE types, which are intrinsically rare or faint have
  $\sim20-70$ galaxies. This is further shown in \fg \ref{zTfrac}
  where we plot the fractions of galaxies grouped by type as
  a function of redshift. At redshifts $z<0.02$, the Sd, Sdm, Sm and Im galaxies
  represent $\sim 41$\% of the sample.

  The EFIGI sample is not a volume-limited sample. \citet{lapparent1} show
  that it is essentially limited in apparent diameter, as measured by $D_{25}$ 
  \citep{rc2}, following the RC3 selection criterion \citep{devaucouleurs}. It is 
  nevertheless interesting to compare the distribution of morphological types in
  the EFIGI catalogue with that expected from a magnitude-limited sample, since
  setting a magnitude cut-off is the most common selection scheme in morphological
  follow-up studies of photometric surveys \citep{lintott,nair}. 

  We compute the fractions of galaxies by morphological 
  type from the luminosity functions (LFs hereafter) listed in \cite{lapparent03a}
  and \cite{lapparent03b}. The LF parameters are extracted from Table 6 of 
  \cite{lapparent03a} for the shape parameters (using $\sum_2$ for ellipticals 
  and Centaurus parameters for dE+dS0 and Im+BCD galaxies) and from Table 1 of 
  \cite{lapparent03b} for the amplitudes $\phi^*$. The mapping made between the 
  LF morphological types and the EFIGI types is given in Table \ref{mixmap}. 
  Colour conversion terms between the Johnson-Cousins system and SDSS
  system are extracted from \cite{jordi}.

  \begin{table}
    \caption{\label{mixmap}Morphological type mapping between
    luminosity function types and EFIGI types.}
    \centering
    \begin{tabular}{lr}
      \hline
      \hline
      LF types & EFIGI types\\
      \hline
      E & cE,E,cD\\
      S0 & S0$^-$,S0$^0$,S0$^+$\\
      \hline
      Sa/Sb & S0/a,Sa,Sab,Sb,Sbc\\
      Sc & Sc,Scd\\
      Sd/Sm & Sd,Sdm,Sm\\
      \hline
      dE+dS0 & dE\\
      Im+BCD & Im\\
      \hline
    \end{tabular}
  \end{table}

  The galaxy counts per morphological type are then calculated as:
  \begin{equation}
    N_{T} = \int_{-\infty}^{-10}\phi_{T}(M)V_{lim}(M)dM ;
  \end{equation}
  $\phi_T(M)$ is the LF for Hubble type T and $V_{lim}$ the
  volume defined as
  \begin{equation}
    V_{lim}(M) = \frac{\Omega}{3} 10^{0.6(g_{lim} - M) - 15},
  \end{equation}
  where $\Omega$ is the solid angle (6670 deg$^2$), and $g_{lim}$ is the
  \textit{g}-band apparent magnitude limit of the sample. Based on \fg\ref{compmag} 
  above, and \fg 3 of \citet{lapparent1}, we use $g_{lim}=15.75$. 

  The differential fractions $\frac{dN_{T}}{dM}$ for the various
  galaxy types are plotted in \fg \ref{lf}, and the integrated
  fractions obtained from the homogenised EFIGI catalogue are shown in
  \fg\ref{mix}. The giant galaxy types (E) dominate at $M - 5 \log
  h \le -22.5$ whereas small spirals (Sd+Sm), Im and dE contribute
  significantly at $M - 5 \log h > -18$. As a result, the difference
  in morphological fractions between the EFIGI catalogue and the LF
  predictions are for Sd+Sm galaxies which are largely missing in the
  (bright) magnitude-limited sample. The opposite occurs for Sa+Sb
  types. The EFIGI sample also recovers three times and twice more E
  and Im galaxies \resp, whereas it undersamples the Sa+Sb and the Sc
  types by $\sim30$\% compared to the magnitude-limited sample.

  \subsection{Redshift distribution}

  \begin{figure}
    \includegraphics[angle=0,width=\columnwidth]{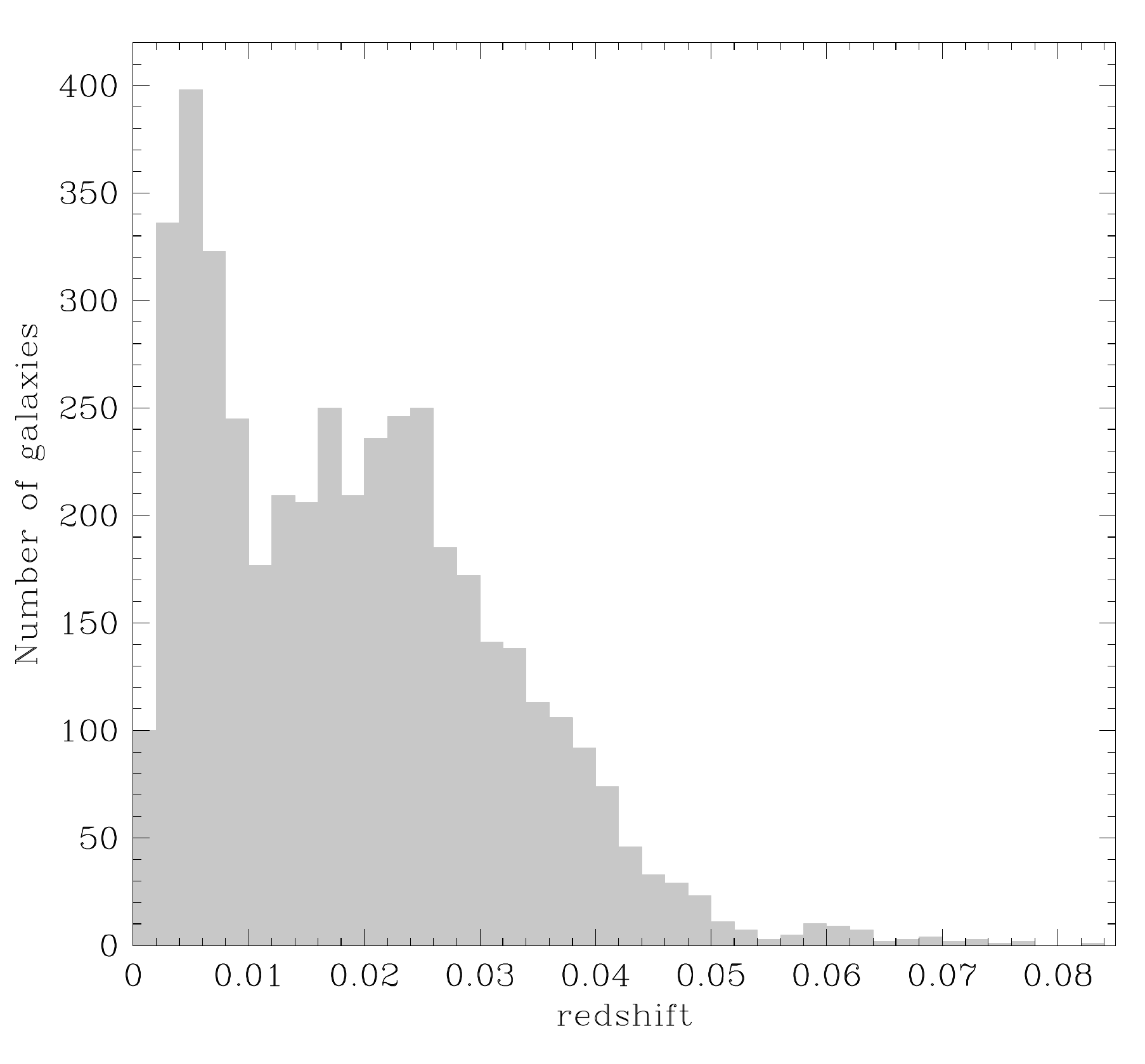}
    \caption{\label{zhist}The redshift distribution of the 4437 EFIGI galaxies
   with a redshift measurement. The peak at $z\simeq0.004$ is due to the Virgo Cluster.}
  \end{figure}

  \begin{figure}
    \centering
    \includegraphics[width=\columnwidth]{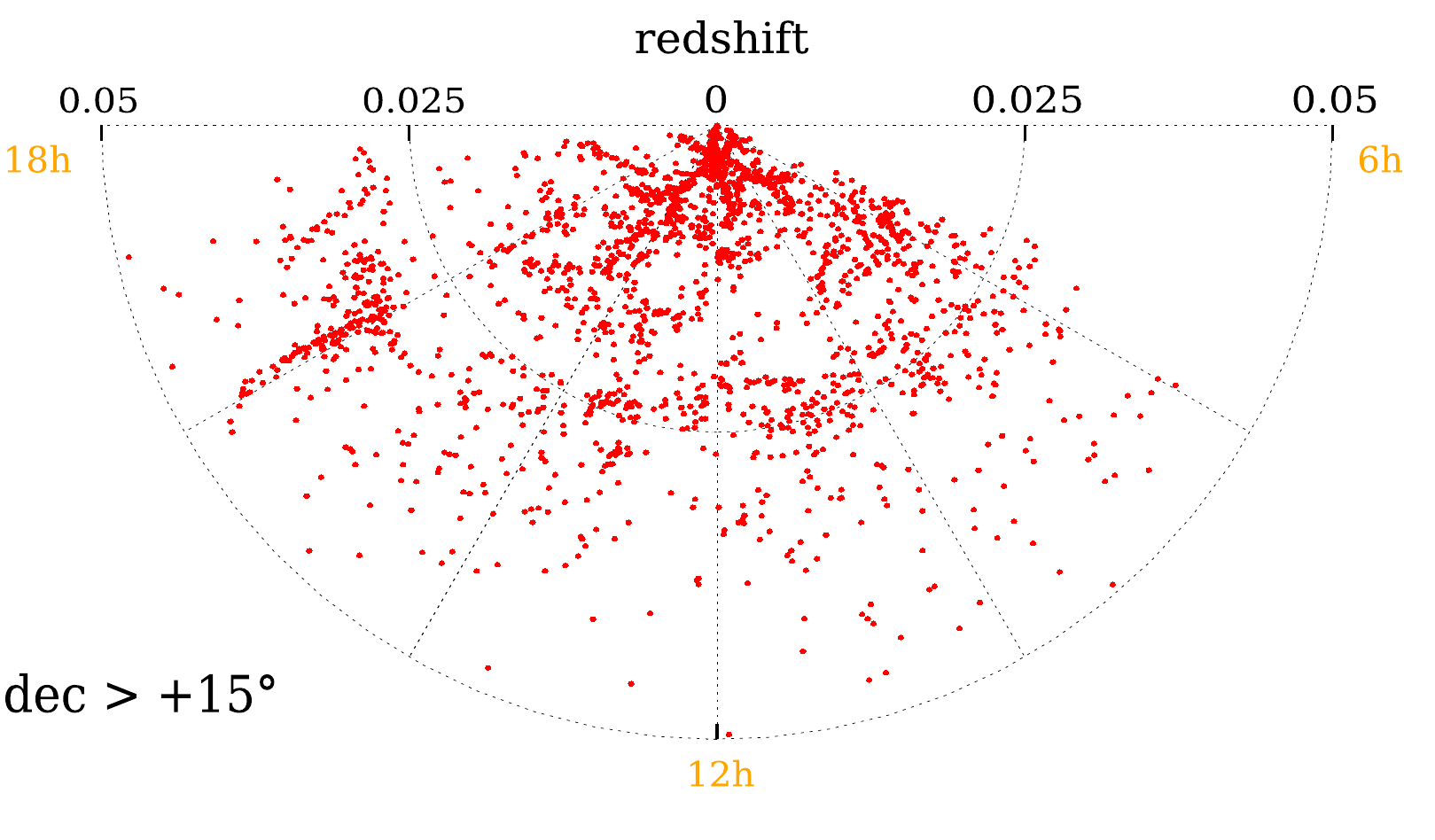}
    \includegraphics[width=\columnwidth]{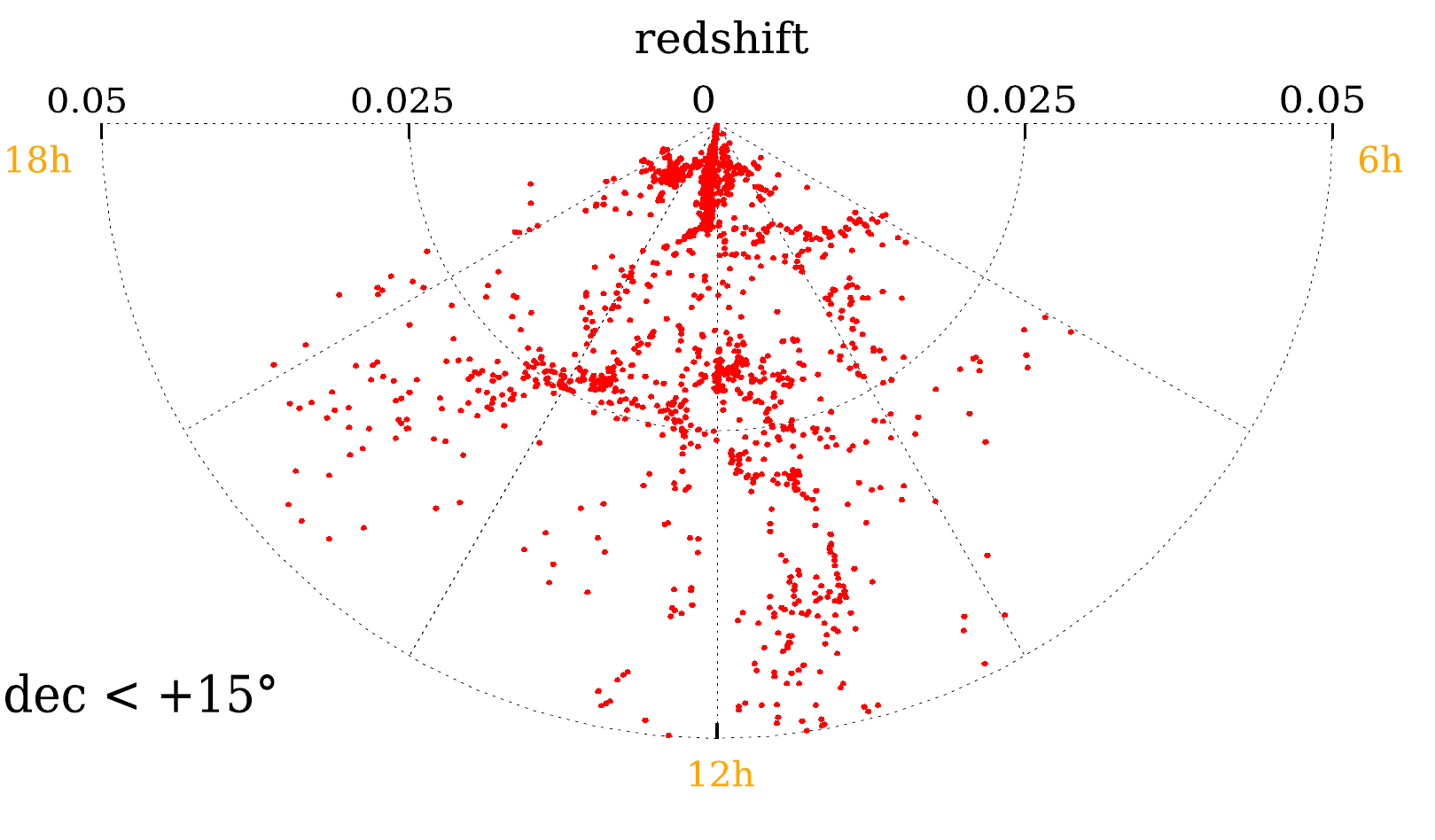}
    \includegraphics[width=\columnwidth]{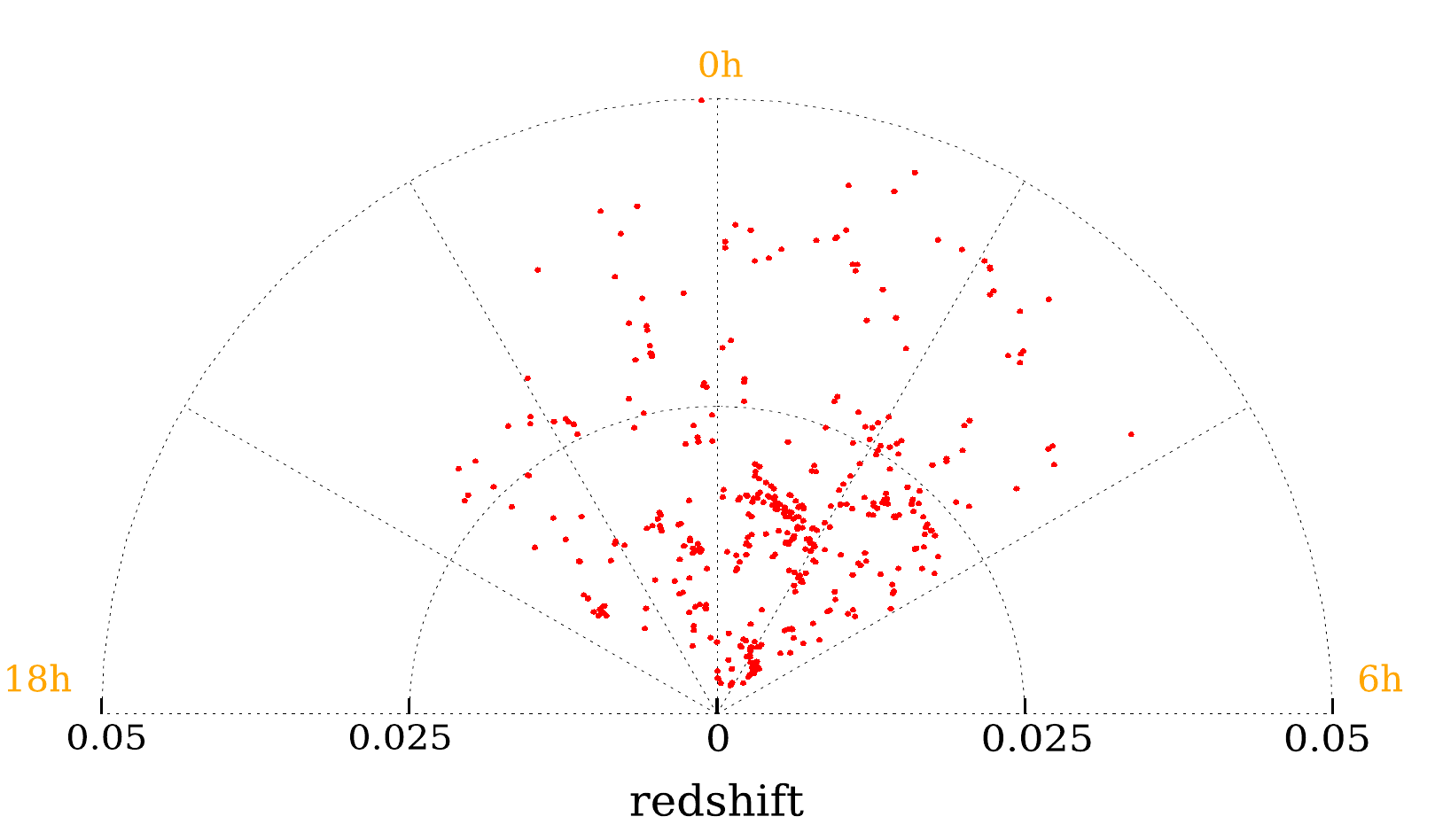}
    \caption{\label{coneplot}Redshift pie-diagrams of EFIGI galaxies. The
    top panel corresponds to $6\,{\rm h} < \alpha < 18\,{\rm h}$
    and $\delta > 15^\circ$. Middle panel: $6\,{\rm h} < \alpha < 18\,{\rm h}$
    and $\delta < 15^\circ$. Bottom panel: $18\,{\rm h} < \alpha < 6\,{\rm h}$.}
  \end{figure}

  The redshift completeness of the EFIGI catalogue is high: 4437 of the 4458 EFIGI galaxies
  have a redshift, that is 99.53\%; $\sim 90$\% of redshifts are from HyperLeda, 
  $\sim 9$\% from NED, the rest from SDSS (see \citealp{lapparent1} for details).
  The majority of galaxies in the catalogue (4365 over 4458) have $z < 0.05$ 
  as shown on \fg \ref{zhist}, which is consistent
  with the RC3 selection process ($cz < 15,000$ km s$^{-1}$). This graph
  also shows that the redshift distribution of the EFIGI catalogue differs from that for
  a magnitude-limited sample by having a denser sampling at redshifts lower
  than the peak of the redshift distrbution ($z\sim0.025$). 
  Moreover, the Virgo Cluster contains
  nearly half of the galaxies with $z\le0.01$.

  \fg \ref{coneplot} displays 3 pie-diagrams, 2 for the northern galactic cap, and 
  one for the 3 strips in the southern galactic cap. 
  Despite the spatially inhomogeneous sampling of the EFIGI catalogue, 
  the large-scale structure of the universe is visible, with clusters,
  voids and walls \citep{lapparent86}.

\section{\label{sec:nair}Comparison with \cite{nair}}

  \begin{figure*}
\centerline{\includegraphics[width=0.27\textwidth,angle=-90]{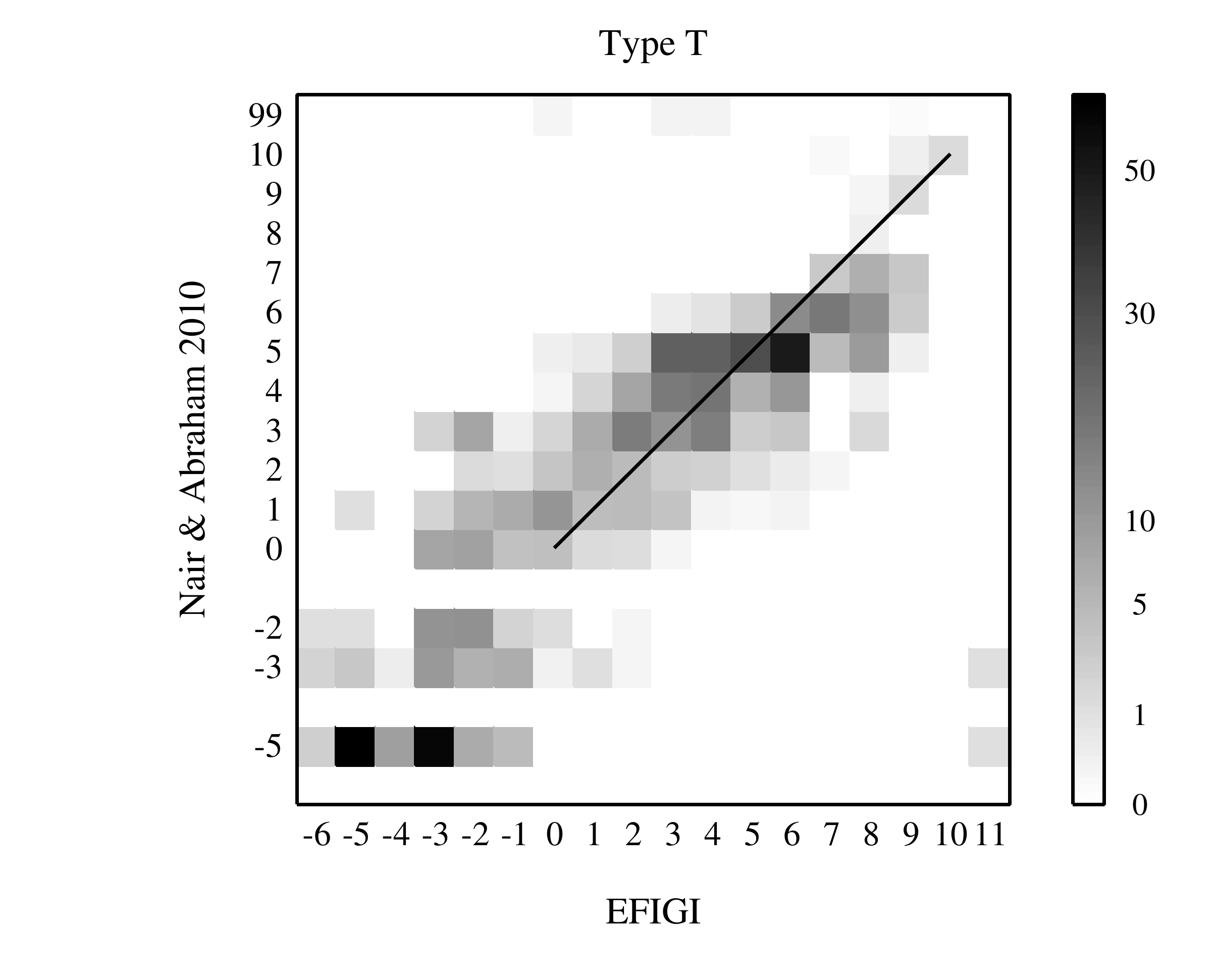}
	\includegraphics[width=0.27\textwidth,angle=-90]{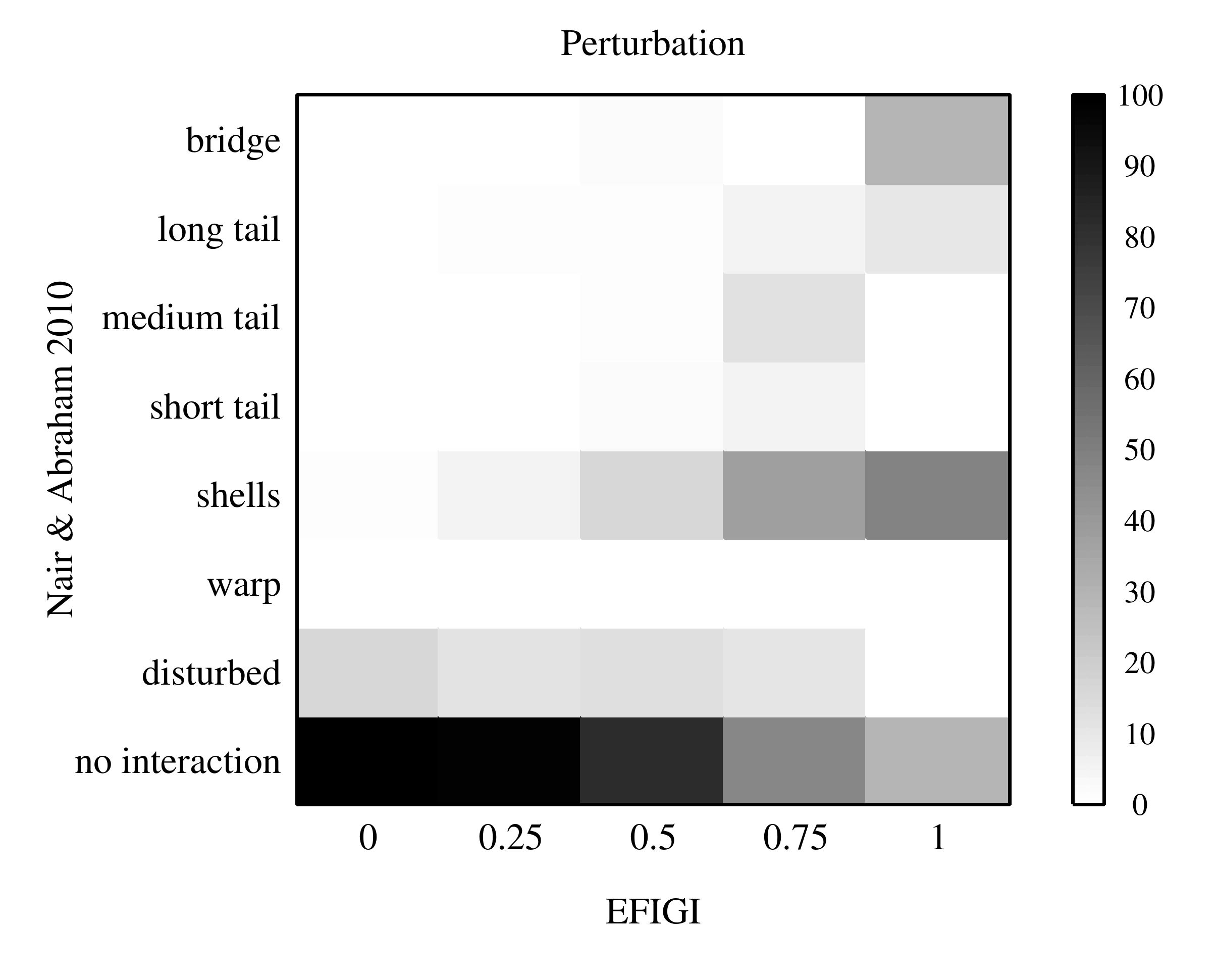}
	\includegraphics[width=0.27\textwidth,angle=-90]{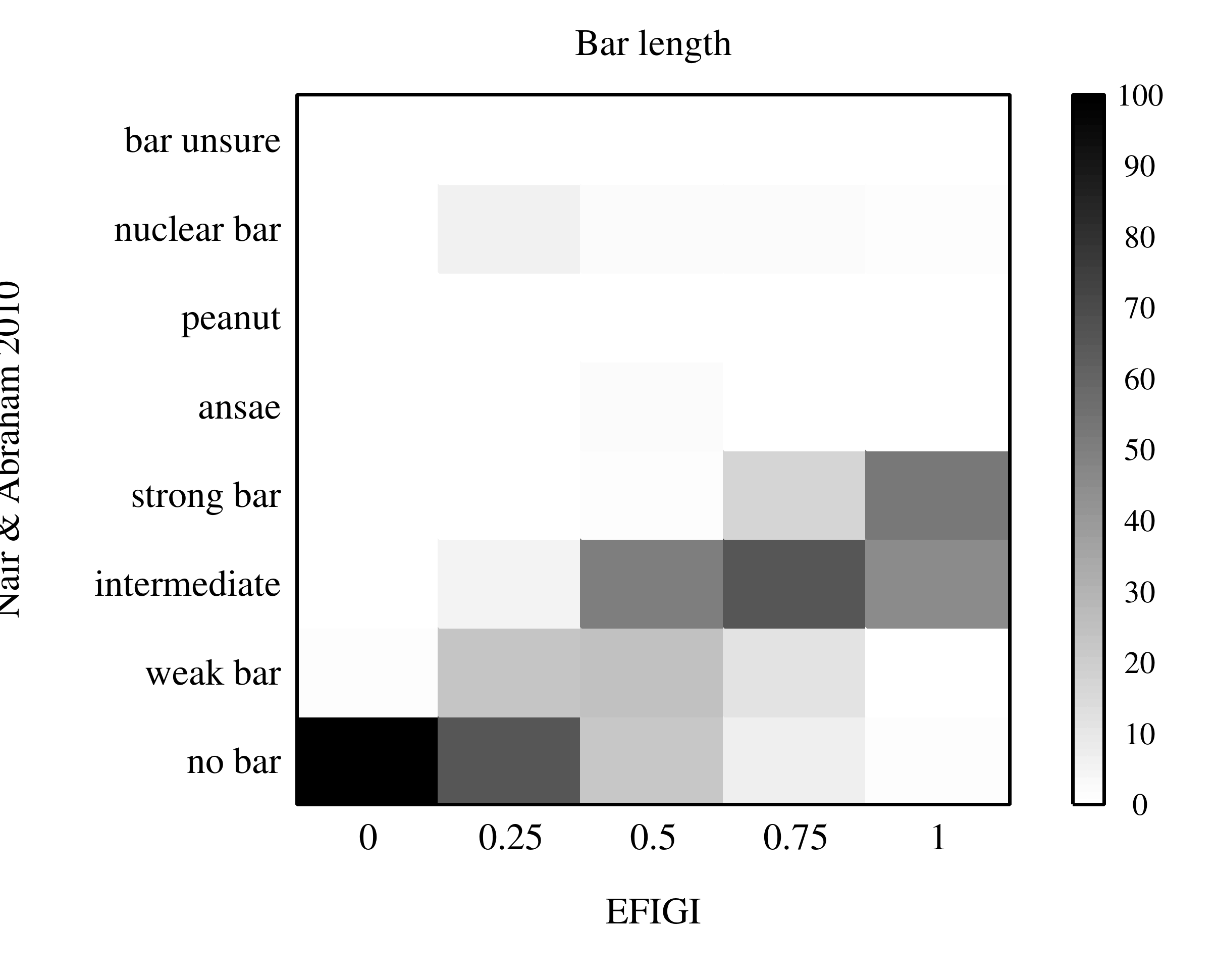}}
\centerline{\includegraphics[width=0.27\textwidth,angle=-90]{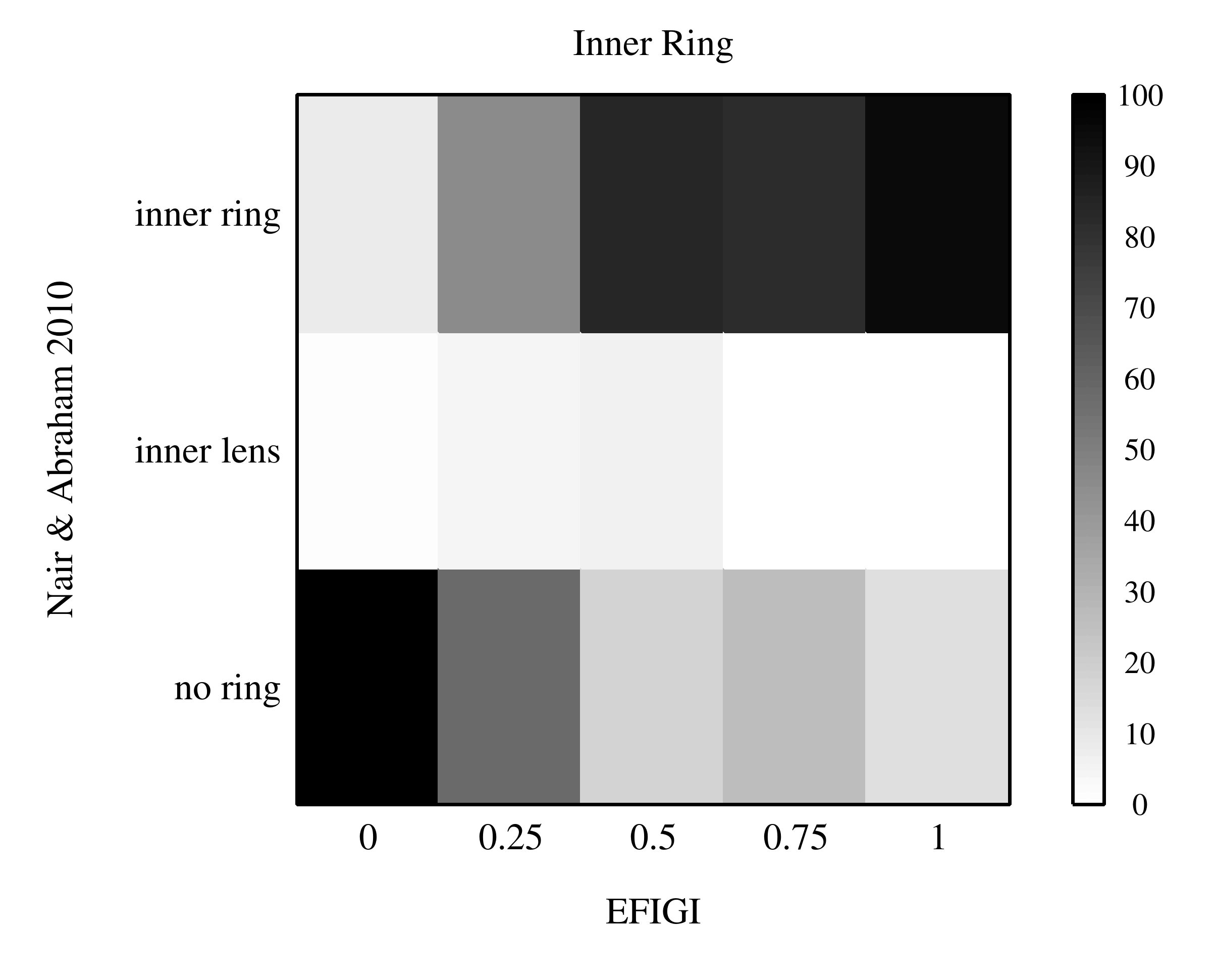}
	\includegraphics[width=0.27\textwidth,angle=-90]{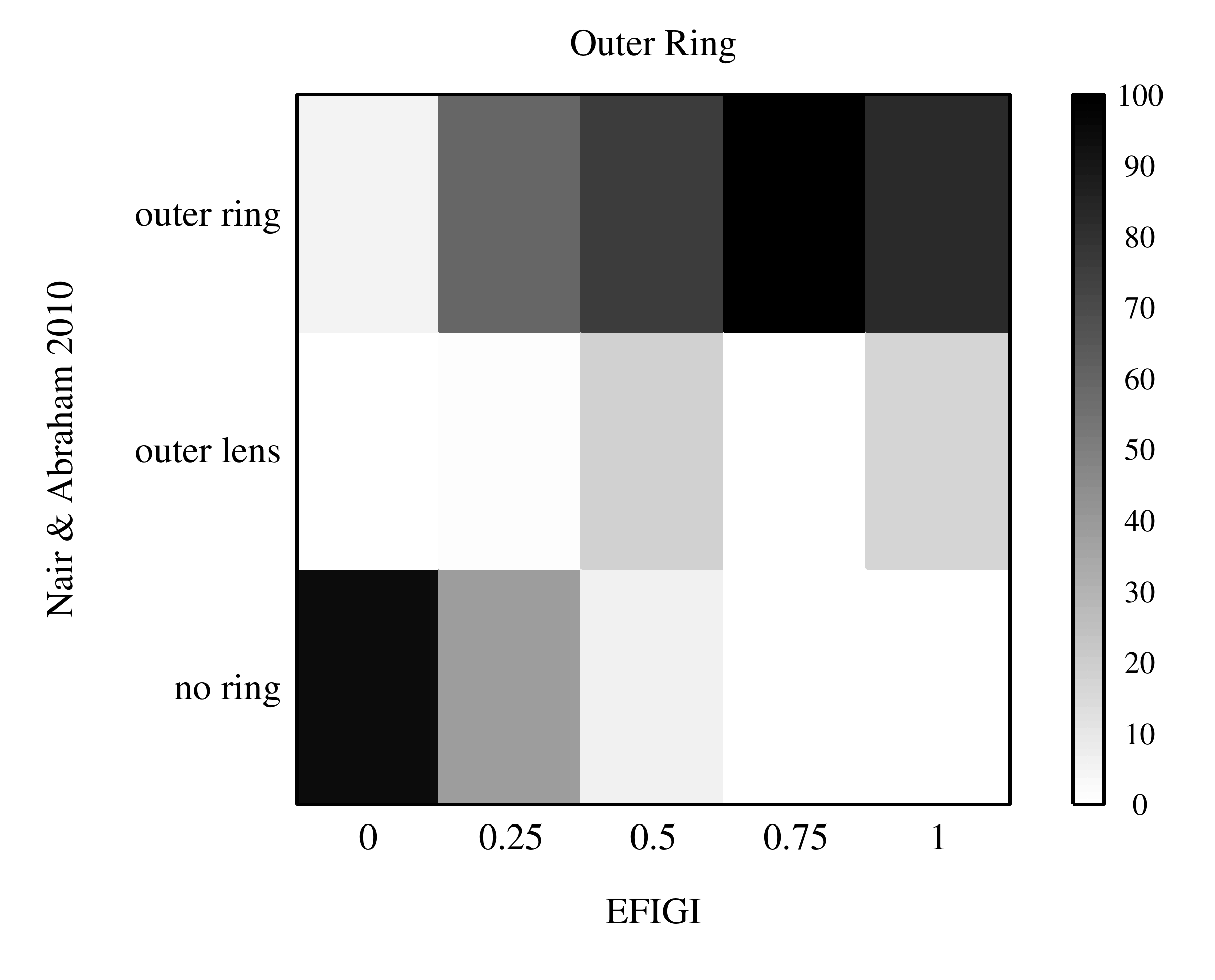}
	\includegraphics[width=0.27\textwidth,angle=-90]{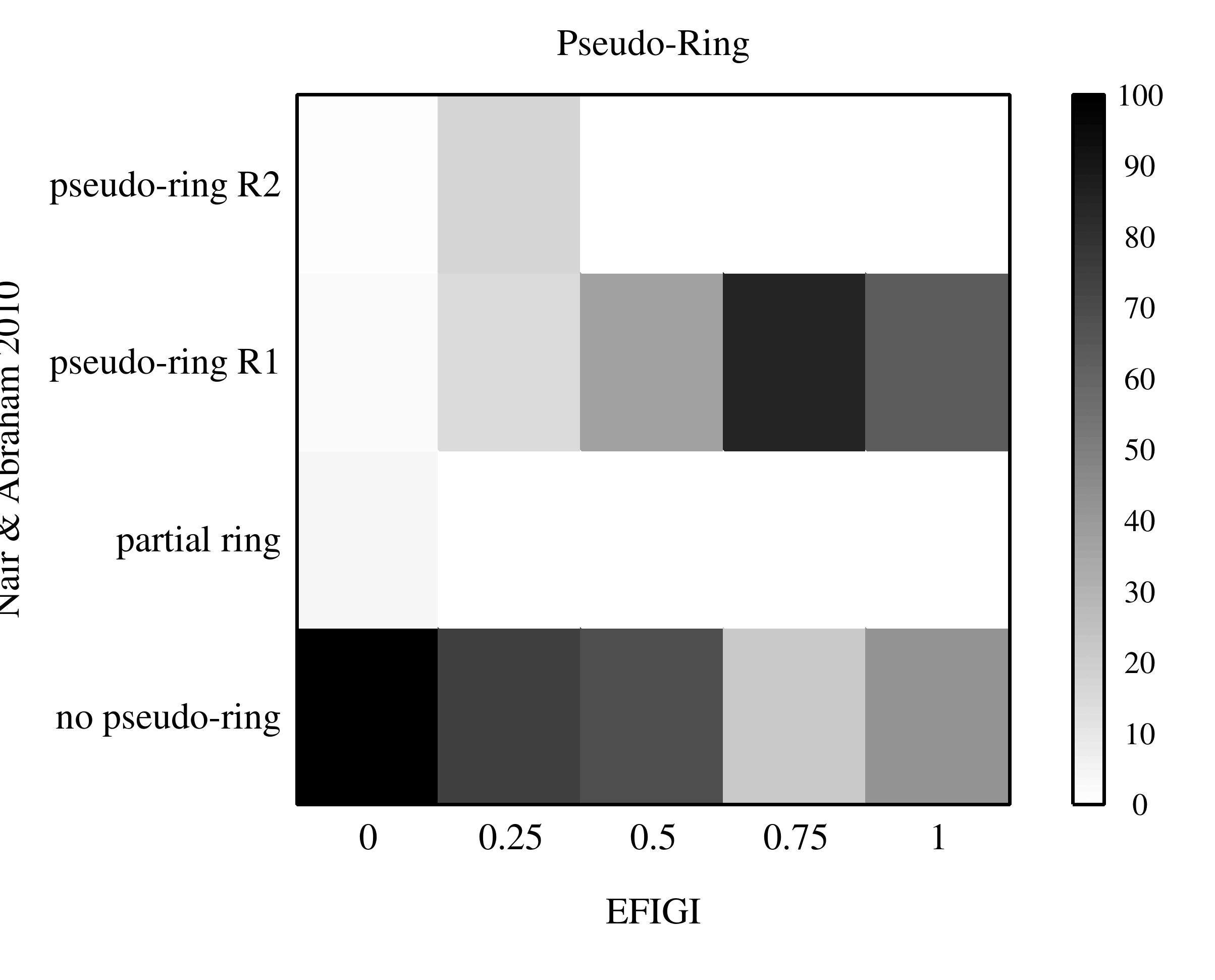}}
\caption{Comparison between the EFIGI and the \citet{nair} classifications for
the 1438 galaxies in common with both catalogues. From left to
right, top to down: EFIGI Hubble type (the straight line indicates perfect
agreement),
{\tt perturbation}, {\tt bar length}, {\tt inner ring}, {\tt outer ring}, and
{\tt pseudo-ring} attributes are compared to the Hubble type and corresponding 
features recorded by \citet{nair}. Results are shown in the form of confusion
matrices; the contribution of each galaxy is weighted as in \fg\ref{trends}.
For the Hubble type, cell values are proportional to the effective
number of galaxies per cell (with the corresponding numbers indicated on the
vertical scale on the right). For the attributes, cell values are expressed in
percentage of the total effective number of galaxies per EFIGI attribute bin.}
    \label{nacomp}
  \end{figure*}

  \citet[ NA2010 hereafter]{nair} have recently released a
  morphological catalogue of 14 034 visually classified SDSS
  galaxies extracted from the SDSS DR4 spectroscopic release.
  The NA2010 catalogue is limited to objects brighter than SDSS \textit{g}-band
  magnitude 16, and lie within a redshift interval $0.01<z<0.1$. In addition to a
  morphological type based on the RC3 classification, a number of flags
  describing the morphological appearance of bars, rings, lenses, and
  perturbation features are provided. Bar strength is divided in four levels:
  no bar, weak, intermediate, strong. Others feature indicators are strictly
  binary.
  The NA2010 galaxy type is based on \textit{g}-band images only, whereas 
  morphological flags rely also on $r$ and $i$ images when necessary, 
  in order to confirm the presence of certain features.

  Despite the significantly different classification criteria used by NA2010, 
  a rough comparison with the EFIGI catalogue can be made for the morphological
  type and some features in common between both catalogues: {\tt perturbation},
  {\tt bar length}, {\tt inner ring}, {\tt outer ring}, and {\tt pseudo-ring}
  (\fg\ref{nacomp}).
  One third of EFIGI galaxies have redshifts below the $z=0.01$ cut-off in NA2010, 
  and 15\% have SDSS (incorrect) \textit{g} magnitudes $>16$, hence only 1438 galaxies 
  are in common within a cross-identification radius of 20 arcsec, that 
  is about 10\% of the NA2010 sample.

  A comparison of Hubble types between both catalogues (\fg\ref{nacomp}, top left) 
  exhibits essentially the same systematic features as
  \fg 14 of NA2010 (comparison with the RC3 catalogue). This is not surprising
  since the EFIGI classification scheme is very close to the RC3 system
  (\fg\ref{confHT}).
  NA2010 do not distinguish between cE, cD, dwarf spheroidal and ``regular''
  ellipticals. Galaxies classified as lenticulars or early spirals (Sa, Sab, Sb,
  Sbc) in the EFIGI catalogue are, in average, shifted towards later stages in
  NA2010. An opposite effect is seen for later spirals (Sc, Scd, Sd).
  The agreement for Sdm, Sm and Im seems better, although this only comes from
  a handful of galaxies; in total these types comprise $\sim 1$\% of all
  NA2010 galaxies, compared to $\sim20$\% in the EFIGI sample.

  Comparing the EFIGI {\tt perturbation} index to NA2010
  ``distortion'' flags sheds light on the staging process. First of
  all, a major fraction of galaxies described as slightly or
  moderately ``perturbed'' in the EFIGI catalogue are not flagged as
  distorted in NA2010, which we interpret as suggesting the EFIGI
  classification might be more sensitive towards relatively ``benign''
  features such as spiral arm asymmetry.  Although there is a clear
  trend towards galaxies with higher {\tt perturbation} indices
  containing a higher fraction of tails, bridges or shells (up to
  $\sim70$\% altogether for galaxies with {\tt perturbation} = 1), the
  fraction of objects classified as ``disturbed'' by NA2010 is
  essentially stable along the perturbation sequence, at about 10\%.

  The {\tt bar length} EFIGI index appears to follow very closely
  the scale set by ``bar strength'' flags in NA2010. We note that bars with 
  EFIGI {\tt bar length} = 0.25 to 0.75 are detected in the EFIGI catalogue 
  for a fraction of objects with ``no bar'' in NA2010, whereas there is essentially
  no case of bar detection in NA2010 for EFIGI objects with {\tt bar length} = 0.
  Finally, galaxies flagged as having a nuclear bar in NA2010 are consistently
  assigned a {\tt bar length} of 0.25 in the EFIGI sample.

  There is also a good agreement between the NA2010 ring flags and the EFIGI
  {\tt inner ring} and {\tt outer ring} attributes. The $\sim30$\% to $\sim60\%$
  of EFIGI galaxies with a value of 0.25 for {\tt inner ring} and {\tt outer ring} \resp\
  maybe be due to tight spiral arms in a galaxy centre resembling a ring for
  the former attribute, and to portions of rings for the latter; the former may
  not have activated the NA2010 inner ring flag, whereas the latter may have 
  activated instead the NA2010 ``distortion'' flag for tails or shells.
  The inner and outer lenses were deliberately graded as 
  {\tt inner ring} and {\tt outer ring} \resp\ in the EFIGI sample. Comparison
  with the NA2010 `` inner lens''  and `` outer lens'' flags indicates that
  lenses in EFIGI galaxies amount to up to 6\% for {\tt inner ring} and up to 
  18\% for {\tt outer ring}.

  The comparison of the EFIGI {\tt pseudo-ring} attribute with the
  corresponding NA2010 flags confirms the inclusion of both
  $R_1^\prime$ and $R_2^\prime$ types into this EFIGI attribute, with
  respective attribute values of 0.75-1.0 and 0.25-0.5 (see
  \sct\ref{chap:att}). The additional EFIGI detections with {\tt
    pseudo-ring} attribute values of 0.25-0.5 whereas the flag is not
  activated in NA2010 are likely to correspond to $R_1R_2^\prime$ 
  types, which are not flagged in NA2010. There is also a number of
  EFIGI galaxies with $R_1^\prime$ rings ({\tt pseudo-ring} attribute
  values of 0.75-1.0) which have not been flagged by NA2010.

  To conclude, EFIGI perturbation, bar and ring attributes are in good
  agreement with the corresponding NA2010 flags. It is unclear whether
  the low level detections in EFIGI testify to higher sensitivity or
  false detections; in principle the combination of a five-level scale
  and a confidence interval for every EFIGI attribute is expected to
  provide the best representation of marginally detectable features.

\section{\label{sec:conclusion}Summary and perspectives}

We present the EFIGI catalogue, a multiwavelength catalogue of 4458 galaxies selected from
the PGC and observed with the SDSS. The catalogue provides
detailed morphological information using 16 morphological attributes:
{\tt B/T ratio}, {\tt arm strength}, {\tt arm curvature}, {\tt rotation},
{\tt visible dust}, {\tt dust dispersion}, {\tt flocculence}, {\tt hot spots},
{\tt bar length}, {\tt inner ring}, {\tt outer ring},
{\tt pseudo-ring}, {\tt perturbation},
{\tt inclination/elongation}, {\tt multiplicity}, {\tt contamination},
These attributes are defined on a five-level scale, 
and describe the internal properties and the different components
of a galaxy: bulge, spiral arms, dynamical features, textural aspect, 
as well as its appearance and environment. All galaxies are also
classified along a morphological sequence based on the RC3 Revised
Hubble Sequence. 

The EFIGI catalogue (attributes and Hubble type), complemented by
identification, morphological and redshift data from the PGC, SDSS,
HyperLeda and NED, is available on the {\tt http://www.efigi.org}
website, along with additional imaging products: FITS image data,
colour images, and PSF models.  The catalogue is also available from
the ``Centre de Donn\'ees Astronomiques de Strasbourg'' (CDS) using
the ViZiER Catalogue Service. Here we show, for the 5 levels of each
attribute, composite colour images of representative and atypical
galaxies over a wide variety of Hubble types.

Various statistical tests demonstrate the reliability of this
morphological description. We find a remarkable level of agreement
between the RC3 Revised Hubble Sequence and the EFIGI Hubble sequence
over the 15 common classes, despite classifications being carried out
by different astronomers based on different imaging
material. Moreover, the sky distribution of the weighted average
attribute strengths shows for all 16 attributes the absence of obvious
trends as a function of right ascension along which EFIGI galaxies
are ordered. The projected sky distribution of the EFIGI catalogue
also shows that it covers the whole area of the SDSS Data Release 4,
with ten known clusters of galaxies identified as concentrations of
more than 5 galaxies in the catalogue.

Confusion matrices of the 16 attributes quantify the strong correlation 
observed between the {\tt B/T ratio} and {\tt arm curvature},
a major criterion for defining the progression of spiral types along the Hubble sequence.
The anti-correlations between the {\tt B/T ratio} and the various
other attributes that increase in strength along the Hubble sequence,
{\tt flocculence}, {\tt hot spots}, {\tt perturbation}, and {\tt arm
strength}, are also highlighted.  These latter attributes, together with 
{\tt visible dust} are correlated among themselves 
because they are related indicators of the star formation in a
galaxy.  The {\tt perturbation} attribute is also strongly correlated
with both {\tt hot spots} and {\tt flocculence}, as tidal and merging
processes contribute to enhance star formation. We also detect
the strong correlation between {\tt internal ring} and {\tt bar
length}.

Contrary to the bell-shape $B_T$ magnitude distribution, the SDSS
\textit{ugriz} distributions show faint extensions (for Petrosian
magnitudes fainter than $\sim$ 20 in \textit{u} and fainter than
$\sim17$ in \textit{g}, \textit{r}, \textit{i}, and \textit{z}). This
reflects the failure of the SDSS photometric pipeline for large
late-type spirals and irregular galaxies which are split into multiple
units. More details are provided in de Lapparent \& Bertin (2011a, in prep{.}) where unbiased
photometric properties are derived for all EFIGI galaxies using a new
version of {\sc SExtractor} \citep{bertin96} incorporating
model-fitting capabilities \citep{bertin10}.

We show that the EFIGI catalogue is a dense sub-sample of the local Universe:
over the area of the SDSS DR4, its photometric completenesses at
bright magnitudes (14.5 in \textit{u} and $12.5-13.5$ in \textit{g}, \textit{r}, \textit{i}, and \textit{z}) 
are $\sim80$\% in the \textit{u} and \textit{g} filters, and $\sim85$\% in the \textit{r}, \textit{i},
\textit{z} filters. This is much higher than the raw completenesses calculated
from the SDSS photometry: $\sim30$\% in \textit{u} and $\sim 60-70$\% in \textit{g},
\textit{r}, \textit{i}, and \textit{z}. When examining the SDSS galaxies not present in the EFIGI catalogue at
bright magnitudes, we found that 90\% of the sources are spurious
and are caused by haloes of a bright saturated stars, or less frequently,
by satellite trails.

Most of EFIGI galaxies (99.53\%) have a redshift (mostly from the HyperLeda database),
and the vast majority have  $z\la0.05$.
Pie-diagram redshift maps show the large-scale clustering of galaxies,
with a strong contribution from the Virgo cluster at $z\la0.01$. 

Comparison of the EFIGI morphological type fractions with the typical
expectations for a magnitude-limited sample of the local Universe show
that the EFIGI catalogue largely oversamples late-type spirals (Sd,
Sdm, Sm) and irregulars. These types are as or more numerous than the
EFIGI early types (E, S0$^-$, S0, S0$^+$, Sa, Sab). The late types
also appear mainly at redshift $z\la0.03$, due to their faint absolute
magnitude, whereas earlier types are present at all redshifts. As
shown by \citet{lapparent1}, this is a consequence of the apparent
diameter limit of the EFIGI catalogue.

As a first application, the EFIGI morphological attributes were used
for supervised learning tasks: using ``Support Vector Machines'' 
\citet{baillard08} showed that one can determine the Hubble type
from a reduced number of EFIGI attributes, which are, in decreasing
order of significance: the bulge-to-total luminosity ratio, the
strength of spiral arms, the curvature of spiral arms; the amount of
visible dust and flocculence play a less significant role in
determining the Hubble type. 

An experimental test of the automatic
determination of morphological attributes has also been performed
within {\sc SExtractor}, using a projection of the residual profiles
onto ring-based functions \citep{baillard08}. More recently, \cite{perret:al}
have applied an MCMC algorithm with multiple temperature simulated annealing
to the EFIGI image data-set to perform multispectral decompositions of galaxy
components. Ultimately, merging an automatic
determination of the morphological estimators with the determination of
the Hubble Type should allow one to perform a fully automatic determination of
the Hubble Type from galaxy images.

The dense sampling of all Hubble types by the EFIGI catalogue, and its
high photometric and spectroscopic completenesses make this catalogue
directly usable for statistical analyses of the galaxy distribution at
low redshift.  A companion article \citep{lapparent1} reports on the
statistical analysis of the EFIGI morphological attributes, and
provides for the first time a quantitative description of the Hubble
Sequence in terms of specific morphological features.

Two other articles further use the EFIGI catalogue to study the 
properties of nearby galaxies: the bulge and disk properties along 
the Hubble sequence (de Lapparent \& Bertin, 2011a, in prep{.}), using the new profile fitting 
tools within {\sc SExtractor} \citep{bertin96,bertin10}; the luminosity
functions as a function of morphological type and galaxy component 
(de Lapparent \& Bertin, 2011b, in prep{.}). The first of these analyses illustrates the usefulness
of a well-defined calibration sample for application of the automatic
tools for measuring galaxy morphometry. Detailed analyses at low
redshift are also indispensable for understanding the evolution
of galaxy morphometry with look-back time using deeper imaging as in
the Canada-France-Hawaii Telescope Legacy survey.

\begin{acknowledgements}

We are grateful to the referee, Ron Buta, 
for his very useful comments.
This work has been supported by grant 04-5500 (``ACI masse de donn\'ees'') from
the French Ministry of Research.

This research made use of the HyperLeda database
(http://leda.univ-lyon1.fr), the VizieR catalogue access tool
\citep{ochsenbein:al} and the Sesame
service at CDS (Strasbourg, France), and the NASA/IPAC Extragalactic
Database (NED), which is operated by the Jet Propulsion Laboratory,
California Institute of Technology, under contract with the National
Aeronautics and Space Administration.

This publication also made use of the Sloan Digital Sky Survey. Funding
for the SDSS and SDSS-II has been provided by the Alfred P. Sloan
Foundation, the Participating Institutions, the National Science
Foundation, the U.S. Department of Energy, the National Aeronautics
and Space Administration, the Japanese Monbukagakusho, the Max Planck
Society, and the Higher Education Funding Council for England. The
SDSS Web Site is http://www.sdss.org/.  The SDSS is managed by the
Astrophysical Research Consortium for the Participating
Institutions. The Participating Institutions are the American Museum
of Natural History, Astrophysical Institute Potsdam, University of
Basel, University of Cambridge, Case Western Reserve University,
University of Chicago, Drexel University, Fermilab, the Institute for
Advanced Study, the Japan Participation Group, Johns Hopkins
University, the Joint Institute for Nuclear Astrophysics, the Kavli
Institute for Particle Astrophysics and Cosmology, the Korean
Scientist Group, the Chinese Academy of Sciences (LAMOST), Los Alamos
National Laboratory, the Max-Planck-Institute for Astronomy (MPIA),
the Max-Planck-Institute for Astrophysics (MPA), New Mexico State
University, Ohio State University, University of Pittsburgh,
University of Portsmouth, Princeton University, the United States
Naval Observatory, and the University of Washington.

\end{acknowledgements}

  \bibliographystyle{aa}
  \bibliography{article}

\appendix

\section{\label{apx:tables}Data products and tables}

The EFIGI catalogue is available at CDS and in the dedicated
  website {\tt http://www.efigi.org}. The catalogue contains for each object the
  EFIGI morphological type and 16 morphological attributes as listed
  in Table \ref{EFIGIatt}; although not indicated in Table \ref{EFIGIatt}
  for clarity reasons, we provide for each attribute the lower and upper bound of
  its confidence interval.  We also provide the centred coordinates (see
  \sct\ref{chap:sdssima}), the selected heliocentric and Virgo-infall corrected 
  redshifts and comoving, luminosity and angular diameter distances, 
  as listed in Table \ref{EFIGIfields}. We convert redshifts into distances
  using a Hubble constant $H_0=70$ km/s/Mpc \citep{freedman01}, and the 
  currently standard cosmological parameters 
  $\Omega_\mathrm{m}=0.3$ and $\Omega_\Lambda=0.7$ \citep{wmap5}.

  We then provide the various PGC parameters
  listed in Table \ref{PGCfields}, the HyperLeda parameters listed in
  Table \ref{Hyperledafields}, the NED parameters listed in Table
  \ref{NEDfields}, and the SDSS parameters listed in Table
  \ref{SDSSfields}.  Additional HyperLeda and NED parameters can be
  retrieved from these tables using the listed PGC identifiers, and
  additional SDSS parameters can be retrieved using the listed SDSS
  identifiers.
 
  For each EFIGI galaxy, we also provide in the {\tt
    http://www.efigi.org} website: \textit{ugriz} FITS images, a
  composited \textit{irg} ``true colour'' image in PNG format (see
  \sct\ref{chap:sdssima}), and the \textit{ugriz} PSF models (see
  \sct\ref{sec:psf}).

\begin{table} 
  \caption{EFIGI attributes}
  \label{EFIGIatt}
  \begin{tabularx}{\columnwidth}{lX}
    PGC\_name   & PGC name		\\
    T         & EFIGI morphological type \\
    Bulge\_to\_Total & Bulge-to-total ratio \\
    Arm\_Strength    & Strength of spiral arms\\
    Arm\_Curvature   & Average curvature of the spiral arms \\
    Arm\_Rotation    & Winding direction of the spiral arms \\
    Bar\_Length    & Length of central bar   \\
    Inner\_Ring   & Strength of inner ring, lens or pseudo-ring\\
    Outer\_Ring   & Strength of outer ring \\
    Pseudo\_Ring     & Type and strength of outer pseudo-ring \\
    Perturbation    & Deviation from a profile with rotational symmetry\\
    Visible\_Dust     & Strength of dust features \\
    Dust\_Dispersion  & Patchiness of dust features\\
    Flocculence      & Strength of scattered HII regions  \\
    Hot\_Spots        & Strength of regions of strong star formation, active nuclei, or stellar nuclei \\
    Inclination    & Inclination of disks or elongation of spheroids \\
    Contamination   & Severity of contamination by stars, galaxies or artifacts\\
    Multiplicity    & Abundance of neighbouring galaxies \\
  \end{tabularx}
\end{table} 
 
\begin{table} 
  \caption{EFIGI fields}
  \label{EFIGIfields}
  \begin{tabularx}{\columnwidth}{lX}
    PGC\_name   & PGC name		\\
    RA       	& Right ascension J2000 (degrees)\\
    DEC   	& Declination J2000 (degrees)\\
    z\_hel          & Selected heliocentric redshift \\
    z\_hel\_err     & Uncertainty in selected heliocentric redshift \\
    z\_dis          & Selected redshift corrected for Local Group infall into Virgo \\
    z\_dis\_err     & Uncertainty in selected redshift corrected for Local Group infall into Virgo \\
    D\_com          & Comoving distance derived from z\_dis or z\_hel (Mpc)\\
    D\_lum          & Luminosity distance derived from z\_dis or z\_hel (Mpc)\\
    D\_diam         & Transverse diameter distance derived from z\_dis or z\_hel (Mpc)\\
  \end{tabularx}
\end{table} 
 
\begin{table}
  \caption{PGC fields}
  \label{PGCfields} 
  \begin{tabularx}{\columnwidth}{lX}
    PGC\_name   & PGC name		\\
    T\_PGC      & RC3 morphological type\\
    e\_T\_PGC	& Mean error on T\_PGC\\
    type	& Expanded morphological type			\\
    D25 & $\log{D_{25}}$, decimal logarithm of mean apparent major
    isophotal diameter measured at or reduced to surface brightness
    level $\mu_B = 25.0\ \mbox{mag.arcsec}^{-2}$ (in units of 0.1
    arcmin)\\
    R25 & $\log{R_{25}}$, decimal logarithm of ratio of mean
    major isophotal diameter, $D_{25}$, to mean minor isophotal
    diameter measured at or
    reduced to the surface brightness level $\mu_B = 25.0\ \mbox{mag.arcsec}^{-2}$\\
    PA		 & Position angle of major axis (degrees)\\
    B\_T\_mag	 & Total B magnitude\\
    e\_B\_T\_mag & Mean error on B\_T\_mag\\
    B\_V\_T	 & Total (B-V)\\
    e\_B\_V\_T	 & Mean error on total (B-V)\\
    cz		 & Heliocentric velocity (km/s)\\
    z            & Redshift  \\
  \end{tabularx}
\end{table}

\begin{table}
  \caption{\label{Hyperledafields}HyperLeda fields}
  \begin{tabularx}{\columnwidth}{lX}
      PGC\_name	        & PGC name		\\
      PGC\_no	        & PGC number		\\
      vrad		& Heliocentric radial velocity from radio measurement (km/s)\\
      e\_vrad		& Actual error on vrad (km/s)\\
      vopt		& Heliocentric radial velocity from optical measurement (km/s)\\
      e\_vopt		& Actual error on vopt (km/s)\\
      v		        & Mean heliocentric radial velocity (km/s)\\
      e\_v		& Actual error on v (km/s)\\
      vvir		& Radial velocity corrected for Local Group infall into Virgo (km/s)\\
      zvir		& Redshift corrected for Local Group infall into Virgo \\
      z\_err            & Redshit error derived from e\_v \\
      type              & Morphological type \\
      objname		& Principal designation\\
      hl\_names	        & List of all object names\\
  \end{tabularx}
\end{table}

 \begin{table}
   \caption{\label{NEDfields}NED fields}
  \begin{tabularx}{\columnwidth}{lX}
    PGC\_name   & PGC name		\\
    cz	        & Heliocentric velocity (km/s) \\
    redshift 	& Redshift\\
    nedname     & Object name 		\\
   \end{tabularx}
\end{table}

\begin{table}
  \caption{\label{SDSSfields}SDSS fields}
  \begin{tabularx}{\columnwidth}{lX}
      PGC\_name	        & PGC name		\\
      objID		& Unique SDSS photometric identifier composed from skyVersion, rerun, run, camcol, field, obj\\
      specObjID	        & Unique SDSS spectroscopic identifier\\
      z  	        & Redshift \\
      zErr  	        & Uncertainty in redshift \\
      zConf             & Confidence level in redshift \\
  \end{tabularx}
\end{table}

\end{document}